\documentclass[11pt]{article}
\usepackage[hmargin=1in,vmargin=1in]{geometry}
\usepackage{hchang}

\usepackage{cleveref}
\usepackage{subfigure}
\usepackage{utfsym}  

\makeatletter
\renewcommand*{\@fnsymbol}[1]{\ensuremath{\ifcase#1\or *\or \dagger\or \ddagger\or
   \mathsection\or \mathparagraph\or \|\or **\or \dagger\dagger
   \or \ddagger\ddagger \else\@ctrerr\fi}}
\makeatother

\usepackage{epstopdf}
\usepackage[sans]{dsfont}
\usepackage{tcolorbox}
\usepackage{multirow}
\usepackage{xspace}
\tcbuselibrary{skins,breakable}
\tcbset{enhanced jigsaw}

\newtheorem{theorem}{Theorem}[section]
\newtheorem{heuristic algorithm}{Heuristic Algorithm}
\newtheorem{lemma}[theorem]{Lemma}

\newtheorem{claim}[theorem]{Claim}

\newtheorem{observation}[theorem]{Observation}

\usepackage[color=red, backgroundcolor=TODOcolor, textcolor=red, linecolor=lightgray, bordercolor=red, textsize=scriptsize]{todonotes}

\makeatletter\@addtoreset{section}{part}\makeatother%

\begin{document}
\lineskip=0pt

\newcommand{\algline}{
	\rule{0.5\linewidth}{.1pt}\hspace{\fill}%
	\par\nointerlineskip \vspace{.1pt}
}
\newenvironment{tbox}{\begin{tcolorbox}[
		enlarge top by=5pt,
		enlarge bottom by=5pt,
		breakable,
		boxsep=0pt,
		left=4pt,
		right=4pt,
		top=10pt,
		boxrule=1pt,toprule=1pt,
		colback=white,
		arc=-1pt,
		]
	}
	{\end{tcolorbox}}



\newenvironment{properties}[2][0]
{\renewcommand{\theenumi}{#2\arabic{enumi}}
	\begin{enumerate} \setcounter{enumi}{#1}}{\end{enumerate}\renewcommand{\theenumi}{\arabic{enumi}}}

\newif\ifnocomments
\nocommentstrue


\ifnocomments

\newcommand{\znote}[1]{}

\else
\newcommand{\znote}[1]{\textcolor{red}{\sc{[ZT: #1]}}}

\fi

\ifnocomments

\newcommand{\snote}[1]{}

\else
\newcommand{\snote}[1]{\textcolor{red}{\sc{[SK: #1]}}}

\fi


\newcommand{\tG}{\textbf{G}}
\newcommand{\tH}{\textbf{H}}
\newcommand{\tE}{\textbf{E}'}
\newcommand{\tC}{\textbf{C}}
\newcommand{\tphi}{\bm{\phi}}
\newcommand{\tpsi}{\bm{\psi}}
\newcommand{\tSigma}{\bm{\Sigma}}
\newcommand{\tB}{\tilde B}
\newcommand{\dout}{D_{\mbox{\tiny{out}}}}
\newcommand{\notF}{\overline{F}}
\newcommand{\St}{Steiner Tree\xspace}
\newcommand{\ST}{Steiner Tree\xspace}

\renewcommand{\P}{\mbox{\sf P}}
\newcommand{\NP}{\mbox{\sf NP}}
\newcommand{\PCP}{\mbox{\sf PCP}}
\newcommand{\ZPP}{\mbox{\sf ZPP}}
\newcommand{\DTIME}{\mbox{\sf DTIME}}
\newcommand{\opt}{\mathsf{OPT}}
\newcommand{\optcro}{\mathsf{OPT}_{\mathsf{cr}}}
\newcommand{\optcrors}{\mathsf{OPT}_{\mathsf{cnwrs}}}
\newcommand{\sse}{\subseteq}
\newcommand{\B}{{\mathcal{B}}}
\newcommand{\tset}{{\mathcal T}}
\newcommand{\uset}{{\mathcal U}}
\newcommand{\iset}{{\mathcal{I}}}
\newcommand{\pset}{{\mathcal{P}}}
\newcommand{\nset}{{\mathcal{N}}}
\newcommand{\dset}{{\mathcal{D}}}
\newcommand{\tpset}{\tilde{\mathcal{P}}}
\newcommand{\qset}{{\mathcal{Q}}}
\newcommand{\tqset}{\tilde{\mathcal{Q}}}
\newcommand{\lset}{{\mathcal{L}}}
\newcommand{\bset}{{\mathcal{B}}}
\newcommand{\tbset}{\tilde{\mathcal{B}}}
\newcommand{\aset}{{\mathcal{A}}}
\newcommand{\cset}{{\mathcal{C}}}
\newcommand{\fset}{{\mathcal{F}}}
\newcommand{\mset}{{\mathcal M}}
\newcommand{\jset}{{\mathcal{J}}}
\newcommand{\xset}{{\mathcal{X}}}
\newcommand{\wset}{{\mathcal{W}}}
\newcommand{\gset}{{\mathcal{G}}}
\newcommand{\oset}{{\mathcal{O}}}
\newcommand{\yset}{{\mathcal{Y}}}
\newcommand{\rset}{{\mathcal{R}}}
\newcommand{\I}{{\mathcal I}}
\newcommand{\hset}{{\mathcal{H}}}
\newcommand{\sset}{{\mathcal{S}}}
\newcommand{\zset}{{\mathcal{Z}}}
\newcommand{\notu}{\overline U}
\newcommand{\nots}{\overline S}
\newcommand{\eint}{E^{\tiny\mbox{int}}}
\newcommand{\event}{{\cal{E}}}

\newcommand{\marcon}{{\mathsf{MC}}}
\newcommand{\cov}{{\mathsf{cov}}}
\newcommand{\mst}{{\mathsf{MST}}}
\newcommand{\coi}{{\mathsf{COI}}}
\newcommand{\setcover}{{\textnormal{\sf SC}}}
\newcommand{\algsetcover}{{\textnormal{\sf AlgSetCover}}}
\newcommand{\stcost}{{\mathsf{ST}}}

\newcommand{\cover}{\textsf{cover}}
\newcommand{\eps}{\varepsilon}
\newcommand{\bfs}{\textnormal{\textsf{BFS}}}
\newcommand{\pbfs}{\textnormal{\textsf{BFS}}}
\newcommand{\lv}{\textsf{lv}}
\newcommand{\tsp}{\mathsf{TSP}}
\newcommand{\gtsp}{\textsf{GTSP}}
\newcommand{\ebt}{\tset}
\newcommand{\eb}{\textsf{EB}}
\newcommand{\optmst}{\textsf{MST}}
\newcommand{\defi}{\textsf{def}}
\newcommand{\ord}{\textsf{ord}}
\newcommand{\rc}{\textnormal{\textsf{rc}}}
\newcommand{\dist}{\textnormal{\textsf{dist}}}
\newcommand{\cost}{\textnormal{\textsf{cost}}}
\newcommand{\bw}{\textsf{bw}}
\newcommand{\local}{\textsf{Local}}
\newcommand{\pseudo}{\textsf{Pseudo-IP}}
\newcommand{\vin}{v^{\textnormal{\textsf{in}}}}
\newcommand{\vout}{v^{\textnormal{\textsf{out}}}}
\newcommand{\expect}{\mathbb{E}}
\newcommand{\proover}{\pi_{\textsf{Overwrite}}}
\newcommand{\promst}{\pi_{\textsf{MST}}}
\newcommand{\protsp}{\pi_{\textsf{TSP}}}
\newcommand{\mstest}{\textsf{MST}_{\textsf{apx}}}
\newcommand{\tspest}{\textsf{TSP}_{\textsf{apx}}}
\newcommand{\proind}{\pi_{\textsf{Index}}}
\newcommand{\ind}{\textsf{Index}}
\newcommand{\distIND}{\mathcal{D}_{\textsf{Index}}}
\newcommand{\distMST}{\mathcal{D}_{\textsf{MST}}}
\newcommand{\ic}{\textnormal{\textsf{IC}}}
\newcommand{\cc}{\textnormal{\textsf{CC}}}
\newcommand{\tvd}[2]{\ensuremath{\Delta_{\textnormal{\texttt{TV}}}(#1,#2)}}
\newcommand{\dkl}[2]{\ensuremath{D_{\textnormal{\textsf{KL}}}(#1 \| #2)}}
\newcommand{\pr}{\textsf{Pr}}

\newcommand{\hel}{h}
\newcommand{\II}{I}
\newcommand{\HH}{H}

\newcommand{\RV}[1]{\mathbf{#1}}
\newcommand{\prot}{\ensuremath{\Pi}}
\newcommand{\Prot}{\ensuremath{\Pi}}
\newcommand{\findmiss}{\sf{FindBit}}
\newcommand{\overwrite}{\sf{Overwrite}}
\newcommand{\distfind}{\mathcal{D}_{\textsf{FindBit}}}
\newcommand{\distover}{\mathcal{D}_{\textsf{Overwrite}}}
\newcommand{\temp}{\textsf{temp}}
\newcommand{\IA}{\textsf{IA}}
\newcommand{\IB}{\textsf{IB}}

\newcommand{\row}{\textsf{Row}}
\newcommand{\col}{\textsf{Col}}
\newcommand{\alg}{\ensuremath{\mathsf{Alg}}\xspace}

\newcommand{\opttsp}{\textnormal{\textsf{TSP}}}

\newcommand{\sep}{\sf{sep}}
\newcommand{\core}{\sf{core}}
\newcommand{\scut}{\sf{Shortcut}}
\newcommand{\adv}{\mathsf{adv}}
\newcommand{\lig}{\sf{light}}
\newcommand{\maxmat}{\mathsf{MM}}
\newcommand{\midd}{\mathsf{mid}}
\newcommand{\bottom}{\mathsf{bot}}
\newcommand{\topp}{\mathsf{top}}
\newcommand{\snfl}{tree\xspace}
\newcommand{\snfls}{trees\xspace}
\newcommand{\inn}{\sf in}
\newcommand{\wD}{w_{\downarrow}}
\newcommand{\wU}{w_{\uparrow}}
\newcommand{\walkcost}{\mathsf{MWC}}

\newcommand{\cutpath}{\textsc{Split}}
\newcommand{\gluepath}{\textsc{Glue}}
\newcommand{\basesizebound}{100(\log r)^2}
\newcommand{\algmerge}{\textsc{Combine}}

\renewcommand{\tau}{{\mathcal T}}

\newcommand{\bd}{{\partial\!}}

\begin{titlepage}
	
  \title{Near-Linear $\eps$-Emulators for Planar Graphs%
    \footnote{This is the full version of the paper ``Almost-Linear $\eps$-Emulators for Planar Graphs'' that appears in STOC 2022.  As indicated in the title change, the main difference is that the emulator size's dependence on $k$ is improved here from $k^{1+o(1)}$ to $k\log^{O(1)}k$. }
  }
	
	\author{%
		Hsien-Chih Chang\thanks{Department of Computer Science, Dartmouth College. Email: {\tt hsien-chih.chang@dartmouth.edu}.  Supported in part by the startup fund at Dartmouth College.}  \and 
		Robert Krauthgamer\thanks{Weizmann Institute of Science. Work partially supported by ONR Award N00014-18-1-2364, the Israel Science Foundation grant \#1086/18, the Weizmann Data Science Research Center, and a Minerva Foundation grant. 
			Email: {\tt robert.krauthgamer@weizmann.ac.il}. }  \and
		Zihan Tan\thanks{Computer Science Department, University of Chicago. Email: {\tt zihantan@uchicago.edu}. Supported in part by NSF grant CCF-2006464.}
	}
	
	\date{\today}
	
	\maketitle
	
	\thispagestyle{empty}
	
	\begin{abstract}
		We study vertex sparsification for distances, 
		in the setting of planar graphs with distortion: 
		Given a planar graph $G$ (with edge weights)
		and a subset of $k$ terminal vertices,
		the goal is to construct an \emph{$\varepsilon$-emulator},
		which is a small planar graph $G'$ that contains the terminals and
		preserves the distances between the terminals up to factor $1+\varepsilon$.
		
		We construct the first $\varepsilon$-emulators for planar graphs
		of near-linear size $\tilde O(k/\eps^{O(1)})$. 
		In terms of $k$, this is a dramatic improvement
		over the previous quadratic upper bound 
		of Cheung, Goranci and Henzinger, 
		and breaks below known quadratic lower bounds for exact emulators (the case when $\varepsilon=0$).
		Moreover, our emulators can be computed in (near-)linear time,
		which lead to fast $(1+\varepsilon)$-approximation algorithms for basic optimization problems on planar graphs, including multiple-source shortest paths, minimum $(s,t)$-cut, graph diameter, and offline dynamic distace oracle.
	\end{abstract}
\end{titlepage}

\tableofcontents

\newpage

\section{Introduction}

Graph compression describes a paradigm of transforming a large graph $G$
to a smaller graph $G'$ that preserves, perhaps approximately,
certain graph features such as distances or cut values.
The algorithmic utility of graph compression is apparent ---
the compressed graph $G'$ may be computed as a preprocessing step,
reducing computational resources for subsequent processing and queries. 
This general paradigm covers famous examples
like spanners, Gomory-Hu trees, and cut/flow/spectral edge-sparsifiers,
in which case $G'$ has the same vertex set as~$G$, but fewer edges. 
Sometimes the compression is non-graphical
and comprises of a small data structure instead of a graph $G'$; famous examples are distance oracles and distance labeling.

We study another well-known genre of compression, called \emph{vertex sparsification},
whose goal is for $G'$ to have a small vertex set.
In this setting, the input graph $G$ has a collection of $k$ designated vertices $T$, called the \emph{terminals}.
The compressed graph $G'$ should contain, besides the terminals in~$T$, a small number of vertices and preserve a certain feature among the terminals.
Specifically, we are interested in preserving the distances between terminals
up to multiplicative factor $\alpha\ge 1$ 
in an edge-weighted graph (where the weights are interpreted as lengths).
Formally, given a graph $G$ with terminals $T\subseteq V(G)$,
an \emph{emulator} for $G$ with \emph{distortion} $\alpha\ge 1$
is a graph $G'$ that contains the terminals, i.e., $T\subseteq V(G')$,
satisfying
\begin{equation} \label{eq:distortion}
\forall x,y\in T,
\quad
\dist_{G}(x,y)  
\leq \dist_{G'}(x,y)  
\leq \alpha\cdot \dist_G(x,y) ,
\end{equation}
where $\dist_G$ denotes the shortest-path distance in $G$ (and similarly for $G'$).
In the important case when $\alpha =  1+\eps = e^{\Theta(\eps)}$ for $0\le\eps\le1$, 
we simply say $G'$ is an \emph{$\eps$-emulator}.%
\footnote{Our definition in Section~\ref{sec:prelim}
	differs slightly (allowing two-sided errors),
	affecting our results only in some hidden~constants.
}
Notice that $G'$ need not be a subgraph or a minor of $G$
(in such two settings $G'$ is known as a \emph{spanner} and a \emph{distance-approximating~minor}).

We focus on the case where $G$ is known to be planar,
and thus require also $G'$ to be planar
(which excludes the trivial solution of a complete graph on $T$).
This requirement is natural and also important for applications,
where fast algorithms for planar graphs can be run on $G'$ instead of on $G$.
Such a requirement that $G'$ has structural similarity to $G$
is usually formalized by assuming that both $G$ and $G'$ belong to $\mathcal{F}$
for a fixed graph family $\mathcal{F}$ (e.g., all planar graphs).
If $\mathcal{F}$ is a minor-closed family,
one can further impose the stronger requirement that $G'$ is a minor of $G$,
and this clearly implies that $G'$ is in $\mathcal{F}$.

Vertex sparsifiers commonly exhibit a tradeoff between accuracy and size,
which in our case of an emulator $G'$,
are the distortion $\alpha$ and the number of vertices of $G'$.
Let us briefly overview the known bounds for planar graphs.
At one extreme of this tradeoff we have the ``exact'' case,
where distortion is fixed to $\alpha=1$
and we wish to bound the (worst-case) size of the emulator $G'$ 
\cite{CGH16,CGMW18,GHP20}. 
For planar graphs, the known size bounds are $O(k^4)$~\cite{KNZ14} and $\Omega(k^2)$~\cite{KZ12,co-pemm-2020}.%
\footnote{For fixed distortion $\alpha=1$, 
	every graph $G$ in fact
	admits a minor of size $O(k^4)$ \cite{KNZ14}, 
	but for some planar graphs (specifically grids) 
	every minor \cite{KNZ14} or just planar emulator \cite{KZ12,co-pemm-2020}
	must have $\Omega(k^2)$ vertices.
}
At the other extreme,
we fix the emulator size to $|V(G')|=k$, i.e., zero non-terminals,
and we wish to bound the (worst-case) distortion~$\alpha$
\cite{BG08,CXKR06,KKN15,Cheung18,FKT19}.
For planar graphs, the known distortion bounds are $O(\log k)$~\cite{Filtser18} 
and lower bound $2$~\cite{Gupta01}.%

Our primary interest is in minimizing the size-bound when the distortion $\alpha$ is $1+\eps$, i.e., $\eps$-emulators,
a fascinating sweet spot of the tradeoff. 
The minimal loss in accuracy is a boon for applications,
but it is usually challenging as one has to control the distortion over iterations or recursion. 
For planar graphs, the known size bounds for a distance-approximating minor 
are $\tilde{O}((k/\eps)^2)$ \cite{CGH16} 
and $\Omega(k/\eps)$ \cite{KNZ14}.
Improving the upper bound from quadratic to linear in $k$ is an outstanding question that offers a bypass to the aforementioned $\Omega(k^2)$ lower bound for exact emulators ($\alpha=1$). 
In fact, no subquadratic-size emulators for planar graphs are known to exist even when we allow the emulators to be arbitrary graphs, except for when the input is unweighted~\cite{CGMW18} or for trivial cases like trees.

\paragraph{Notation.}
Throughout the paper, we consider undirected graphs with non-negative edge weights,
and denote $n=|V(G)|$ and $k=|T|$.
A \emph{plane graph} refers to a planar graph
together with a specific embedding in the plane. 
We suppress poly-logarithmic terms by writing $\tilde{O}(t) = t\cdot\poly\log t$,
and multiplicative factors that depend on $\eps$ by writing $O_\eps(t) = O(f(\eps)\cdot t)$.
We write $\log^* t$ for the iterated logarithm of $t$.

\subsection{Main Result}
\label{sec:results}

We design the first $\eps$-emulators for planar graphs that have near-linear size;
furthermore, these emulators can be computed in near-linear time.
These two efficiency parameters
can be extremely useful,
and we indeed present a few applications in \Cref{sec:applications}. 

\begin{theorem}
	\label{thm:main}
	For every $n$-vertex planar graph $G$ with $k$ terminals and parameter $0<\eps<1$,
	there is a planar $\eps$-emulator graph $G'$
	of size $|V(G')|=\tilde O(k/\eps^{O(1)})$.
	Furthermore, such an emulator can be computed deterministically
	in time $\tilde O(n /\eps^{O(1)})$. 
\end{theorem}

The result dramatically improves over the previous $\tilde{O}((k/\eps)^2)$ upper bound
of Cheung, Goranci and Henzinger~\cite{CGH16}.
Moreover, it breaks below the aforementioned lower bound $\Omega(k^2)$ 
for exact emulators ($\alpha=1$)~\cite{KZ12,KNZ14,co-pemm-2020}.
Unsurprisingly, our result is unlikely to extend to all graphs,
because for some (bipartite) graphs,
every minor with fixed distortion $\alpha<2$ must have $\Omega(k^2)$ vertices~\cite{CGH16}. 
See \Cref{tab:PlanarEmulators} for comparison to prior work.

\begin{table}[h!]\small
\centering
\smallskip
\def\arraystretch{1.3}
\begin{tabular}{c:cc:cc}
  \multicolumn{1}{c}{Distortion} & \multicolumn{2}{c}{Size (lower/upper)} & Requirement & Reference \\ 
  \hline
  $1$ & $\Omega(k^2)$ &
  & planar & \cite{KZ12,co-pemm-2020} 
  \\
  $1$ & & $O(k^4)$
  & minor & \cite{KNZ14} 
  \\
  \hdashline
  $1+\eps$ & $\Omega(k/\eps)$ & 
  & minor & \cite{KNZ14} 
  \\
  $1+\eps$ & & $\tilde{O}((k/\eps)^2)$ 
  & minor & \cite{CGH16} 
  \\
  \rowcolor{Highlight}
  $1+\eps$ & & $\tilde{O}(k /\poly\eps)$
  & planar & Theorem~\ref{thm:main} 
  \\
  \hdashline
  $O(\log k)$ & & $k$ 
  & minor & \cite{Filtser18} 
\end{tabular}
\label{tab:PlanarEmulators}
\caption{Distance emulators for planar graphs.}
\end{table}

\subsection{Algorithmic Applications}
\label{sec:applications}

We present a few applications of our emulators
to the design of fast $(1+\eps)$-approximation algorithms for standard optimization problems on planar graphs. 

\medskip
Our first application is to construct an approximate version of the multiple-source shortest paths data structure, called \emph{$\e$-MSSP}: 
Preprocess a plane graph $G$ and a set of terminals $T$ on the outerface of $G$,
so as to quickly answer distance queries between terminal pairs
within $(1+\eps)$-approximation.
The preprocessing time of our data structure is $O_\e(n)$,
which for any fixed $\e>0$ is faster than
Klein's $O(n \log n)$-time algorithm~\cite{kle-msppg-2005}
for the exact setting when $\e=0$.
Both algorithms have the same query time $O(\log n)$. 

\begin{theorem}
	\label{Th:mssp}
	Given a parameter $0<\e<1$, an $n$-vertex plane graph $G$ with the range of edge weights bounded by $n^{O(1)}$,\footnote{Our algorithm can also handle general weights with a slightly slower $O_\e(n \poly(\log^* n))$ preprocessing time.}
	and a set of terminals $T$ all lying on the boundary of $G$ with $|T| \le O(n/\log^C n)$ for some large enough constant $C$,
	one can preprocess an $\e$-MSSP data structure on~$G$ with respect to $T$ in time $O_\e(n)$,
	that answers queries in time $O(\log n)$.
\end{theorem}

Our second application is an $O_\eps(n)$-time algorithm
to compute $(1+\e)$-approximate minimum $(s,t)$-cut in planar graphs,
which for fixed $\e>0$ is faster than
the $O(n \log\log n)$-time exact algorithm 
by Italiano, Nussbaum, Sankowski, and Wulff-Nilsen~\cite{insw-iamcm-2011}.

\begin{theorem}
	\label{Th:minimum-cut}
	Given an $n$-vertex planar graph $G$ with two distinguished vertices $s,t\in V(G)$ 
	and a parameter $0<\e<1$, 
	computing $(1+\e)$-approximate minimum $(s,t)$-cut in $G$ takes $O_\e(n)$ time.
\end{theorem}

Our third application is an $O_\eps(n \log n)$-time algorithm
to compute a $(1+\e)$-approximate diameter in planar graphs,
which for fixed $0<\e<1$ is faster than
the $O(n \log^2 n + \e^{-5} n\log n)$-time algorithm
of Chan and Skrepetos~\cite{cs-faddo-2019}
(which itself improves over Weimann and Yuster~\cite{wy-adpgl-2016}). 

\begin{theorem}
\label{Th:diameter}
Given an $n$-vertex planar graph $G$ and a parameter $0<\e<1$, 
one can compute a $(1+\e)$-approximation to its diameter
in time $O_\e(n \log n)$. 
\end{theorem}

Finally, one important open problems in the field of dynamic algorithms is the existence of efficient $(1+\e)$-approximate distance oracle on planar graphs.
Abboud and Dahlgaard~\cite{ad-pcbdp-2016a} provided an $\Omega(n^{1/2-o(1)})$ lower bound on the query and update time for such oracles in the exact setting.
Recently, Chen \etal~\cite{cgh+-fdcde-2020a} showed that if one can efficiently construct a $(1+\e)$-\emph{distance-approximating minor} of size $\tilde{O}(k)$ for a planar graph with $n$ nodes and $k$ terminals in $O(n \poly(\log n, \e^{-1}))$ time, then there is an offline dynamic $(1+\e)$-approximate distance oracle with $O(\poly\log n)$ query and update time.

Here we show that while our $\e$-emulator is not strictly a $(1+\e)$-distance-approximating minor, the same distance oracle can still be constructed.  This demonstrates that an efficient $(1+\e)$-approximate distance oracle on planar graphs exists. 

\begin{theorem}
\label{Th:dynamic-oracle}
There is an offline dynamic $(1+\e)$-approximate distance oracle for any planar graph of size $n$ with $O(\poly\log n)$ query and update time.  
\end{theorem}

\subsection{Technical Contributions}

A central technical contribution of this paper is to carry out a \emph{spread reduction} for the all-terminal-pairs shortest path problem when the input graph can be embedded in the plane and the terminals all lie on the outerface; the \emph{spread} is defined to be the ratio between the largest and the smallest distances between terminals.  
Spread reduction is a crucial preprocessing step for many optimization problems,
particularly in Euclidean spaces or on planar graphs~\cite{sa-atpgs-2012,bg-ltase-2013,KKN15,cfs-ntasc-2019,fl-ntasg-2020},
that replaces an instance with a large spread 
with one or multiple instances with a bounded spread. 
In many cases, one can reduce the spread to be at most polynomial in the input size.
However, we are not aware of previous work that achieves such a reduction 
in our context, where many pairs of distances have to be preserved all at once. 
In fact, even after considerable work we only managed to reduce the spread to be sub-exponential.

We now provide a bird-eye's view of our emulator construction.
The emulator problem on plane graphs with an arbitrary set of terminals can be reduced to the same problem on plane graphs, but with the strong restriction that all the terminals lies on a constant number of faces, known as \emph{holes} (cf.\ Section~\ref{sec: general graph}), 
using a separator decomposition that splits the number of vertices and terminals evenly; 
such a decomposition (called the \emph{$r$-division}) can be computed efficiently~\cite{fre-faspp-1987,kms-srsdp-2013}.
From there we can further slice the graph open into another plane graph with all the terminals on a single face, which without loss of generality we assume to be the outerface.
We refer to it as a \emph{one-hole instance}. 

\begin{figure}[h]
	\centering
	\subfigure[A one-hole instance, a set of paths (shown in red, green and purple curves), and portals (shown as red boxes). Slicing the instance open along these paths gives us smaller pieces.]{\scalebox{0.5}{\includegraphics[scale=0.19]{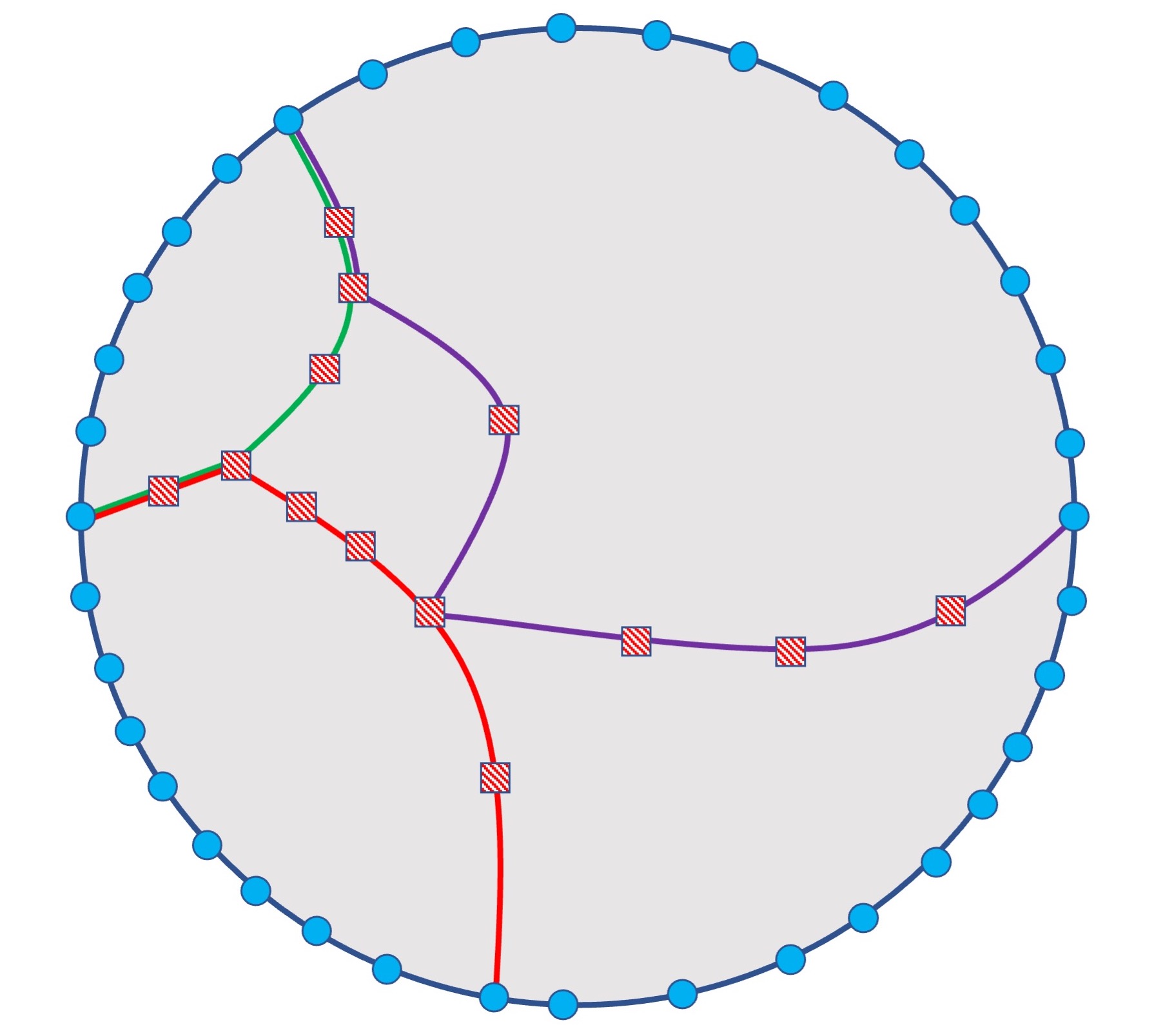}}}
	\hspace{0.4cm}
	\subfigure[The one-hole instance obtained from gluing together the emulators for the small pieces at the portals (shown as red boxes).]{\scalebox{0.5}{\includegraphics[scale=0.19]{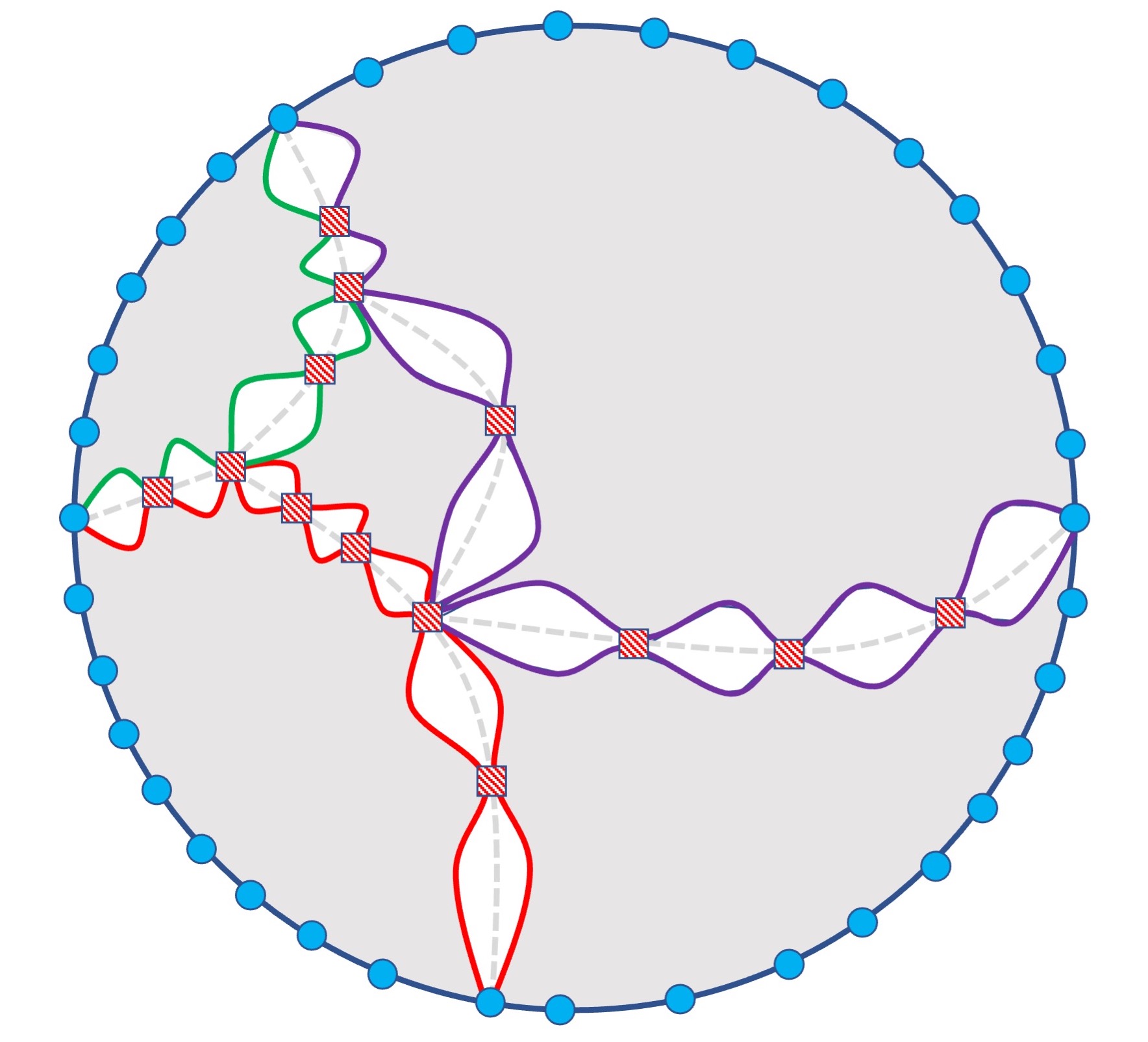}}}
	\caption{Illustration of the split-and-combine process for a one-hole instance.}
	\label{fig: intuition}
\end{figure}

To construct an emulator for a one-hole instance $G$ we adapt a recursive \emph{split-and-combine} strategy (cf.\ Section~\ref{sec: planar emulator}).
We will attempt to split the input instance into multiple one-hole instances along some shortest paths that distribute the terminals evenly (cf.\ Lemma~\ref{lem: decomposing step}).
Every time we slice the graph $G$ open along a shortest path~$P$, we compute a small collection of vertices on $P$ called the \emph{portals}, that approximately preserve the distances from terminals in $G$ to the vertices on~$P$.
The portals are duplicated during the slicing along $P$ and added to the terminal set 
(i.e., become terminals) at each piece incident to $P$, 
to ensure that further processing will (approximately) preserve their distances as well.
We emphasize that the naive idea of placing portals at equally-spaced points along $P$ 
is not sufficient, as some terminals in $G$ might be arbitrarily close to $P$.
Instead, we place portals at exponentially-increasing intervals from both ends of $P$.
After splitting the original instance into small enough pieces by recursively slicing along shortest paths and computing the portals, 
we compute exact emulators for each piece using any of the polynomial-size construction~\cite{KNZ14,co-pemm-2020}.  
Next we glue these small emulators back along the paths by identifying multiple copies of the same portal into one vertex. 
See Figure~\ref{fig: intuition}.

Let $U$ be the set of terminals in the current piece, and let $r \coloneqq |U|$.
We need the portals to be dense enough
so that only a small error term, of the form $r^{-\delta}$
(meaning that the distortion increases multiplicatively by $1+r^{-\delta}$)
will be added to the distortion of the emulator after the gluing,
as this will eventually guarantee (through more details like the stopping condition of the recursion) 
that the final distortion is $1+\eps$ 
and the final emulator size has polynomial dependency on $\e^{-1}$.
At the same time, the number of portals cannot be too large, 
as they are added to the terminal set,
causing the number of terminals per piece to go down slowly and creating too many pieces, and in the end the size of the combined emulator might be too big.
It turns out that the sweet spot is to take roughly $L_r \coloneqq r/\log^2 r$ portals.  
Calculations show that in such case the portals preserve distances up to
an additive error term $\log\Phi/L_r$,
where $\Phi$ is the \emph{spread} of the terminal distances (cf.\ Claim~\ref{clm: ratio loss for contracting to portals}).
When $\Phi \leq \smash{2^{r^{0.9}}}$,
we will get the polynomially-small $\tilde{O}(r^{-0.1})$ error term
needed for the gluing (cf.\ Section~\ref{SSS:small-spread}).
However, even when the original input has a polynomial spread to start with, in general we cannot control the spread of all the pieces occurring during the split-and-combine process, 
because portals are added to the terminal sets.
Therefore a new idea is needed.

When $\Phi > 2^{r^{0.9}}$, we need to tackle the spread directly.
We perform a \emph{hierarchical clustering} of the terminals (cf.\ Section~\ref{SSS:large-spread}).  At each level $i$, we connect two clusters of terminals from the previous level $i-1$ using an edge if their distance is at most $r^{2i}$; then we group each connected component into a single cluster.
The key to the spread reduction is the idea of \emph{expanding clusters}.  
A cluster $S$ is \emph{expanding} if its parent cluster $\hat S$ is at least \smash{$\sim\!e^{r^{-0.7}}$}-factor bigger.
Intuitively, if all clusters are expanding, then the number of levels in the hierarchical clustering must be at most $r^{0.7}$, 
and therefore the spread must be at most sub-exponential.
So in the high-spread case some non-expanding cluster must exist.
\begin{itemize}
	\item
	If such non-expanding cluster $S$ is of moderate size (that is, in between $r/5$ and $4r/5$) (cf.\ Section~\ref{SSS:balanced}),
	we construct a collection of \emph{non-crossing} shortest paths between terminals in $S$
	(non-crossing means that no two paths 
	with endpoint pairs $(s_1,s_2)$ and $(t_1,t_2)$
	have their endpoints in an interleaving order $(s_1,t_1,s_2,t_2)$ on the outerface)
	in which no two paths intersect except at their endpoints.
	Again compute portals on the paths from every terminal in $\hat S \setminus S$, 
	but now using $\e_r$-covers~\cite{tho-corad-2004} for $\e_r\coloneqq r^{-0.1}$, 
	and split along the paths to create sub-instances.
	Because the cluster is non-expanding and has moderate size, 
	the number of terminals in $\hat S \setminus S$ is at most $(e^{r^{-0.7}} - 1)|S| \le r^{0.3}$,
	and thus the number of portals is $O(r^{0.3}/\e_r) \le O(r^{0.4})$, 
	which is a gentle enough increase in the number of terminals.
	The hard part is to argue that the portals created are sufficient for the recombined instance to be an emulator.
	This can be done by observing that terminal pairs among $U\setminus \hat S$ are far apart, and similarly when one terminal is from $S$ and the other is from $U\setminus \hat S$; 
	hence only terminal pairs involving $\hat S \setminus S$ have to be dealt with using properties of $\e_r$-covers (cf.\ Claim~\ref{clm: ratio loss for gluepathset}).
	
	\item
	If there are no non-expanding clusters with moderate size (cf.\ Section~\ref{SSS:unbalanced}), 
	we find a non-expanding cluster $\tilde S$ of lowest level that contains most of the terminals, and construct a collection of non-crossing shortest paths between terminals in $\tilde S$ like the previous case.  
	However this time, after computing the $r^{-0.1}$-covers and splitting along the paths, there might be one instance containing too many terminals.
	In this case, we find \emph{every} non-expanding cluster $S$ of \emph{maximal level}; such clusters must all lie within $\tilde{O}(r^{0.7})$ levels from $\tilde S$ because we cannot have nested expanding clusters for $\tilde{O}(r^{0.7})$ consecutive levels.  
	The Monge property 
	guarantees that the shortest paths generated by the union of these maximal-level non-expanding clusters must be non-crossing because all such clusters are disjoint (cf.\ Observation~\ref{obs: sets non-crossing}).
	Now if we split the graph based on the path set generated, each resulting instance either has moderate size, or must have small spread, and we safely fall back to the earlier cases.
\end{itemize}

\paragraph{Applications.}
A widely adopted pipeline in designing efficient algorithms for distance-related optimization problems on planar graphs in recent years consists of the following steps:
\begin{enumerate}
	\item Decompose the input planar graph into small pieces each of size at most $r$ with a small number of boundary vertices and $O(1)$ holes, called an \emph{$r$-division} (see Frederickson~\cite{fre-faspp-1987} and Klein-Mozes-Sommer~\cite{kms-srsdp-2013}; 

	\item Process each piece so that all-pairs shortest paths between boundary vertices within a piece can be extracted efficiently by the \emph{multiple-source shortest paths} algorithm for planar graphs (Klein~\cite{kle-msppg-2005});

	\item Further process each piece into a \emph{compact data structure} that supports efficient min-weight-edge queries and updates (SMAWK~\cite{akm+-gama-1987}, Fakcharoenphol and Rao~\cite{fr-pgnwe-2006});
	
	\item Compute shortest paths in the original graph in a problem-specific fashion, now with each piece replaced with the compact data structure, using a \emph{modified Dijkstra algorithm} (Fakcharoenphol and Rao~\cite{fr-pgnwe-2006}).
\end{enumerate}
The conceptual role of our planar emulators is an alternative to Step~3.  
%
The reason for the development of the aforementioned machinery and complex algorithms is to get around the size lower bound in representing the all-pairs distances for the pieces.
The benefit of replacing the data structure with a single planar emulator is that the whole graph stays planar.
One can then simply replace Step 4 with the standard Dijkstra algorithm (or even better, with the $O(n)$-time algorithm for planar graphs by Henzinger~\etal~\cite{hkrs-fsapg-1997}).  
More importantly, one can \emph{recurse} on the resulting graph when appropriate,
and compress the graph further and further with small additive errors slowly accumulated (cf.\ Section~\ref{SS:bootstrapping}).
This allows us to construct near-linear-size $\e$-emulator in $O_\e(n \poly\log^* n)$ time and even $O_\e(n)$ time using a precomputed look-up table for pieces that are tiny compared to $n$ when the spread of the input graph is bounded by a polynomial, which can easily be achieved by standard spread reduction techniques for many optimization problems.

\subsection{Related Work}


In addition to emulators, there are other lines of research on graph compression preserving distance information. 
Among them the most studied objects are \emph{spanners} and \emph{preservers} (when the sparsifier is required to be a subgraph of the input graph) and \emph{distance oracles} (a data structure that reports exact or approximate distances between pairs of vertices). 
We refer the reader to the excellent survey \cite{ahmed2020graph}.

There are also rich lines of works for constructing vertex sparsifiers that preserve cut/flow values (known as \emph{cut/flow sparsifiers})
exactly \cite{HKNR98,CSWZ00,KR13,KR14,KPZ17,GHP20,KR20}
or approximately \cite{Moitra09,CLLM10,Chuzhoy12,AGK14,EGKRTT14,MM16,GR16,goranci2021expander}.




\section{Preliminaries}
\label{sec:prelim}








All logarithms are to the base of $2$.  All graphs are simple and undirected.
%
Let $G$ be a connected graph. A vertex $v\in V(G)$ is called a \EMPH{cut vertex} of $G$ if the graph $G\setminus \set{v}$ is disconnected.
The cut vertices of a plane graph $G$ can be computed in time $O(|V(G)|+|E(G)|)$.
Let $G$ be a graph with an edge-weight function
$\EMPH{$w$}\colon E(G) \to \mathbb{R}_{+}$.
The weight of a path $P$ is defined as $w(P) \coloneqq\sum_{e\in E(P)} w(e)$.
The shortest-path distance between two vertices $u$ and $v$ is denoted by \EMPH{$\dist_G(u,v)$}.
For a subset $S$ of vertices in $G$, we define
$\EMPH{$\diam_G(S)$} \coloneqq \max_{u, u'\in S}\dist_G(u,u')$.
For a pair of disjoint subsets of vertices $(S,S')$ in $G$, we define $\EMPH{$\dist_G(S,S')$} \coloneqq \min_{u\in S, u'\in S'}\dist_G(u,u')$.

\paragraph{Emulators.}
Throughout, we consider graph $G$ equipped with a special set of vertices $T$, called \EMPH{terminals}.
We refer to the pair $(G,T)$ as an \EMPH{instance}.
Let $(G,T)$ and $(H,T)$ be a pair of instances with the same set of terminals,
and let $\e\in[0,1]$.
We say that $H$ is an \EMPH{$\e$-emulator} for $G$ with respect to $T$, or equivalently, instance $(H,T)$ is an $\e$-emulator for instance $(G,T)$ if
\begin{equation} \label{eq:distortion2}
\forall x,y\in T,
\quad 
e^{-\e}\cdot\dist_G(x,y) \le
\dist_H(x,y) \le
e^{\e}\cdot\dist_G(x,y).
\end{equation}
Throughout, we use Equation~\eqref{eq:distortion2} as the definition of 
an $\e$-emulator instead of Equation~\eqref{eq:distortion};
but since we restrict our attention to $\e<1$,
the two definitions are equivalent up to scaling $\eps$ by a constant factor.
By definition, if $(H,T)$ is an $\e$-emulator for $(G,T)$, then $(G,T)$ is also an $\e$-emulator for $(H,T)$.
Moreover, 
if $(G,T)$ is an $\eps$-emulator for $(G',T)$ and $(G',T)$ is an $\eps'$-emulator for $(G'',T)$, then $(G,T)$ is an $(\eps+\eps')$-emulator for $(G'',T)$.


Most instance $(G,T)$ considered in this paper are \EMPH{planar instances} where graph $G$ is a connected plane graph. 
We say that a planar instance $(G,T)$ is an \EMPH{$h$-hole instance} for an integer $h>0$
if the terminals lie on at most $h$ faces in the embedding of $G$.
The faces incident to some terminals are called \EMPH{holes}.
Notice that in a one-hole instance $(G,T)$, we can safely assume all the terminals in $T$ lie on the outerface $G$. 
By definition, a $0$-emulator preserves distances exactly, 
i.e., $\dist_G(x,y)=\dist_{G'}(x,y)$ for all $x,y\in T$.

\begin{theorem}[Chang-Ophelders~{\cite[Theorem~1]{co-pemm-2020}}]
	\label{thm: quartergrid enumator}
	Given one-hole instance $(G,T)$ with $n \coloneqq |V(G)|$ and $k \coloneqq |T|$, 
	one can compute a $0$-emulator $(G',T)$ for $(G,T)$ of size $|V(G')| \le k^2$. 
	The running time of the algorithm is $O((n+k^2)\log n)$.
\end{theorem}

\paragraph{Crossing pairs and the Monge property.}
Let $(G,T)$ be a one-hole instance.
Assume that no terminal in $T$ is a cut vertex of $G$, every terminal appears exactly once as we traverse the boundary of the outerface.
Let $(t_1,t_2), (t'_1,t'_2)$ be two terminal pairs
whose four terminals are all distinct.
We say that the pairs $(t_1,t_2),(t'_1,t'_2)$ are \EMPH{crossing} if the clockwise order in which these terminals appear on the boundary is either $(t_1,t'_1,t_2,t'_2)$ or $(t_1,t'_2,t_2,t'_1)$; otherwise we say that they are \EMPH{non-crossing}. 
A collection $\mset$ of pairs of terminals in $T$ is called \emph{non-crossing}
if every two pairs in $\mset$ is non-crossing.
Sometimes we abuse the language and say that a set of shortest paths $\pset$ in $G$ is \emph{non-crossing} when the collection of endpoint pairs for the paths is non-crossing.
The \EMPH{Monge property}%
\footnote{Technically, this is known as the \EMPH{cyclic Monge property}~\cite{co-pemm-2020}. }
states that, for every one-hole instance $(G,T)$ and every crossing pairs of terminals $(t_1,t_2)$ and $(t'_1,t'_2)$, 
\[
\dist_G(t_1,t_2)+\dist_G(t'_1,t'_2) \ge \dist_G(t'_1,t_2)+\dist_G(t_1,t'_2).
\]

\paragraph{Well-structured sets of shortest paths.}
Consider a graph $G$ and a collection $\pset$ of shortest paths in $G$. 
We say that the set $\pset$ is \EMPH{well-structured} if for every pair of paths $(P,P')$ in $\pset$, $P\cap P'$ is a single subpath of both $P$ and $P'$. 
It is not hard to see that every collection of shortest paths in~$G$ is well-structured if the shortest path between any two vertices in $G$ is unique.
Such condition can be enforced with high probability if we perturb the edge-weights in $G$ slightly and apply the \emph{isolation lemma}~\cite{mvv-memi-1987}.
If randomization is to be avoided, one can use a \emph{lexicographic perturbation} by redefining the edge weights to be a vector~\cite{cha-odlp-1952,dow-gsmml-1955,hm-amcpm-1994}, or the \emph{leftmost rule} when choosing a shortest path~\cite{ek-lamfm-2013} when $G$ is a plane graph.
A deterministic lexicographic perturbation scheme that guarantees the uniqueness of shortest paths in an $n$-vertex plane graph can be computed in $O(n)$ time~\cite{efl-hmpfs-2018}.
Therefore from here on we assume that all the planar graphs we consider have unique shortest path between every pair of vertices, and every collection of shortest paths is well-structured.
The proof of the following lemma is provided in Appendix~\ref{apd: Proof of well-structured path set}.

\begin{lemma}
	\label{lem: well-structured path set}
	Given a one-hole instance $(G,T)$ and a non-crossing collection $\mset$ of pairs of terminals in $T$, one can compute a well-structured set $\pset$ of shortest paths, one for each pair of terminals in $T$ in $O(|E(G)|\cdot \log|\mset|)$ time.
\end{lemma}

\paragraph{\boldmath{$\eps$}-covers.}
We use the notion of \emph{$\eps$-covers}~\cite{ks-fdass-1998,tho-corad-2004}. 
Let $\e \in (0,1)$ be a parameter.
Let $G$ be a graph and let $P$ be a shortest path in $G$ connecting some pair of vertices. Consider now a vertex $v$ in $G$ that does not belong to path $P$. 
An \EMPH{$\eps$-cover} of $v$ on $P$ is a subset $S$ of vertices in $P$ such that, for each vertex $x\in V(P)$, taking the detour from $v$ to some $y\in S$ then to~$x$ is a $(1+\e)$-approximation to the shortest path from $v$ to $x$,
i.e., there exists $y\in S$ for which
$\dist_G(v,y)+\dist_G(y,x)\le (1+\eps)\cdot \dist_G(v,x)$.
Small $\eps$-cover of size $O(1/\eps)$ is known to exist.
\begin{theorem}[%
	Thorup~{\cite[Lemma~3.4]{tho-corad-2004}}]
	\label{thm: eps_cover}
	Let $\e \in (0,1)$ be a constant.
	For every shortest path $P$ in some graph $G$ and every vertex $v\notin P$, there is an $\eps$-cover of $v$ on $P$ with size $O(1/\eps)$. Moreover, such an $\eps$-cover can be computed in $O(|E(G)|)$ time.
\end{theorem}
%
%
We emphasize that choosing $O(1/\e)$ ``portals'' at equal distance on the path $P$ as in Klein-Subramanian~{\cite{ks-fdass-1998}} is not sufficient, because the distance from $v$ to $P$ might be much smaller than the length of $P$.
The linear-time construction is not stated in Lemma 3.4 of~\cite{tho-corad-2004},
but it can be inferred from their proof. 
In fact, we will use the following construction that allows us to efficiently compute the union of $\e$-covers of a subset $Y$ of vertices along the boundary of plane graph; the proof is a simple divide-and-conquer similar to Reif~\cite{rei-mscpu-1981}, which we omit here.



\begin{lemma}
	\label{lem: eps_cover_subset}
	Let $\e \in (0,1)$ be a constant and $G$ is a plane graph.
	Given a subset $Y$ of vertices that lie on the same face of $G$ and a shortest path $P$ connecting a pair of vertices in $G$, 
	we can compute the union of $\e$-covers of each vertex in $Y$ on $P$ in $O(|E(G)| \cdot \log |Y|)$ time.
\end{lemma}



\section{Emulators for One-Hole Instances}
\label{sec: planar emulator}

In this section and the next one we design a near-linear time algorithm for constructing $\eps$-emulators for one-hole instances,
as stated in Theorem~\ref{Th:emulator-1hole}. 
We say that an $\e$-emulator $(G',T)$ for a one-hole instance $(G,T)$
is \EMPH{aligned} if $(G',T)$ is also a one-hole instance,
and the circular orderings of the terminals on the outerfaces of $G$ and of $G'$ are identical.

\begin{theorem}
\label{Th:emulator-1hole}
Given a parameter $\e \in (0,1)$ and a one-hole instance $(G,T)$ with $|T|=k$, one can compute an aligned $\e$-emulator for $(G,T)$ of size $|V(G')|=\tilde O(k/\eps^{O(1)})$
in $\tilde O\big((n+k^2)/\eps^{O(1)}\big)$~time.
\end{theorem}

We complement the upper bound in Theorem~\ref{Th:emulator-1hole} with an $\Omega(k/\eps)$ lower bound on the size of aligned $\eps$-emulators for one-hole instances.
This lower bound generalizes the $\Omega(k/\eps)$ lower bound of~\cite{KNZ14},
which holds for one-hole instances too, 
but only when the emulator is a minor of $G$ (and is thus clearly an aligned emulator).

\begin{theorem}
\label{thm: one-hole lower bound}
For every $k\ge 2$ and $(4/k)<\eps <1$,
there is a one-hole instance $(G,T)$ with $|T|=k$, 
such that every aligned $\eps$-emulator $(G',T)$ for $(G,T)$ 
must have size $\Omega(k/\eps)$.
\end{theorem}

All emulators we consider are aligned and therefore we omit the word ``aligned'' from now on.
We describe the algorithm and proof for Theorem~\ref{Th:emulator-1hole} in Section~\ref{sec: 1hole_algo}, with the help of the core decomposition lemma (cf.\ Lemma~\ref{lem: decomposing step}).
The proof to Lemma~\ref{lem: decomposing step} itself is deferred to Section~\ref{sec: Proof of decomposing step}.
%
The proof of Theorem~\ref{thm: one-hole lower bound} is provided in \Cref{apd: Proof of one-hole lower bound}, since it is not relevant to the proof of Theorem~\ref{thm:main}.

\subsection{The Algorithm and its Analysis}
\label{sec: 1hole_algo}


Let $(G,T)$ be the input one-hole instance. 
The algorithm for Theorem~\ref{Th:emulator-1hole} consists of two stages. 
In the first stage, we iteratively decomposes $(G,T)$ into smaller one-hole instances; and  in the second stage, we compute emulators for these small instances and then combines them into an emulator for $(G,T)$.

Throughout the algorithm we maintain a collection \EMPH{$\hset$} of one-hole instances, that is initialized to be $\hset=\set{(G,T)}$. 
Set $\EMPH{$\lambda^*$} \coloneqq c^*\log^2 k/\eps^{20}$, where $k \coloneqq |T|$ and $c^*>0$ is a large enough constant.
In the first stage, we repeatedly replace a one-hole instance $(H,U)\in \hset$ where $|U|>\lambda^*$ with smaller one-hole instances obtained by applying the algorithm from Lemma~\ref{lem: decomposing step} to $(H,U)$, until every one-hole instance $(H,U)$ in $\hset$ satisfy $|U|\le \lambda^*$.
The core of our construction is the following lemma. 

\begin{lemma}
\label{lem: decomposing step}
Given one-hole instance $(H,U)$ with $r \coloneqq |U|$,
one can compute a collection of one-hole instances $\set{(H_1,U_1), \ldots, (H_s,U_s)}$, such that
\begin{itemize}
    \item $U\subseteq \big(\bigcup_{1\le i\le s} U_i\big)$;
    \item $|U_i|\le 9r/10$ for each $1\le i\le s$; 
    \item $\sum_{1\le i\le s} |U_i| \le O(r)$; and
    \item for any parameter $100 < \lambda \le \log^2 r$, $\sum_{i: |U_i|> \lambda} |U_i| \le r\cdot\big(1+O(1/\lambda)\big)$.
\end{itemize}
Moreover, given an $\e$-emulator $(Z_i,U_i)$ for each $(H_i,U_i)$, 
algorithm $\algmerge$ computes for $(H,U)$ 
an \smash{$\big(\e+O(\frac{\log^4 r}{r^{0.1}})\big)$}-emulator $(Z,U)$ of size
\(
|V(Z)|\le \sum_{1\le i\le s} |V(Z_i)|.
\)
The running time of both algorithms is at most 
$O\big( (|V(H)|+r^2)\cdot\log r \cdot \log |V(H)| \big)$.
\end{lemma}
We prove this lemma in Section~\ref{sec: Proof of decomposing step},
and in the remainder of this subsection we use it to complete the proof of Theorem~\ref{Th:emulator-1hole}.




\bigskip
We associate with the decomposition process a \EMPH{partitioning tree~$\tau$}.
Its nodes are all the one-hole instances that ever appear in the collection $\hset$. Its root node is the initial one-hole instance $(G,T)$,
and every tree node $(H,U)$ has children nodes corresponding to the new instances $(H_1,U_1),\ldots,(H_s,U_s)$ generated by Lemma~\ref{lem: decomposing step}.  
The leaves of $\tau$ are those that are in $\hset$ at the end of the first stage.
(To avoid ambiguity, we refer to elements in $V(\tau)$ as \emph{nodes} and elements in $V(H)$ as \emph{vertices}.)

We now describe the second stage of the algorithm.
For each one-hole instance $(H,U)$ in $\hset$ at the end of the first stage, compute a $0$-emulator $(Z,U)$ for $(H,U)$ using the algorithm from Theorem~\ref{thm: quartergrid enumator}.%
\footnote{This step can use any $0$-emulator that has size $\poly k$
  and can be constructed in time $\tilde{O}(n+\poly k)$,
  and we conveniently use Theorem~\ref{thm: quartergrid enumator}.
}
We then iteratively process the non-leaf nodes in $\tau$ inductively in a bottom-up fashion:
Given a non-leaf node $(H,U)$ with children $(H_1,U_1),\ldots,(H_s,U_s)$,
let $(Z_i,U_i)$ be the emulator computed for $(H_i,U_i)$ by induction.
Apply algorithm $\algmerge$ from Lemma~\ref{lem: decomposing step} to the emulators $(Z_1,U_1),\ldots,(Z_s,U_s)$ to obtain an emulator $(Z,U)$ for instance $(H,U)$.
After all nodes in $\tau$ have been processed, output the emulator $(G',T)$ constructed for the root node $(G,T)$.

We proceed to show that the instance $(G',T)$ computed by the above algorithm satisfies all the properties required in Theorem~\ref{Th:emulator-1hole}.

\paragraph{Size Bound.}
We first show that $|V(G')|=\tilde O(k/\eps^{O(1)})$.
We denote by $\hset$ the collection obtained at the end of the first stage.
Note that
$|V(G')|
  \le \sum_{(H,U)\in \hset} O(|U|^2)
  \le O(\max_{(H,U)\in \hset} |U|) \cdot \sum_{(H,U)\in \hset} |U|$.
As $\max_{(H,U)\in \hset} {|U|}\le \lambda^*= O(\log^2 k/\eps^{O(1)})$,
it now suffices to bound the total number of terminals in all resulting one-hole instances in $\hset$ by $\tilde O(k/\eps^{O(1)})$, which we do next via a charging scheme.
Let $(H,U)$ be a node in $\tau$ with children $(H_1,U_1),\ldots,(H_s,U_s)$.
\begin{itemize}\itemsep=0pt
	\item
	For instances $(H_i,U_i)$ with $|U_i|\le \lambda^*$ (which will all be in $\hset$ at the end of the first stage), charge every vertex in $U_i$ to vertices in $U$.
	Since $\sum_i |U_i|\le O(|U|)$, each vertex of $U$ gets a charge of $\smash{O(1)}$ this way.  
	We call these charge \EMPH{inactive}.
	
	\item
	For instances $(H_i,U_i)$ with $|U_i|> \lambda^*$, let $U'$ be the set of all new vertices, i.e., they appear in some set $U_i$ but not in $U$; we have $|U'|\le O(|U|/\log^2 |U|)$ by Lemma~\ref{lem: decomposing step}.
	Charge every vertex in $U'$ uniformly to vertices in $U$, so each vertex gets $O(1/\log^2 |U|)$ charge. 
	We call these charge \EMPH{active}.
\end{itemize}

The total inactive charge on each vertex of $T$ is 
$O(\log k)$ because $\tau$ has height $O(\log k)$.
%
%
As for the total active charge to each vertex in $T$,
a quick calculation 
shows that it is at most
$O(1/(\log_{(10/9)} \lambda -1))\le 1/2$.
%
(For a complete proof see Appendix~\ref{apd: calculations}.)
Note that this only accounts for the \EMPH{direct} active charge. 
For example, some terminal does not belong to the initial one-hole instance $(G,T)$, that was first actively charged to the terminals in $T$, can in turn be actively charged by some other terminals later. 
We call such charge \EMPH{indirect} active charge. 
The total direct and indirect active charge for each terminal in $T$ is at most $1/2+(1/2)^2+\cdots \le 1$. 

Altogether, each terminal in $T$ is charged $O(\log k)$. Therefore, the total number of terminals in all resulting instances in $\hset$ is bounded by $O(k\log k)$,
which, combined with previous discussion, implies that $|V(G')|\le \tilde O(k/\eps^{O(1)})$.

\paragraph{Correctness.}
It remains to show that $(G',T)$ is an $\eps$-emulator for $(G,T)$. 
%
Recall that we have associated with the algorithm in first stage a partitioning tree $\tau$. 
We now define, for each tree node ${(H,U)}$, a value \EMPH{$\eps_{(H,U)}$} as follows.
If ${(H,U)}$ is a leaf node, we define $\eps_{(H,U)} \coloneqq 0$. 
Otherwise, ${(H,U)}$ is a non-leaf node with child nodes in $\tau$ be ${(H_1,U_1)},\ldots,{(H_s,U_s)}$. 
Denote $\EMPH{$r$} \coloneqq |U|$, and let $c>0$ be a large enough constant that is greater than the constants hidden in all big-O notations in Lemma~\ref{lem: decomposing step} and $c<(c^*)^{1/20}$. We define 
\[
\eps_{(H,U)} \coloneqq \frac{c\log^4 r}{r^{0.1}}+\max\set{\eps_{(H_1,U_1)},\ldots,\eps_{(H_s,U_s)}}.
\]
From the properties of the algorithm $\algmerge$, it is easy to verify that for each node ${(H,U)}$ in $\tau$, the one-hole instance $(Z,U)$ we construct is an $\eps_{(H,U)}$-emulator for $(H,U)$. 

We now show that $\eps_{(G,T)}\le \eps$. 
Observe that there are integers $r_1,\ldots,r_t$ with $r_1\le k$, $r_t\ge \lambda^*$, such that for each $1\le i\le t-1$, $r_i\ge (10/9)\cdot r_{i+1}$, $\eps_{(G,T)}=\sum_{1\le i\le t}c\log^4 r_i/(r_i^{0.1})$.
A quick calculation gives us $\eps_{(G,T)}\le c\cdot (\log \lambda^*)^{4}/(\lambda^*)^{0.1}$.
(For a complete proof see Appendix~\ref{apd: calculations}.) 
Since $c$ is a constant, and recall that $\lambda^*=c^*/\eps^{20}$ where $c^*>c^{20}$ is large enough, so $\eps_{(G,T)}\le c\cdot (\log \lambda^*)^{4}/(\lambda^*)^{0.1}<\eps$, and therefore $(G',T)$ is an  $\eps$-emulator for $(G,T)$.

\paragraph{Running Time.}
Every time we implement the algorithm from Lemma~\ref{lem: decomposing step} to split some instance in $(H,U)\in \hset$ with $n' \coloneqq |H|$ and $r \coloneqq |U|$, the running time is $O\big((n'+r^2)\log r\log n'\big)$.
We charge its running time (and also the time for $\algmerge$) to vertices in $H$ as follows: 
\begin{itemize}
\item charge the $O(n' \log r \log n')$ term uniformly to vertices in $H$ (each gets $O(\log k \log n)$ charge);
\item charge the $O(r^2\log r \log n')$ term uniformly to terminals in $U$ (each gets $O(k\log k \log n)$ charge). 
\end{itemize}
Since the depth of the partitioning tree $\tset$ is at most $O(\log k)$, each non-terminal vertex in $G$ gets in total $O(\log^2 k\log n)$ charge, and each terminal in the resulting collection $\hset$ at the end of the first stage gets in total $O(k\log^2 k\log n)$ charge. 
Therefore, the total running time of the algorithm is
\[
O(\log^2 k \log n)\cdot n + O(k\log^2 k\log n)\cdot \tilde{O}(k/\eps^{O(1)})
= \tilde{O}\big( (n+k^2) / \eps^{O(1)} \big).
\]

\section{Construct Emulator using $\cutpath$ and $\gluepath$: Proof of Lemma~\ref{lem: decomposing step}}
\label{sec: Proof of decomposing step}

In this subsection we provide the proof of Lemma~\ref{lem: decomposing step}. 
We first introduce the basic graph operations $\cutpath$ and $\gluepath$ in Section~\ref{SS:split-and-glue}.
Then we describe the algorithm and its analysis.
%

\subsection{Splitting and Gluing}
\label{SS:split-and-glue}

In this subsection we introduce building blocks for the divide-and-conquer: procedures $\cutpath$ and $\gluepath$.
We will decompose a single one-hole instance $(H,U)$ into many small one-hole instances using procedure $\cutpath$, compute emulators for each of them, and then glue the collection of small emulators together into an emulator for $(H,U)$ using procedure $\gluepath$. 
We now introduce the procedures in more detail.

\paragraph{Splitting.}
The input to procedure \EMPH{$\cutpath$} consists of
\begin{itemize}
    \item a one-hole instance $(H,U)$;
    \item a non-crossing set $\pset$ of shortest paths in $H$ connecting pairs of terminals in $U$; and 
    \item a subset $Y$ of vertices on the union of shortest paths in $\pset$; set $Y$ must contain all endpoints of paths in $\pset$ and all vertices with degree at least three in the graph $\bigcup_{P\in \pset} P$ (we call them \EMPH{branch vertices}).
\end{itemize}

\begin{figure}[h]
	\centering
	\subfigure{\scalebox{0.55}{\includegraphics[scale=0.2]{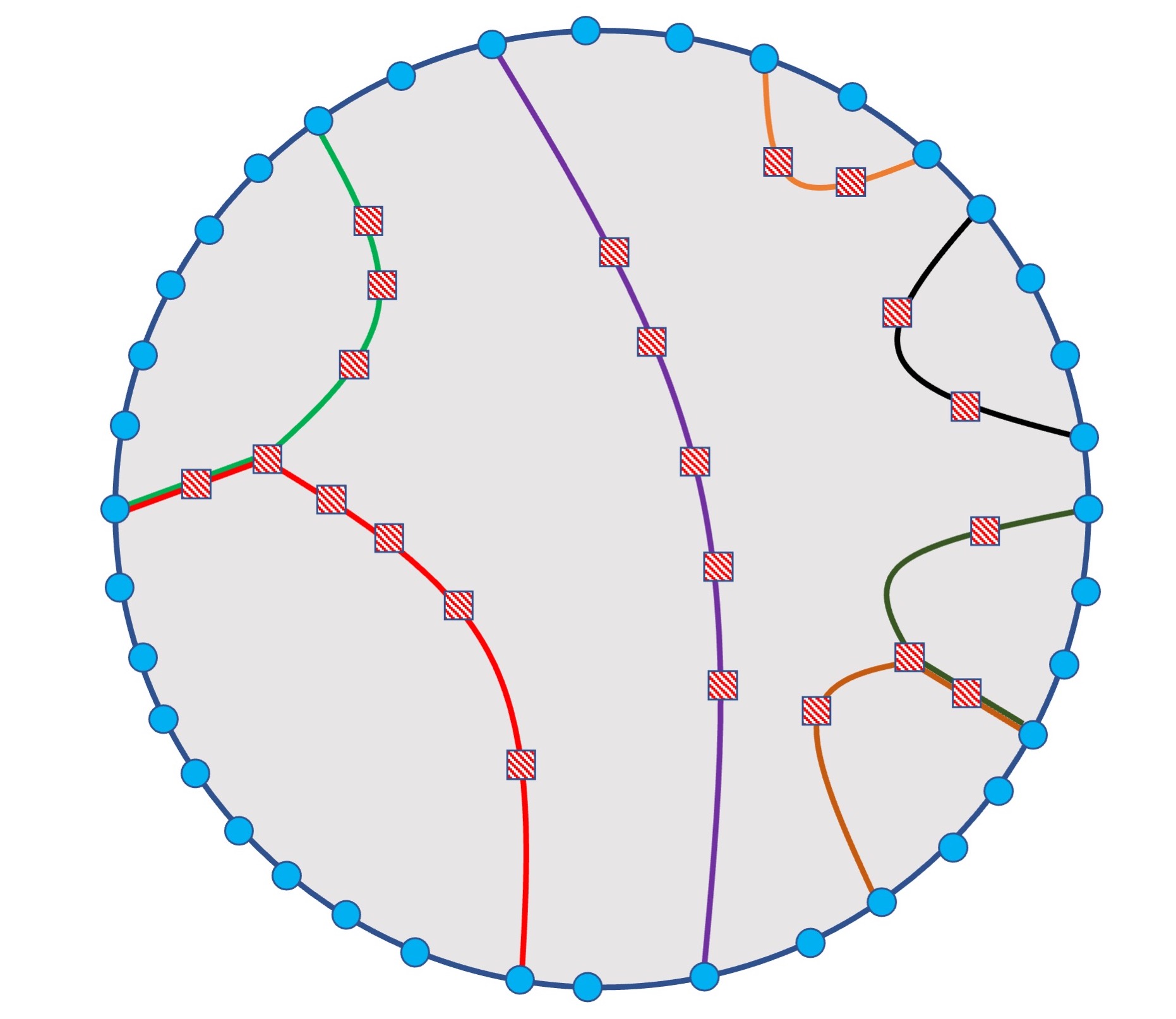}}}
	\hspace{0.7cm}
	\subfigure{\scalebox{0.4}{\includegraphics[scale=0.4]{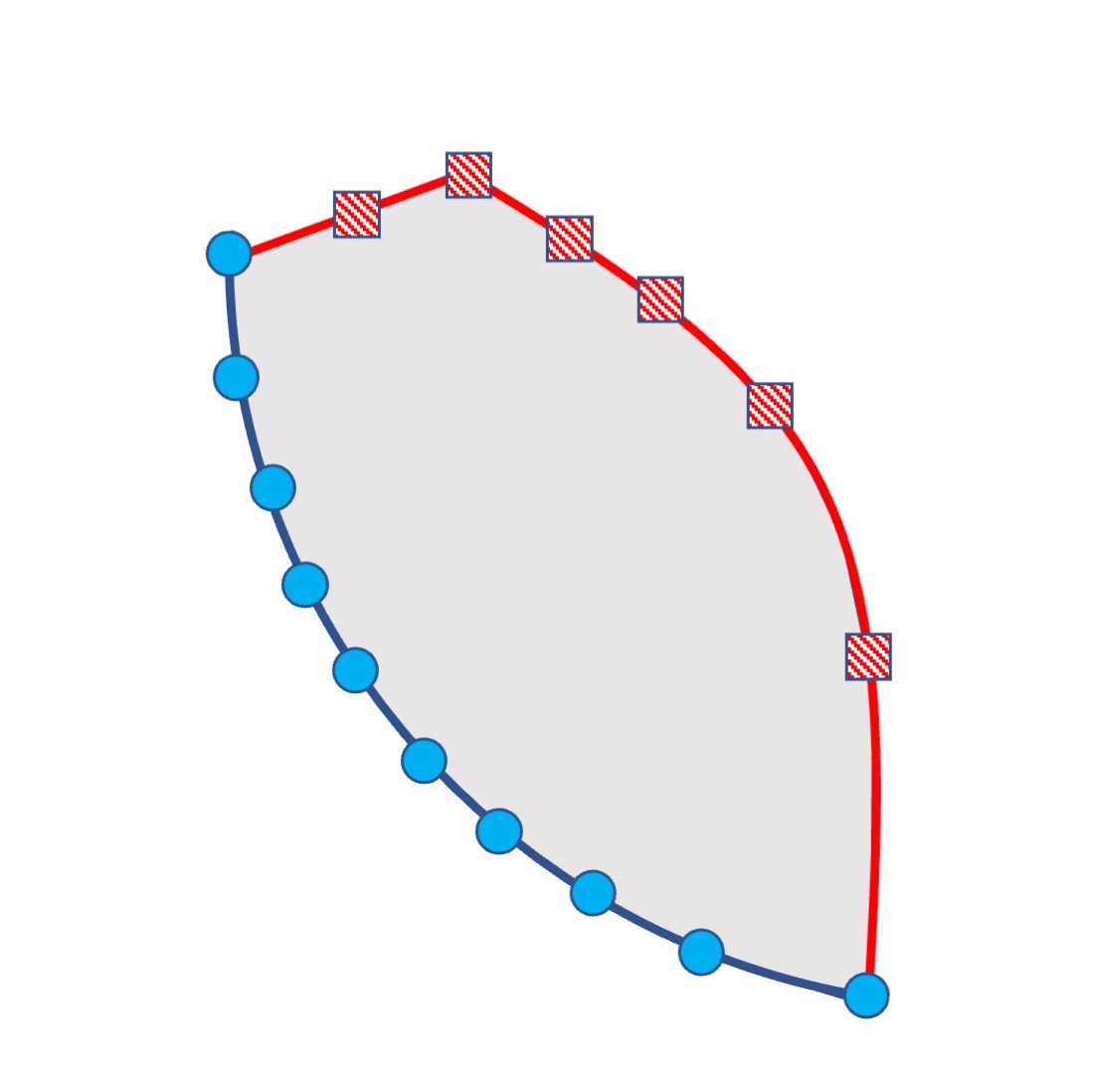}}}
	\caption{An illustration of splitting a one-hole instance along a path set $\pset$.
	\textit{Left}: Graph $H$, together with terminals in set $U$ (in blue), paths in set $\pset$ (in different colors), and vertices of $Y$ (red boxes).
	\textit{Right}: An output instance (that corresponds to the left bottom region of $H$) by procedure $\cutpath$.}
	\label{fig: cut_noncrossing}
\end{figure}

The output of procedure $\cutpath$ is a collection of one-hole instances constructed as follows.
Consider a plane embedding of $H$ where all the terminals in $U$ lying on the outerface of $H$.
We \emph{slice}%
\footnote{The slicing operation, which can be traced back to Reif~\cite{rei-mscpu-1981} (when describing the minimum-cut algorithm by Itai-Shiloach~\cite{is-mfpn-1979}), is sometimes referred to as \emph{cutting}~\cite{efn-gmcse-2012} or \emph{incision}~\cite{mnnw-mcdpg-2018} in the literature.}
$H$ open along each path $P$ in $\pset$ by duplicating every vertex and edge of $P$ to create another path $P'$ identical to $P$.
The set of edges incident to each vertex on $P$ are split into two sides naturally based on their cyclic order around the vertex.
We index the collection of subgraphs of $H$ obtained by the slicing of $H$ along $\pset$ by \EMPH{$\rset$}.
Let $R$ be an index in $\rset$ that corresponds to the subgraph $H_R$. 
The plane embedding of $H$ naturally induces a planar embedding of $H_R$.
Define \EMPH{$U_R$} to be the set of all vertices of $H_R$ that is either a terminal in $H_R\setminus P$ or a vertex in $Y$. 
All vertices of $U_R$ appear on the outerface of $H_R$, and so $(H_R,U_R)$ is a one-hole instance. 
The output of procedure $\cutpath$ is simply the collection $\set{(H_R,U_R)\mid R\in \rset}$ that contains, for each subgraph $H_R$ obtained by slicing $H$,
a one-hole instance defined in the above way.
See \Cref{fig: cut_noncrossing} for an illustration. 
Note that each vertex $y\in Y$ may now belong to multiple instances in $\hset$. We call them \emph{copies} of~$y$.

\begin{figure}[h]
	\centering
	\includegraphics[scale=0.12]{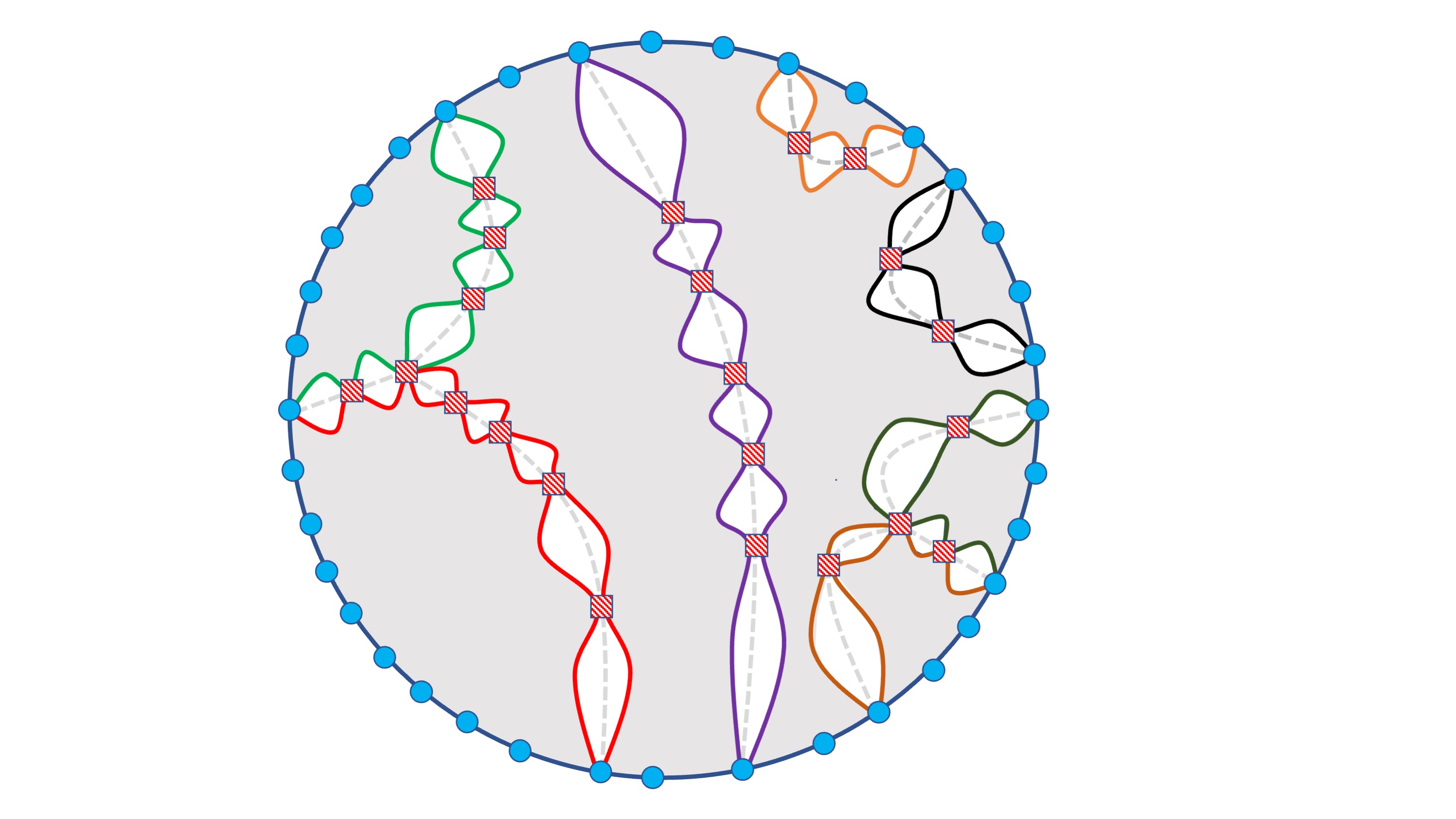}
	\caption{An illustration of gluing one-hole instances at outer-boundaries. Identified vertices of $U$ are shown in blue) and identified vertices of $Y\setminus U$ are shown in red boxes. 
	}
	\label{fig: glue}
\end{figure}

\paragraph{Gluing.}
We now describe procedure \EMPH{$\gluepath$}.
%
Assume that we have applied procedure $\cutpath$ to a one-hole instance $(H,U)$, a non-crossing set $\pset$ of shortest paths, and a vertex subset $Y$ to obtain a collection $\hset = \set{(H_R,U_R)\mid R\in \rset}$ of one-hole instances.
The input to procedure \emph{$\gluepath$} consists of
\begin{itemize}
    \item one emulator $(Z_R,U_R)$ for each one-hole instance $(H_R,U_R)$ in $\hset$; and
    \item the same vertex subset $Y$ given as the input to procedure $\cutpath$.
\end{itemize}
The output of procedure $\gluepath$ is an emulator $(Z,U)$ for $(H,U)$, which is constructed as follows.
%
%
Graph~$Z$ is obtained by taking the union of all graphs in $\set{Z_R\mid R\in \rset}$, and identifying, for each vertex $y\in Y$, all copies of $y$. 
Graph $Z$ is naturally a plane graph by inheriting the embeddings of all $Z_R$s.
(See \Cref{fig: glue} for an illustration.)
By the assumption that $Y$ contains all the endpoints of paths in $\pset$, every vertex in $U$ shows up uniquely on the outerface of $Z$.
Therefore, $(Z,U)$ is a one-hole instance. 
Moreover, it is easy to observe that $|V(Z)|\le \sum_{R\in \rset}|V(Z_R)|$.

One can verify that both procedures $\cutpath$ and $\gluepath$ can be implemented in $O(|V(H)|)$ time.
Now we now summarize the behavior of the procedures with the following claims.
The proofs of Claim~\ref{clm: branch pts} and Claim~\ref{clm: glueset_emulators} are deferred to
Appendix~\ref{apd: Proof of branch pts} and~\ref{apd: Proof of glueset_emulators} respectively. 



\begin{claim}
\label{clm: branch pts}
Let $\hset$ be the output of procedure $\cutpath$ applied to a valid input $((H,U),\pset,Y)$, then 
	\begin{enumerate}
		\item the number of branch vertices is at most $O(|U|)$; and \label{p1}
		\item if we denote by $Y^*$ the subset of all branch vertices in $Y$, then for every parameter $\lambda \ge 100$,\\
		$\sum_{(H_R,U_R)\in \hset:\text{ } |U_R|\ge \lambda}|U_R|\le |U|\cdot\big(1+O(1/\lambda)\big)+O(|Y\setminus Y^*|).$ \label{p2}
	\end{enumerate}
\end{claim}

\begin{claim}
\label{clm: glueset_emulators}
Let $\hset$ be output collection of procedure $\cutpath$ when applied to a valid input $((H,U),\pset,Y)$.
Let $(\hat H, U)$ be output of procedure $\gluepath$ when applied to the collection $\hset$ and set $Y$.
For each instance $(H_R,U_R)\in \hset$, let $(Z_R,U_R)$ be an $\eps$-emulator for $(H_R,U_R)$, and let $(Z,U)$ be the output of procedure $\gluepath$ when applied to the collection $\set{(Z_R,U_R)}_R$ and set $Y$. Then $(Z,U)$ is an $\eps$-emulator for $(\hat H,U)$.
\end{claim}





\subsection{Remove All Cut Vertices in $U$}
\label{subsec: Remove All Cut Vertices}

Before we proceed with the main ingredient for proving Lemma~\ref{lem: decomposing step}, first we describe a reduction on the input instance $(H,U)$ so that no vertex in $U$ is a cut vertex of graph $H$. 
The impatient readers may skipped ahead to Section~\ref{SSS:small-spread}.

We first compute the set \EMPH{$U'$} of all cut vertices of $H$ in $U$,
and along the way the maximal 2-vertex-connected subgraphs $\hat H_1,\ldots, \hat H_t$ of $H$ that each contains at least two terminals of $U$.
For each $i \in \set{1, \dots, t}$, we denote $\hat U_i \coloneqq U\cap V(\hat H_i)$, so $(\hat H_i, \hat U_i)$ is a one-hole instance. 
Moreover, from Claim~\ref{clm: glueset_emulators}, if we are given an $\eps$-emulator for instance $(\hat H_i, \hat U_i)$ for each $i$, then by simply 
gluing them at terminals in $U'$, we can obtain an $\eps$-emulator for instance $(H,U)$.
We use the following claim in order to bound $\sum_{1\le i\le t}|\hat U_i|$ and $\sum_{|\hat U_i|\ge \lambda}|\hat U_i|$.

\begin{claim}
\label{clm: removing separators}
$\sum_{1\le i\le t}|\hat U_i|\le O(|U|)$, and $\sum_{|\hat U_i|\ge \lambda}|\hat U_i|\le |U| \cdot (1+O(1/\lambda))$.
\end{claim}

\begin{proof}
Recall that $r \coloneqq |U|$.
Consider the following tree \EMPH{$\tau'$}:
The node set of tree $\tau'$ is $U'\cup V'$, where $V' \coloneqq \set{v_i\mid 1\le i\le t}$.
The edge set of tree $\tau'$ contains, for each $1\le i\le t$ and each node $u'\in U'$, an edge $(u',v_i)$ if $u'\in \hat U_i$.
Since vertices of $U'$ are cut vertices of $H$, it is easy to verify that the graph $\tau'$ constructed above is a tree,
and moreover, all leaves of $\tau'$ lie in $V'$.

We partition set $V'$ into three subsets: \EMPH{$V'_1$} contains all leaf nodes of $\tau'$, \EMPH{$V'_2$} contains all nodes of degree $2$ in $\tau'$, and \EMPH{$\smash{V'_{\ge 3}}$} contains all nodes of degree at least $3$ in $\tau'$. 
%
Observe that, for each node $v_i\in V'_1$, since $|\hat U_i|\ge 2$, at least one terminal in $\hat U_i$ does not belong to any other set in $\set{\hat U_1,\ldots,\hat U_t}$.
Therefore, $|V'_1|\le r$. Since $\tau'$ is a tree, $|V'_{\ge 3}|\le |V'_1| \le r$. 
Since for every node in $V'_2$, both its neighbors lie in $U'$, we get that $|V'_2|\le |U'|\le r$.
Altogether, $|V(\tau')|\le O(r)$.
Note that for every terminal $u'\in U'$, the number of sets $\hat U_i$ that contains $u$ is exactly $\deg_{\tau'}(u')$. Therefore,
\[
\sum_{1\le i\le t}|\hat U_i|\le |U\setminus U'|+\sum_{u'\in U'}\deg_{\tau'}(u')\le |U|+O(|V(\tau')|)=O(r).
\]
We now upper bound $\sum_{|\hat U_i|\ge \lambda}|\hat U_i|$ via a charging scheme. 
We root the tree $\tau'$ at an arbitrary node of $V'$, and process the nodes in $U'$ one-by-one as follows. 
Consider a node $u'\in U'$ such that all its child nodes are leaves in $\tau'$.
We denote by $v_1,\ldots,v_s$ the child nodes of $u'$. 
For each $1\le i\le s$, if $|\hat U_i|\ge \lambda$, we charge $u'$ (as one unit) uniformly 
to vertices of $\hat U_i\setminus \set{u'}$, so each terminal in $\hat U_i\setminus \set{u'}$ is charged at most $2/\lambda$ units. 
We delete nodes $u'$ and $v_1,\ldots,v_s$ from $\tau'$ and recurse on the remaining tree, until the tree contains no nodes of $U'$. 
It is easy to observe that the value of $\smash{\sum_{|\hat U_i|\ge \lambda} |\hat U_i|}$ is at most $r$ plus the total charge. We now show that the total charge is $O({1}/\lambda)$. 
In fact, every terminal in $U$ is directly charged at most $2/\lambda$. Note that it is possible that some terminal in $U'$ was first charged to some other terminals in $U'$, and was later (indirectly) charged for other terminals in $U'$. 
It is easy to observe that, the total direct and indirect charge is bounded by $2/\lambda+(2/\lambda)^2+\cdots\le 4/\lambda$. 
Therefore, 
$\sum_{|\hat U_i|\ge \lambda}|\hat U_i|\le r\cdot (1+O(1/\lambda))$.
\end{proof}

Note that we can simply return the collection $\set{(\hat H_i,\hat U_i)\mid 1\le i\le t}$ of one-hole instances as the output, and it is easy to verify from the algorithm and Claim~\ref{clm: removing separators} that the output satisfies all properties required in Lemma~\ref{lem: decomposing step} (where the algorithm $\algmerge$ is simply the procedure $\gluepath$), unless some set $\hat U_i$ contains more than $(9/10)r$ terminals. 
However, from Claim~\ref{clm: removing separators}, there is at most one such large instance.
Assume without loss of generality that $(\hat H_1,\hat U_1)$ is the unique large instance.
We claim that, if Lemma~\ref{lem: decomposing step} holds for instance $(\hat H_1,\hat U_1)$, then Lemma~\ref{lem: decomposing step} holds for the input instance $(H,U)$.
In fact, we apply the algorithm from Lemma~\ref{lem: decomposing step} to instance $(\hat H_1,\hat U_1)$ and obtain a collection $\tilde\hset'$, and we can simply return the collection $\tilde \hset \coloneqq \tilde\hset'\cup \set{(\hat H_i,\hat U_i)\mid 2\le i\le t}$. It is easy to verify from the above discussion that all conditions of Lemma~\ref{lem: decomposing step} hold for the collection $\tilde \hset$ as an output for the original instance $(H,U)$.

From now on we focus on proving Lemma~\ref{lem: decomposing step} for the unique large instance $(\hat H_1,\hat U_1)$. 
For convenience, we rename this large instance by $(H,U)$, denote $r \coloneqq |U|$, and treat it as the original input instance. 
From our algorithm, no vertex in $U$ is a cut vertex of graph $H$, so if we traverse the outerface of $H$, then every terminal of $U$ appears exactly once.

\subsection{The Small Spread Case}
\label{SSS:small-spread}

Let $(H,U)$ be a planar instance. 
The \EMPH{spread}\footnote{sometimes also referred to as \emph{aspect ratio}} of the instance $(H,U)$ is defined to be 
\[
\EMPH{$\Phi(H,U)$} \coloneqq \frac{\max_{u,u'\in U}\dist_H(u,u')}{\min_{u,u'\in U}\dist_H(u,u')}.
\]
For convenience, we denote $\EMPH{$\Phi$} \coloneqq \Phi(H,U)$.
We distinguish between the following two cases, depending on whether $\Phi$ is small or large.
In this subsection we assume \EMPH{$\Phi\le 2^{r^{0.9}\log^{2} r}$}.
The large spread case will be discussed in Section~\ref{SSS:large-spread}.

We will employ the procedure $\cutpath$ in order to decompose the one-hole instance $(H,U)$ into smaller instances. Throughout this case, we use parameters 
\[
\EMPH{$L_r$} \coloneqq r/100\log^2 r 
\quad \text{and} \quad 
\EMPH{$\eps_r$} \coloneqq \log \Phi/L_r,
\]
so $\eps_r = O((\log r)^4/r^{0.1})$.

\paragraph{Balanced terminal pairs.} 
Denote $U \coloneqq \set{u_1,\ldots,u_r}$, where the terminals are indexed according to the order in which they appear on the outerface. 
We say that a pair of terminals $(u_i,u_j)$ (with $i<j$) is a \EMPH{$c$-balanced pair} for some parameter $1/2<c<1$, if and only if $j-i\le c\cdot r$ and $i+r-j\le c\cdot r$. In other words, the terminals $u_i$ and $u_j$ separate the outer boundary into two segments, each contains at most $c$-fraction (and therefore at least $(1-c)$-fraction) of the terminals.

We first compute the $(3/4)$-balanced pair $u,u'$ of terminals in $U$ that, among all $(3/4)$-balanced pairs of terminals in $U$, minimizes the distance between them in $H$. 
We compute the $u$-$u'$ shortest path $P$ in $H$.
Let the set $Y$ contain the endpoints of $P$, together with the following vertices of $P$: for each $1\le i\le L_r$, 
\begin{enumerate}
\item among all vertices $v$ of $P$ with $\dist_P(v,u)\le e^{i\eps_r}$, the vertex that maximizes its distance to $u$;
\item among all vertices $v$ of $P$ with $\dist_P(v,u)\ge e^{i\eps_r}$, the vertex that minimizes its distance to $u$; 
\item among all vertices $v$ of $P$ with $\dist_P(v,u')\le e^{i\eps_r}$, the vertex that maximizes its distance to $u'$;
\item among all vertices $v$ of $P$ with $\dist_P(v,u')\ge e^{i\eps_r}$, the vertex that minimizes its distance to $u'$.
\end{enumerate}
In other words, if we think of path $P$ as a line, and then mark, for each $1\le j\le L_r$, the point on the line that is at distance $e^{i\eps_r}$ from $u$, and the point on the line that is at distance $e^{i\eps_r}$ from $u'$, then set $Y$ contains, for all marked points, the vertices of $P$ that are closest to it from both sides.
By definition, $|Y|\le 4L_r$. 

We apply the procedure $\cutpath$ to the one-hole instance $(H,U)$, the path set $\set{P}$ and the vertex set $Y$ defined above. Let $(H_1,U_1)$ and $(H_2,U_2)$ be the instances we get. We then simply return the collection $\set{(H_1,U_1),(H_2,U_2)}$ as the output of our algorithm.

\paragraph{Analysis of the small spread case.}
We now show that the output of the algorithm in this case satisfies the properties required in Lemma~\ref{lem: decomposing step}.
First, from the definition of procedure $\cutpath$, every terminal in $U$ continues to be a terminal in at least one instance in $\set{(H_1,U_1),(H_2,U_2)}$.
Moreover, since the pair $(u,u')$ of terminals is $(3/4)$-balanced, and $|Y| \le 4L_r = r/(25\log^2 r)$, so $|U_1|\le (3/4)r+r/(25\log^2 r)\le (9/10)r$, and similarly $|U_2|\le (9/10)r$. 
%
Second, note that $|U_1|+|U_2|\le |U|+2|Y|\le r\cdot (1+O(L_r/r))=r\cdot (1+O(\frac{1}{\log^2 r}))=r\cdot (1+O(1/\lambda))$, as $\lambda\le \log^2 r$.  

We now construct an algorithm $\algmerge$ that satisfies the required properties. 
Let $(H'_1,U_1)$ be an $\eps$-emulator for $(H_1,U_1)$ and let $(H'_2,U_2)$ be an  $\eps$-emulator for $(H_2,U_2)$. 
The algorithm $\algmerge$ simply applies the procedure $\gluepath$ to the collection $\set{(H'_1,U_1),(H'_2,U_2)}$ and set $Y$. Let $(H', U')$ be the one-hole instance that it outputs. It is easy to verify that $U'=U$. 
The algorithm $\algmerge$ simply returns the instance $(H', U)$.
It remains to show that the output of algorithm $\algmerge$ satisfies the required properties.
Note that the collection $\set{(H_1,U_1),(H_2,U_2)}$ and the set $Y$ also constitute a valid input for procedure $\gluepath$. 
Let $(\hat H, \hat U)$ be the instance output by $\gluepath$ when applied to $\set{(H_1,U_1),(H_2,U_2)}$ and $Y$. It is easy to verify that $\hat U=U$.  
%
We use the following claim. 

\begin{claim}
\label{clm: ratio loss for contracting to portals}
Instance $(\hat H, U)$ is a $(3\eps_r)$-emulator for instance $(H, U)$.
\end{claim}

We provide the proof of Claim~\ref{clm: ratio loss for contracting to portals} right after we complete the analysis for the small spread case.
From Claim \ref{clm: glueset_emulators}, $(H',U)$ is an $\eps$-emulator for $(\hat H,U)$. 
From Claim~\ref{clm: ratio loss for contracting to portals}, instance $(\hat H, U)$ is a $(3\eps_r)$-emulator for instance $(H, U)$. Altogether, $(H',U)$ is an  $(\eps+3\eps_r)=(\eps+O(\smash{\frac{\log^4 r}{r^{0.1}}}))$-emulator for $(H,U)$. 
This completes the proof of Lemma~\ref{lem: decomposing step} in the small spread case.
 
\begin{proofof}{Claim~\ref{clm: ratio loss for contracting to portals}}
We will show that, for each pair $u_1,u_2$ of terminals in $U$, 
$$\dist_{H}(u_1,u_2)\le \dist_{\hat H}(u_1,u_2)\le e^{3\eps_r}\cdot\dist_{H}(u_1,u_2).$$

From the procedure $\cutpath$, $H_1$ is the subgraph of $H$ whose image lies in the region surrounded by the image of $P$ and the segment of outer-boundary of $H$ from $u$ clockwise to $u'$ (including the boundary), and $H_2$ is the subgraph of $H$ whose image lies in the region surrounded by the image of $P$ and the segment of outer-boundary of $H$ from $u$ anti-clockwise to $u'$ (including the boundary), and path $P$ is entirely contained in both $H_1$ and $H_2$. 
We denote by $\hat H_1$ the copy of $H_1$ in graph $\hat H$, and we define graph $\hat H_2$ similarly, so $V(\hat H_1)\cap V(\hat H_2)=Y$. We denote by $P^1, P^2$ the copies of path $P$ in graphs $\hat H_1$ and $\hat H_2$, respectively.
See \Cref{fig: proof of case 1} for an illustration.

\begin{figure}[h]
	\centering
	\subfigure{\scalebox{0.45}{\includegraphics[scale=0.20]{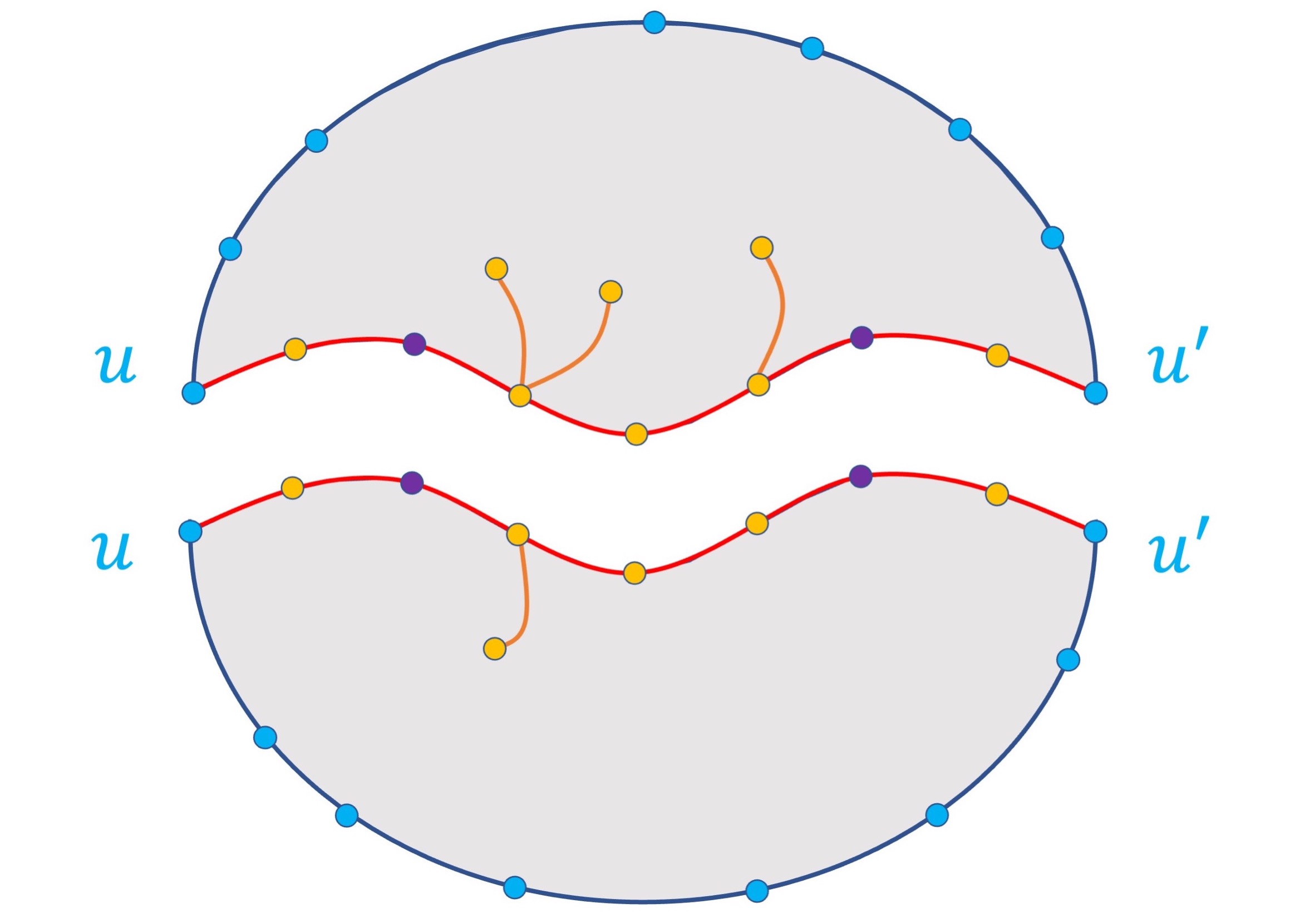}}}
	\hspace{0.3cm}
	\subfigure{\scalebox{0.45}{\includegraphics[scale=0.19]{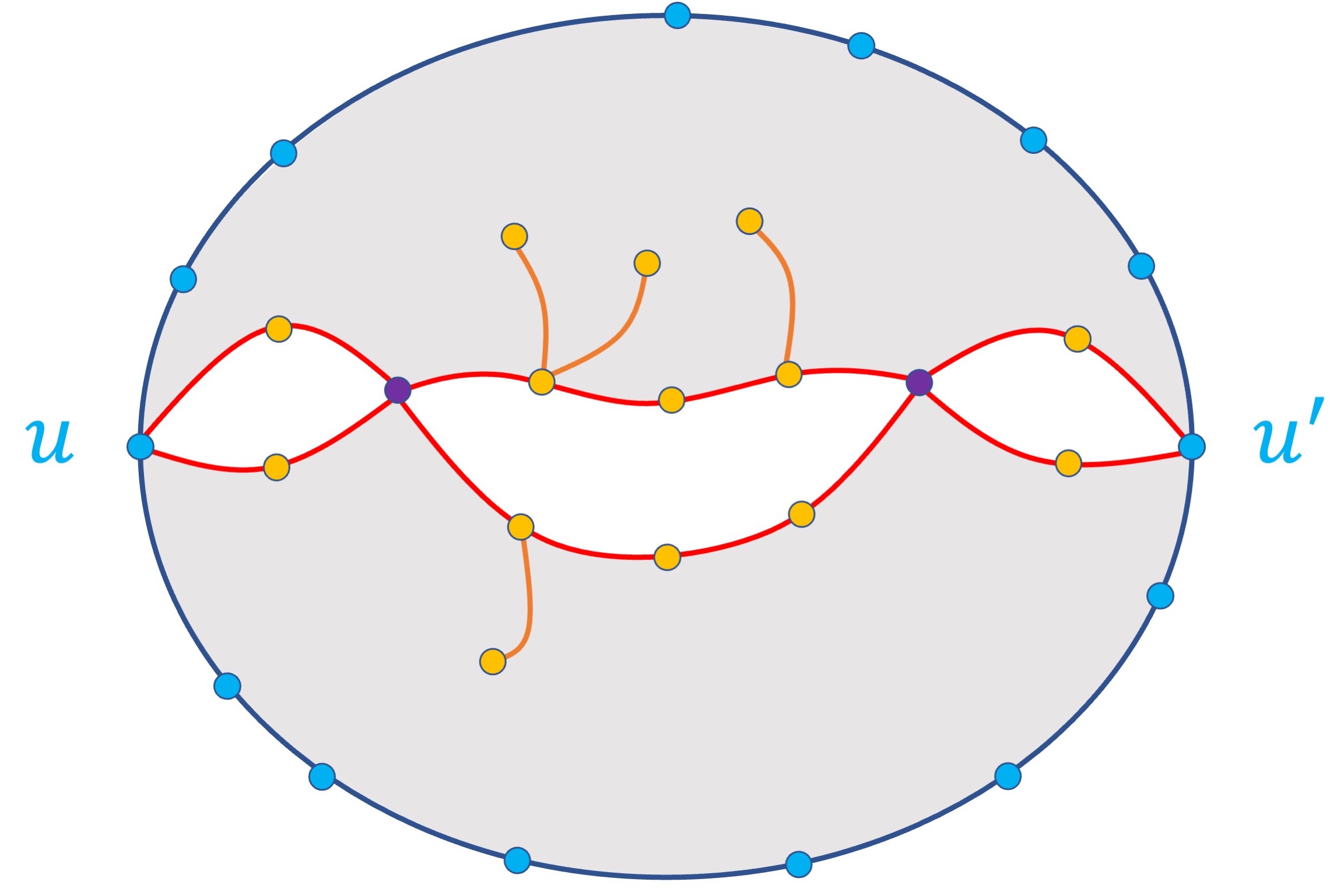}}}
\caption{An illustration graphs $\hat H$, $H_1$, and $H_2$.
\textit{Left:} Graphs $H_1$ (top) graph $H_2$ (bottom) viewed as individual graphs.
\textit{Right:} Subgraphs $\hat H$ obtained by gluing graphs $H_1$ and $H_2$. Vertices in $Y\setminus \set{u,u'}$ are shown in purple.
}
\label{fig: proof of case 1}
\end{figure}

We first show that for each pair $u_1,u_2\in U$, $\dist_{H}(u_1,u_2)\le \dist_{\hat H}(u_1,u_2)$. 
Consider a pair $u_1,u_2\in U$. Assume first that $u_1,u_2$ both belong to $H_1$ (the case where $u_1,u_2$ both belong to $H_2$ is symmetric). Clearly, in graph $\hat H$, there is a $u_1$-$u_2$ shortest path $Q$ that lies entirely in $\hat H_1$. From the construction of $\hat H$, the same path belongs to $H_1$, and therefore $\dist_H(u_1,u_2)\le \dist_{\hat H}(u_1,u_2)$. 
Assume now that $u_1\in V(H_1)\setminus \set{u,u'}$ and $u_2\in V(H_2)\setminus \set{u,u'}$ (the case where  $u_2\in V(H_1)\setminus \set{u,u'}$ and $u_1\in V(H_2)\setminus \set{u,u'}$ is symmetric). It is easy to see that, in graph $\hat H$, there exists a $u_1$-$u_2$ shortest path that is the sequential concatenation of 
\begin{enumerate}
\item a path $Q_1$ in $\hat H_1$ connecting $u_1$ to some vertex $x_1\in V(P^1)$, that is internally disjoint from $P^1$; 
\item a subpath $R^1$ of $P^1$ connecting $x_1$ to a vertex $y\in Y$;
\item a subpath $R^2$ of $P^2$ connecting $y$ to a vertex $x_2$; and
\item a path $Q_2$ in $\hat H_2$ connecting $x_2$ to $u_2$, that is internally disjoint from $P^2$.
\end{enumerate}
Consider the path in $H$ formed by the sequential concatenation of (i) the copy of $Q_1$ in $H_1$; (ii) the subpath $R$ of $P$ connecting the copy of $x_1$ in $P$ to the copy of $x_2$ in $P$; and (iii) the copy of $Q_2$ in $H_2$. Clearly, this path connects $u_1$ to $u_2$ in $P$. Moreover, since the weight of $R$ is at most the total weight of paths $R^1$ and $R^2$, this path in $H$ has weight at most the weight of the $u_1$-$u_2$ shortest path in $\hat H$. Therefore, $\dist_H(u_1,u_2)\le \dist_{\hat H}(u_1,u_2)$.

From now on we focus on showing that, for each pair $u_1,u_2\in U$, $\dist_{\hat H}(u_1,u_2)\le e^{3\eps_r}\cdot\dist_{H}(u_1,u_2)$.
Assume first that $u_1,u_2$ both belong to $H_1$ (the case where $u_1,u_2$ both belong to $H_2$ is symmetric). Similar to the previous discussion, the $u_1$-$u_2$ shortest path in $H$ is entirely contained in $H_1$, and so $\dist_{\hat H}(u_1,u_2)=\dist_{H}(u_1,u_2)$.
Assume now that $u_1\in V(H_1)\setminus \set{u,u'}$ and $u_2\in V(H_2)\setminus \set{u,u'}$ (the case where $u_1\in V(H_2)\setminus \set{u,u'}$ and $u_2\in V(H_1)\setminus \set{u,u'}$) is symmetric.
Let $Q$ be the $u_1$-$u_2$ shortest path in $H$.
The intersection between $Q$ and $P$ is a subpath of $P$. Let $x_1,x_2$ be the endpoints of this subpath, so vertices $u_1,x_1,x_2,u_2$ appear on path $Q$ in this order.
Let $Q_1$ denote the subpath of $Q$ between $u_1$ and $x_1$, 
$Q_2$ the subpath of $Q$ between $u_2$ and $x_2$, and
$Q'$ the subpath of $Q$ between $x_1$ and $x_2$.
%
We consider the following possibilities, depending on the locations of vertices $x_1,x_2$ and vertices in $Y$.

\bigskip
\noindent\textit{Possibility 1. There is a vertex in $Y$ between $x_1$ and $x_2$.} 
Let $y$ be a vertex of $Y$ between vertices $x_1$ and $x_2$. Consider the path $\hat Q$ of $\hat H$ formed by the sequential concatenation of (i) the copy of $Q_1$ in $\hat H_1$ connecting $u_1$ to the copy of $x_1$; (ii) the subpath $R^1$ of $P^1$ connecting the copy of $x_1$ to $y$; (iii) the subpath $R^2$ of $P^2$ connecting $y$ to the copy of $x_2$; and (iv) the copy of $Q_2$ in $\tilde H_2$ connecting the copy of $x_2$ to $u_2$.
Since vertex $y$ lies between $x_1$ and $x_2$ on path $P$, from the construction of $\hat H$, the path $\hat Q$ in $\hat H$ constructed above has weight at most the weight of $Q$ in $H$. Therefore, $\dist_{\hat H}(u_1,u_2)\le \dist_{H}(u_1,u_2)$.

\bigskip
\noindent\textit{Possibility 2. There is no vertex of $Y$ between $x_1$ and $x_2$.} 
Assume without loss of generality that $|V(H_1)\cap U|\ge |U|/2$, and that $x_1$ is closer to $u$ than to $u'$ in $P$. We use the following observation.

\begin{observation}
\label{obs: length to charge to}
$\dist_H(x_1,u_1)\ge \dist_H(x_1,u)$.
\end{observation}
\begin{proof}
Assume not, then 
\(
\dist_H(u_1,u)\le  \dist_H(x_1,u_1)+ \dist_H(x_1,u)< 2\cdot \dist_H(x_1,u)\le \dist_H(u,u')\text{, and}
\)
\(
\dist_H(u_1,u')\le  \dist_H(x_1,u_1)+ \dist_H(x_1,u')< \dist_H(x_1,u)+ \dist_H(x_1,u')\le \dist_H(u,u').
\)
So both $\dist_H(u_1,u)$ and $\dist_H(u_1,u')$ is less than $\dist_H(u,u')$. However, since $|U|/2\le |V(H_1)\cap U|\le (3/4)\cdot|U|$, it is easy to verify that at least one of the pairs $(u_1,u)$, $(u_1,u')$ is $(3/4)$-balanced, a contradiction to the fact that $u,u'$ is the closest $(3/4)$-balanced terminal pair in $H$.
\end{proof}

Think of path $P$ as a line connecting $u$ to $u'$. We now mark, for each $1\le j\le L_r$, the point on the line that is at distance $e^{i\eps_r}$ from $u$, and the point on the line that is at distance $e^{i\eps_r}$ from $u'$, and call these marked points \EMPH{landmarks}.
It is easy to observe that there is no landmark between vertices $x_1$ and $x_2$. This is because, if there is landmark between vertices $x_1$ and $x_2$, since set $Y$ contains, for all landmark, the vertices of $P$ that are closest to it from both sides, either $x_1$ or $x_2$ or some other vertices of $P$ that lie between $x_1$ and $x_2$ will be added to vertex set $Y$, a contradiction.
Let $x$ be the landmark closest to $x_1$ that lies between $u$ and $x_1$, and assume $\dist_P(x,u)=e^{i\eps_r}$. 
Let $y$ be the vertex of $Y$ closest to the landmark $x$ that lies between $x$ and $x_1$.
From the construction of portals,
$e^{i\eps_r}\le \dist_P(y,u)<\dist_P(x_1,u),\dist_P(x_2,u)<e^{(i+1)\eps_r}$. 
Therefore,
$\dist_P(x_1,y),\dist_P(x_2,y)\le (e^{\eps_r}-1)\cdot e^{i\eps_r}$. 
Consider now the $u_1$-$u_2$ path in $\hat H$ formed by concatenation of (i) the copy of $Q_1$ in $\hat H_1$ connecting $u_1$ to the copy $x^1_1$ of $x_1$; (ii) the subpath of $P^1$ connecting $x^1_1$ to $y$; (iii) the subpath of $P^2$ connecting $y$ to the copy $x^2_2$ of $x_2$; and (iv) the copy of $Q_2$ in $\hat H_2$ connecting $x^2_2$ to $u_2$. 
The total weight of this path is at most 
\[
\begin{split}
& \dist_{\hat H_1}(u_1,x^1_1)+\dist_{\hat H_1}(x^1_1,y)+\dist_{\hat H_2}(x^2_2,y)+\dist_{\hat H_2}(u_2,x^2_2)\\
= \text{ } & \dist_{H}(u_1,x_1)+\dist_{P}(x_1,y)+\dist_{P}(x_2,y)+\dist_{H}(u_2,x_2)\\
= \text{ } & \dist_{H}(u_1,x_1)+\dist_{H}(u_2,x_2)+\dist_{P}(x_1,x_2)+\big(\dist_{P}(x_1,y)+\dist_{P}(x_2,y)-\dist_{P}(x_1,x_2)\big)\\
\le  \text{ } & \dist_{H}(u_1,u_2)+ 2\cdot (e^{\eps_r}-1)\cdot e^{i\eps_r}\\
\le  \text{ } & \dist_{H}(u_1,u_2)+ 2\cdot (e^{\eps_r}-1)\cdot \dist_{H}(u,x_1)\\
\le  \text{ } & \dist_{H}(u_1,u_2)+ 2\cdot (e^{\eps_r}-1)\cdot \dist_{H}(u_1,x_1) \quad\quad(\textnormal{from Observation~\ref{obs: length to charge to}})\\
\le \text{ } & e^{3\eps_r}\cdot\dist_{H}(u_1,u_2).
\end{split}
\]
Therefore, $\dist_{\hat H}(u_1,u_2)\le e^{3\eps_r}\cdot\dist_{H}(u_1,u_2)$.
This completes the proof of Claim~\ref{clm: ratio loss for contracting to portals}.
\needqedtrue
\end{proofof}

\subsection{The Large Spread Case}
\label{SSS:large-spread}

Now we assume \emph{$\Phi > 2^{r^{0.9}\log^2 r}$}.
Without loss of generality, we assume that $\min_{u,u'\in U}\dist_H(u,u')=1$ and $\max_{u,u'\in U}\dist_H(u,u')=\Phi$.
In the algorithm for this case, we use the following parameters: 
\[
\EMPH{$\mu$} = r^2,  \quad  
\EMPH{$L$} = \ceil{\log_{\mu} \Phi}, \quad 
\EMPH{$\eps_r$} = \frac{\log^4 r}{r^{0.1}}, \quad  
\EMPH{$\eps'_r$} = \frac{1}{r^{0.7}}.
\]

We first compute a hierarchical partitioning $(\sset_0,\sset_1,\ldots,\sset_L)$ of terminals in $U$ in a bottom-up fashion as follows. We proceed in $L$ iterations. In the $i$th iteration, we compute a collection $\sset_i$ of subsets of $U$ that partition $U$. 
\begin{itemize}
\item
We start by letting collection \EMPH{$\sset_0$} contain, for each terminal $u\in U$, a singleton set $\set{u}$. That is, $\sset_0 \coloneqq \set{\set{u}\mid u\in U}$. 
\item
Consider an index $1\le i\le L$.
Assume we have already computed the collection $\sset_{i-1}$ of subsets, we now describe the computation of collection $\sset_{i}$, as follows. 
First, let graph $W_{i-1}$ be obtained from $H$ by contracting each subset $S\in \sset_{i-1}$ into a single \EMPH{supernode}, that we denote by \EMPH{$v_S$}, and we define $\EMPH{$V_{i-1}$} \coloneqq \set{v_S\mid S\in \sset_{i-1}}$. 
Recall that $H$ is an edge-weighted graph, and we let every edge of $W_{i-1}$ have the same weight as the corresponding edge in $H$.
Then we construct another auxiliary graph \EMPH{$R_{i-1}$} as follows. Its vertex set is $V_{i-1}$, and it contains an edge connecting $v_S$ to $v_{S'}$ if $\dist_{W_{i-1}}(v_S,v_{S'})\le \mu^i$, or equivalently $\dist_{H}(S,S')\le \mu^i$.
Finally, we define \EMPH{$\sset_i$} to be the collection that contains, for each connected component $C$ of graph $R_{i-1}$, the set $\bigcup_{v_S\in V(C)}S$.
It is easy to verify that the sets in $\sset_i$ partition $U$.
\end{itemize}
This completes the description of the hierarchical partitioning $(\sset_0,\sset_1,\ldots,\sset_L)$.
Clearly, collection $\sset_L$ contains a single set $U$.
We denote $\EMPH{$\sset$} \coloneqq \bigcup_{0\le i\le L}\sset_i$.
So collection $\sset$ is a laminar family. That is, for every pair $S,S'\in \sset$, either $S\cap S'=\varnothing$, or $S\subseteq S'$, or $S'\subseteq S$.


\begin{observation}
\label{obs: diameter}
For each set $S$ in collection $\sset_i$, $\diam_H(S)\le 2r\cdot\mu^{i}$.
\end{observation}
\begin{proof}
We prove the observation by induction on $i$. 
The base case is when $i=0$. From the construction, the collection $\sset_0$ contains only single-vertex sets, so the diameter of each such set is at most $0\le 2r\cdot\mu^0$. Assume that the observation holds for $0,1,\ldots,i-1$. Consider now a cluster $\hat S\in \sset_i$. From the construction, it is the union of a collection of sets in $\sset_{i-1}$. Consider any pair $u,u'$ of vertices in $\hat S$. If they belong to the same set of in $\sset_{i-1}$, then from the induction hypothesis, $\dist_H(u,u')\le 2r\cdot \mu^{i-1}\le 2r\cdot \mu^{i}$. Assume now that $u\in S$ and $u'\in S'$ where $S, S'$ are distinct sets in $\sset_{i-1}$. 
Since supernodes $v_{S}$ and $v_{S'}$ lie in the same connected component of graph $R_{i-1}$, there exists a path connecting $v_{S}$ to $v_{S'}$ in $R_{i-1}$, and we denote it by $(v_{S},v_{S_{1}},\ldots,v_{S_{b}},v_{S'})$, where $b\le r-2$ (since the number of supernodes is at most $r$).
If we further denote $S_0=S$ and $S_{b+1}=S'$, then there exist, for each $0\le j\le b+1$, a pair $\hat u_j, \hat u'_j$ of vertices in $S_{j}$, such that 
\begin{itemize}
    \item $u=\hat u_0$, $u'=\hat u'_{b+1}$;
    \item for each $0\le j\le b+1$, $\dist_H(\hat u_j, \hat u'_j)\le 2r\cdot\mu^{i-1}$; and
    \item for each $0\le j\le b$, $\dist_H(\hat u'_j, \hat u_{j+1})\le \mu^{i}$.
\end{itemize}
Therefore, $\dist_H(u,u')\le r\cdot (2r\cdot\mu^{i-1})+r\cdot \mu^i\le 2r\cdot\mu^{i}$, since $\mu=r^2$.
\end{proof}

In order to describe and analyze the algorithm, it would be convenient for us to compute a partitioning tree $\tau$ with the hierarchical partitioning $(\sset_0,\sset_1,\ldots,\sset_L)$, in a natural way as follows. 
The vertex set of $\tau$ is $V(\tau) \coloneqq V_0\cup \ldots \cup V_L$ (recall that for each $i$, $V_{i}=\set{v_S\mid S\in \sset_{i}}$, that is, $V_{i}$ contains, for each set $S\in \sset_{i}$, the supernode $v_S$ representing $S$). 
We call nodes in $V_i$ \EMPH{level-$i$ nodes} of tree $\tau$, and we call sets in $\sset_i$ \EMPH{level-$i$ sets}.
Since $\sset_L=\set{U}$, there is only one level-$L$ node in $\tau$, that we view as the root of $\tau$.
The edge set $E(\tau)$ contains, for each pair $S, \hat S$ of sets such that $S\in \sset_i, \hat S\in \sset_{i+1}$ for some $i$ and $S\subseteq \hat S$, an edge connecting $v_S$ to $v_{\hat S}$, so $v_S$ is a child node of $v_{\hat S}$, and in this case we also say that $S$ is a \EMPH{child set} of $\hat S$ and $\hat S$ is a \EMPH{parent set} of $S$. 
It is easy to verify from the construction that $\tau$ is indeed a tree.



\begin{observation}
\label{obs: sets non-crossing}
Let $S,S'$ be disjoint sets in $\sset$. Let $u_1,u_2$ be any pair of vertices in $S$, and let $u'_1,u'_2$ be any pair of vertices in $S'$. Then the pairs $(u_1,u_2)$ and $(u'_1,u'_2)$ of terminals are non-crossing in $H$.
\end{observation}
\begin{proof}
Assume for contradiction that the pairs $(u_1,u_2)$ and $(u'_1,u'_2)$ are crossing in $H$.
Assume that $S$ is a level-$i$ set and $S'$ is a level-$i'$ set, and assume without loss of generality that $i\ge i'$.

We first find another two pairs $(u_3,u_4), (u'_3,u'_4)$ of terminals such that $\dist_H(u_3,u_4)\le \mu^i$, $\dist_H(u'_3,u'_4)\le \mu^{i'}$ and the pairs $(u_3,u_4)$ and $(u'_3,u'_4)$ are crossing.
We start by finding the pair $(u_3,u_4)$.
In fact, if we denote by $\gamma_1$ the boundary segment clockwise from $u'_1$ to $u'_2$ around the outerface of $H$, and denote by $\gamma_2$ the boundary segment clockwise from $u'_2$ to $u'_1$ around the outerface of $H$, then since we have assumed that $(u_1,u_2)$ and $(u'_1,u'_2)$ are crossing, one of $u_1,u_2$ lies on $\gamma_1$ and the other lies on $\gamma_2$. Assume without loss of generality that $u_1$ lies on $\gamma_1$ and $u_2$ lies on $\gamma_2$. 

From the construction of graphs $R_1,\ldots,R_{i-1}$ and collections $\sset_1,\ldots,\sset_{i}$. It is easy to observe that, for every pair $u,u'$ of terminals that belong to the same level-$i$ set, there exists a sequence $u^1,\ldots, u^t$ of terminals in $U$ that all belong to the same level-$i$ set as $u$ and $u'$, such that, if we denote $u=u^0$ and $u'=u^{t+1}$, then for each $0\le j\le t$, $\dist_H(u^j,u^{j+1})\le \mu^i$; and for every pair $u,u'$ of terminals do not belong to the same level-$i$ set, $\dist_H(u,u')> \mu^i$.

Consider now the pair $u_1,u_2$ of terminals. Note that they belong to the same level-$i$ set. From the above discussion, there exists a sequence of terminals in $S$ starting with $u_1$ and ending with $u_2$, such that the distance between every pair of consecutive terminals in the sequence is less than $\mu^i$. Since $u_1$ lies on $\gamma_1$ and $u_2$ lies on $\gamma_2$, there must exist a pair $(u_3,u_4)$ of terminals appearing consecutively in the sequence, such that $u_3$ lies on $\gamma_1$ and $u_4$ lies on $\gamma_2$, so pairs $(u_3,u_4)$ and $(u'_1,u'_2)$ are crossing and $\dist_H(u_3,u_4)\le \mu^i$. 

We can then use similar arguments to find another pair $(u'_3,u'_4)$, such that the pairs $(u_3,u_4)$ and $(u'_3,u'_4)$ are crossing and $\dist_H(u'_3,u'_4)\le \mu^{i'}$. 
Note that, since $u_3,u_4\in S$ and $u'_3,u'_4\notin S$, $\dist_H(u_3,u'_3)> \mu^i$ and $\dist_H(u_4,u'_4)> \mu^i$.
Altogether, we get that
\[
\dist_H(u'_3,u'_4)+\dist_H(u_3,u_4)\le \mu^i+\mu^{i'}\le \mu^i+\mu^i<\dist_H(u_3,u'_3)+\dist_H(u_4,u'_4),
\]
a contradiction to the Monge property on the crossing pairs $(u_3,u_4)$ and $(u'_3,u'_4)$.
\end{proof}

\paragraph{Expanding sets.}  The central notion in the algorithm for the large spread case is the \emph{expanding sets}. 
Recall that $\eps'_r=r^{-0.7}$.
We say that a set $S\in \sset$ is \EMPH{expanding} if $|\hat S|\ge e^{\eps'_r}\cdot |S|$, where $\hat S$ is the parent set of $S$ (or equivalently, $v_{\hat S}$ is the parent node of $v_S$ in $\tau$); otherwise it is \EMPH{non-expanding}.
We now distinguish between two cases, depending on whether $\sset$ contains a non-expanding set with moderate size.

\subsubsection{The Balanced Case: there is a non-expanding set $S$ with $r/5 \le |S| \le 4r/5$}
\label{SSS:balanced}

We let $\hat S$ be the parent set of $S$. We denote $\EMPH{$S^*$} \coloneqq \hat S\setminus S$, and $\EMPH{$S'$} \coloneqq U\setminus \hat S$, so the sets $S^*$, $S$, and $S'$ partition set $U$. 
Moreover, we have $r/6\le |S|, |S'|\le 5r/6$ and $|S^*|\le (e^{\eps'_r}-1)r$.
We will employ the procedure $\cutpath$ in order to decompose the instance $(H,U)$ into smaller instances, for which we need to compute a non-crossing path set and a set of vertices in the path set, as the input to the procedure, as follows.

We say that an ordered pair $(u,u')$ of terminals in $S$ is a \EMPH{border pair} if the segment on the outer-boundary of $H$ from $u$ clockwise to $u'$ contains no other vertices of $S$ but at least one vertex of $S^*\cup S'$.
We compute the set \EMPH{$\mset$} of all border pairs in $S$, and then apply the algorithm from Lemma~\ref{lem: well-structured path set} to graph $H$ and the set of border pairs $\mset$, to obtain a set $\pset$ of shortest paths connecting pairs in $\mset$. 
We call $\pset$ the \EMPH{border path set} of $S$. 
It is easy to verify that set $\mset$ is non-crossing, and so path set $\pset$ is also non-crossing.

Consider now a border pair $(u,u')$ of terminals and let \EMPH{$P_{u,u'}$} be the $u$-$u'$ shortest path that we have computed.
We apply the algorithm from Lemma~\ref{lem: eps_cover_subset} to graph $H$, path $P_{u,u'}$ and each vertex $u^*\in S^*$ that lies on the segment of the outer-boundary of $H$ from $u$ clockwise to $u'$, with parameter $\eps_r$, and compute an $\eps_r$-cover of $u^*$ on $P_{u,u'}$. 
We then let \EMPH{$Y_{u,u'}$} be the union of all vertices in these $\eps_r$-covers and the endpoints of $P_{u,u'}$, so $Y_{u,u'}$ is a vertex set of $P_{u,u'}$.
Let \EMPH{$Y^*$} be the set of all vertices that are either an endpoint of a path in $\pset$ or have degree at least $3$ in the graph $\bigcup_{P\in \pset}P$.
We then define $\EMPH{$Y$} \coloneqq Y^*\cup (\bigcup_{(u,u')\in \mset}Y_{u,u'})$. From \Cref{thm: eps_cover}, 
\[
|Y\setminus Y^*|\le O\bigg(\frac{|S^*|}{\eps_r}\bigg)\le O\bigg(\frac{(e^{\eps'_r}-1)\cdot r}{\eps_r}\bigg)=
O\bigg(\frac{(1/r^{0.7})\cdot r}{\log^4 r/r^{0.1}}\bigg)=
O\bigg(\frac{r^{0.4}}{\log^4 r}\bigg).
\]

We then apply the procedure $\cutpath$ to the one-hole instance $(H,U)$, the non-crossing path set $\pset$, and the vertex set $Y$. We return the collection $\hset$ of one-hole instances output by the procedure $\cutpath$ as the output of our algorithm in this case.

\paragraph{Analysis of the Balanced Case.}

We now show that the output collection of one-hole instances of the above algorithm satisfies the properties required in Lemma~\ref{lem: decomposing step}.


\medskip
First, we show in the following claim that each instance in $\hset$ contains at most $(9/10)r$ terminals.
\begin{claim}
Each instance in $\hset$ contains at most $(9/10)r$ terminals.
\end{claim}

\begin{proof}
From the construction of the border path set $\pset$, the one-hole instances in $\hset$ can be partitioned into two subsets: 
$\hset_1$ contains all instances that corresponds to a region in $H$ surrounded by a segment of outer-boundary of $H$ and the image of some path $P\in \pset$; and set $\hset_2$ contains all other instances. 

Each instance in $\hset_1$ contains at most two terminals in $S$, and so it contains at most $r-|S|+2+|Y\setminus Y^*|\le (9/10)r$ terminals (note that such an instance does not need to contain branch vertices that are not $\eps_r$-cover vertices on its boundary). On the other hand, each instance in $\hset_2$ does not contain terminals in $S'$, and so it contains at most $r-|S'|+|Y|\le (9/10)r$ terminals. 
\end{proof}

Second, note that $|Y\setminus Y^*|\le O(r^{0.4}/\log^4 r)$, then from Claim~\ref{clm: branch pts}, we get that $\sum_{(H_i,U_i)\in \hset}|U_i|\le O(r)$ and $\sum_{(H_i,U_i)\in \hset : |U_i|> \lambda}|U_i|
\le r\cdot\big(1+O(1/\lambda)\big)$.


We now construct an algorithm $\algmerge$ that satisfies the required properties in Lemma~\ref{lem: decomposing step}. Recall that we are given, for each instance $(H_i,U_i)\in \hset$, an $\eps$-emulator $(Z_i,U_i)$.
The algorithm $\algmerge$ simply applies $\gluepath$ to instances $(Z_1,U_1),\ldots,(Z_s,U_s)$ and returns instance $(Z,U)$ output by $\gluepath$. It remains to show that the algorithm $\algmerge$ satisfies the required properties. Note that the one-hole instances $(H_1,U_1),\ldots,(H_s,U_s)$ also form a valid input for procedure $\gluepath$. Let \EMPH{$(\hat H, \hat U)$} be the one-hole instance that the procedure $\gluepath$ outputs when it is applied to instances $(H_1,U_1),\ldots,(H_s,U_s)$. It is easy to verify that $\hat U=U$. 
We use the following claim, 
whose proof is similar to the proof of Claim~\ref{clm: ratio loss for contracting to portals}, and is deferred to \Cref{apd: Proof of ratio loss for gluepathset}.

\begin{claim}
\label{clm: ratio loss for gluepathset}
Instance $(\hat H, U)$ is an $O(\eps_r)$-emulator for instance $(H, U)$.
\end{claim}

Now we complete the proof of Lemma~\ref{lem: decomposing step} for the Balanced Case using Claim~\ref{clm: ratio loss for gluepathset}.
In fact, since for each $1\le i\le t$, $(Z_i,U_i)$ is an  $\eps$-emulator for $(H_i,U_i)$, from Claim \ref{clm: glueset_emulators}, $(Z,U)$ is an  $\eps$-emulator for $(\hat H,U)$. Then from Claim~\ref{clm: ratio loss for contracting to portals} and Claim~\ref{clm: ratio loss for gluepathset}, we get that $(Z,U)$ is an  $(\eps+O(\eps_r))=(\eps+O(\frac{\log^4 r}{r^{0.1}}))$-emulator for $(H,U)$.
Moreover, from the algorithm $\gluepath$, it is easy to verify that the instance $(Z,U)$ output by the algorithm $\algmerge$ satisfies that $|V(Z)|\le\sum_{(H_i,U_i)\in \hset}|V(Z_i)|$.


\subsubsection{The Unbalanced Case: every set $S$ is either expanding, or $|S|<r/5$,  or $|S|> 4r/5$}
\label{SSS:unbalanced}

The algorithm in this case consists of two steps. Eventually, we will reduce to the Small Spread Case, and use the algorithm there to complete the decomposition of the instance $(H,U)$.

\paragraph{Step 1:}
We say that a set $S\in \sset$ is \EMPH{heavy} if $|S|>4r/5$, and in this case we also say that the node $v_S$ is heavy. Clearly, every level of $\tau$ contains at most one heavy node, and all heavy nodes form a path in $\tau$ which ends at the root node of $\tau$. Let \EMPH{$\hat S$} be the non-expanding heavy set that lies on the lowest level. 
We denote by \EMPH{$\hat L$} the level that $\hat S$ lies in and let \EMPH{$\check S$} be its parent set. 
Define $\EMPH{$\hat S^*$} \coloneqq \check S\setminus \hat S$ and $\EMPH{$\hat S'$} \coloneqq U\setminus \check S$. 
So sets $\hat S^*, \hat S, \hat S'$ partition set $U$, and $|\hat S^*|\le (e^{\eps'_r}-1)r$. We perform the same operations as in the Balanced Case (Section~\ref{SSS:balanced}) to graph $H$ with respect to the partition $(\hat S, \hat S^*, \hat S')$. 
Let \EMPH{$\hat \hset$} be the collection we obtain. 
From similar analysis as in Section~\ref{SSS:balanced}, we get that $\sum_{(H_i,U_i)\in \hat\hset}|U_i|\le O(r)$, and $\sum_{(H_i,U_i)\in \hat\hset : |U_i|> \lambda}|U_i|
=r\cdot \big(1+O(1/\lambda)\big)$.
If additionally we have, for each $(H_i,U_i)\in \hat \hset$, $|U_i|\le (9/10)r$, then we simply return the collection $\hat\hset$ as the output. Assume now that there exists some instance $(H_{i^*},U_{i^*})\in \hat \hset$ with $|U_{i^*}|> (9/10)r$. 
Note that we may  have only one such instance.
It is easy to see from the algorithm $\cutpath$ that no terminal of $U_{i^*}$ is a cut vertex in graph $H_{i^*}$. 
Note that it is now enough to prove Lemma~\ref{lem: decomposing step} for the instance $(H_{i^*}, U_{i^*})$, which we do in the next step. 
Indeed, if Lemma~\ref{lem: decomposing step} holds for instance $(H_{i^*},U_{i^*})$, then we simply apply the algorithm from Lemma~\ref{lem: decomposing step} to instance $(H_{i^*}, U_{i^*})$ and obtain a collection $\hset^*$ instances. 
We simply return the collection $\tilde \hset \coloneqq (\hat\hset\setminus \set{(H_{i^*}, U_{i^*})})\cup \hset^*$. It is easy to verify that the output collection $\tilde \hset$ satisfies all conditions in Lemma~\ref{lem: decomposing step} for the original input instance $(H,U)$ (where again we simply set $\algmerge$ to be $\gluepath$).


\paragraph{Step 2:}
The goal of this step is to further modify and decompose the instance $(H_{i^*}, U_{i^*})$ into instances with small spread, and eventually apply the algorithm from the Small Spread Case to them.
Consider the instance $(H_{i^*}, U_{i^*})$. 
From the algorithm $\cutpath$, the instance $(H_{i^*}, U_{i^*})$ corresponds to a region of $H$, that is surrounded by shortest paths connecting terminals in $U$. Therefore, for every pair $v,v'$ of vertices in $H_{i^*}$ (that are also vertices in $H$), $\dist_H(v,v')=\dist_{H_{i^*}}(v,v')$.
Note that set $U_{i^*}$ can be partitioned into two subsets: set $\tilde S$ contains all terminals in $\hat S$ that lies in $U_{i^*}$, and set $Y_{i^*}$ contains all new terminals (which are vertices in $\eps_r$-covers of vertices of $\hat S^*$ on paths of $\pset$ and the branch vertices) added in Step~1 that lie on the boundary of graph $H_{i^*}$. Note that the distances between a pair of terminals in $Y_{i^*}$ and the distances between a terminal in $Y_{i^*}$ and a terminal in $\tilde S$ could be very small (even much smaller than $\min_{u,u'}\dist_H(u,u')$) at the moment, which makes it hard to bound the spread from above. 
Therefore, we start by modifying the instance $(H_{i^*},U_{i^*})$ as follows. 

\begin{figure}[h!]
	\centering
	\subfigure[Before: the instance $(H_{i^*},U_{i^*})$.]{\scalebox{0.45}{\includegraphics[scale=0.25]{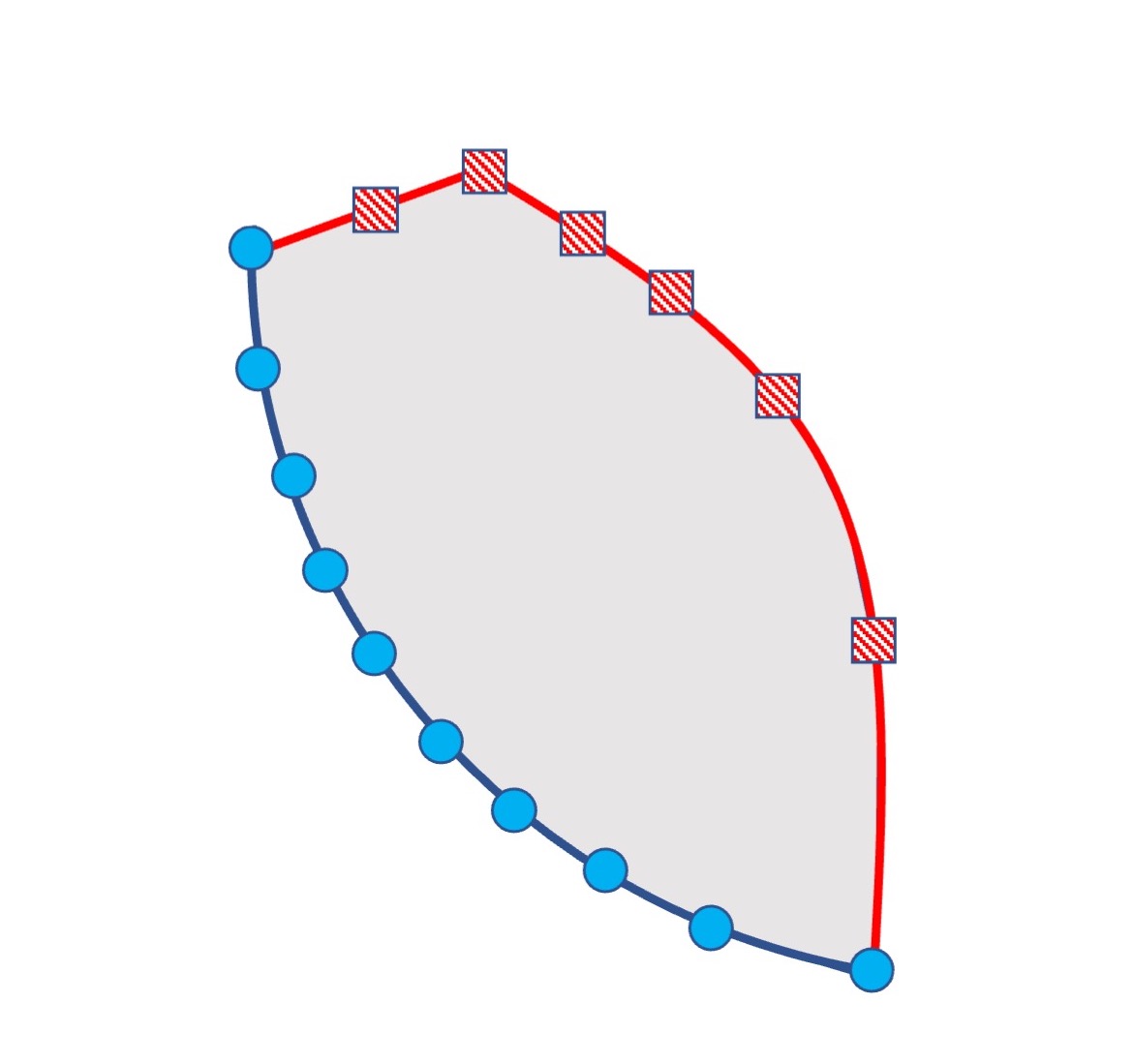}}}
	\hspace{0.8cm}
	\subfigure[After: the instance $(\tilde H,\tilde U)$.]%
	{\scalebox{0.45}{\includegraphics[scale=0.25]{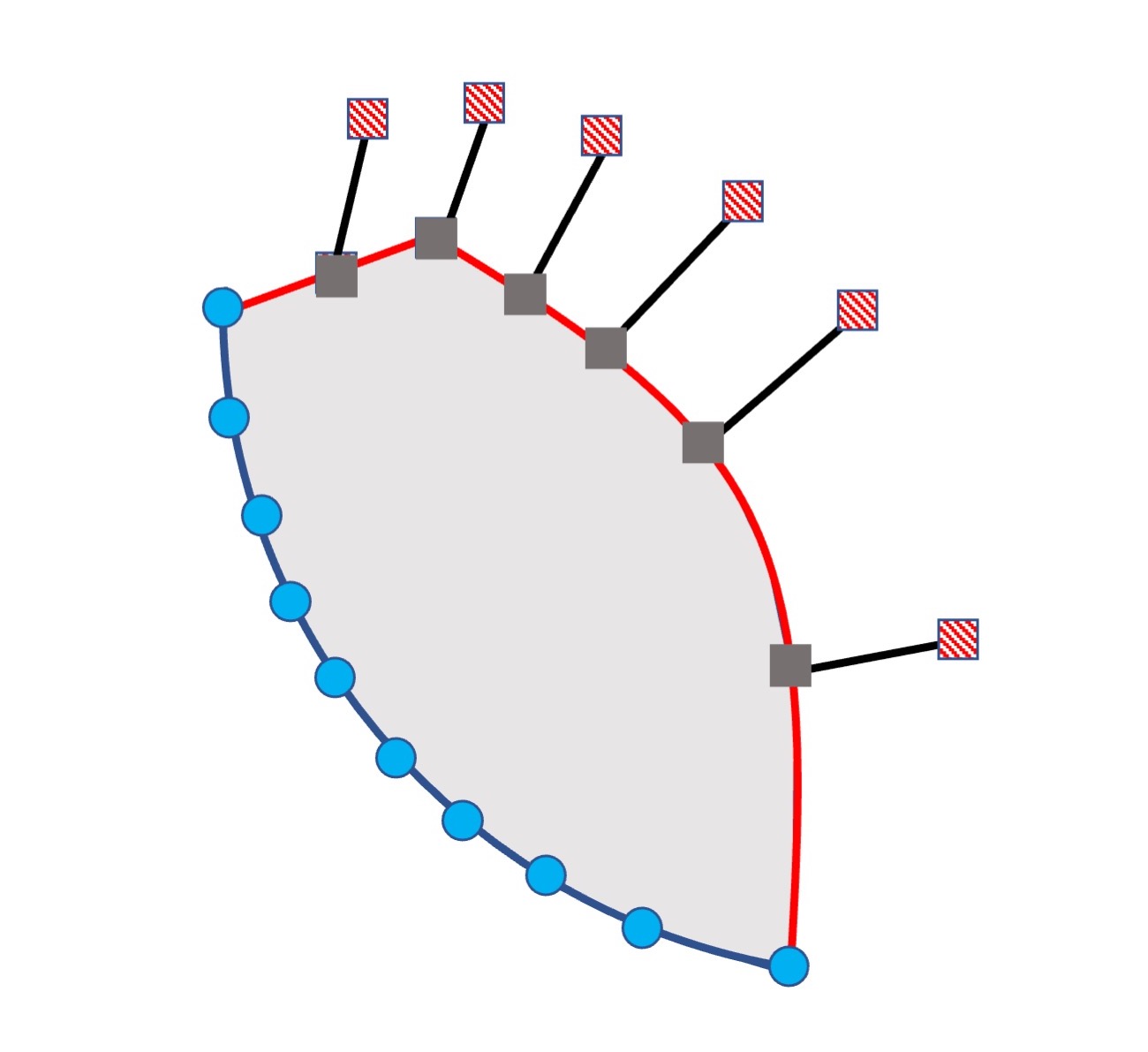}}}
	\caption{An illustration of modifying the instance $(H_{i^*}, U_{i^*})$.\label{fig: central}}
\end{figure}

\bigskip
We let graph \EMPH{$\tilde H$} be obtained from $H_{i^*}$ by adding, for each terminal $u\in Y_{i^*}$, a new vertex $\tilde u$ and an edge $(\tilde u, u)$ with weight $\mu^{\hat L-1}$. We then define $\EMPH{$\tilde U$} \coloneqq \tilde S\cup \set{\tilde u\mid u\in Y_{i^*}}$. This completes the construction of the new instance $(\tilde H, \tilde U)$. 
We call this operation \EMPH{terminal pulling}. 
See \Cref{fig: central} for an illustration. 
It is easy to verify that $(\tilde H, \tilde U)$ is a one-hole instance, and moreover, for each new terminal $\tilde u$ in $\tilde U\setminus \tilde S$, the distance in $\tilde H$ from $\tilde u$ to any other terminal in $\tilde U$ is at least $\mu^{\hat L-1}$. 
We will show later in the analysis that it is now sufficient to prove Lemma~\ref{lem: decomposing step} for the instance $(\tilde H, \tilde U)$.

We now construct the hierarchical clustering \EMPH{$\tilde \sset$} for instance $(\tilde H, \tilde U)$, in the same way as the hierarchical clustering $\sset$ for instance $(H,U)$, that is described at the beginning of the large spread case. 
Let $\tilde \tau$ be the partitioning tree associated with $\tilde \sset$. 
Recall that for every pair of vertices in $H_{i^*}$, the distance between them in $H_{i^*}$ is identical to the distance between them in $H$.
From the construction of instance $(\tilde H,\tilde U)$, it is easy to verify that both $\tilde\sset$ and $\tilde \tau$ has depth $\hat L$, and in levels $\hat L-1,\ldots, 1$, new terminals in $\tilde U\setminus \tilde S$ only form singleton sets as each of them is at distance at least $\mu^{\hat L-1}$ from any other terminal in $\tilde U$. Therefore, every non-singleton set in $\tilde \sset$ is also a set in $\sset$.


\medskip
\noindent We say that a set $S$ is \EMPH{good} if 
\begin{enumerate}
[label=(\roman*), ref=(\roman*)]
\item $|S|>1$; 
\item $S$ lies on level at most $\hat L-2\log r/\eps'_r$; 
\item $S$ is non-expanding; and 
\item for any other set $S'\in \tilde \sset$ that lies on level at most $\hat L-2\log r/\eps'_r$ and $S\subseteq S'$, $S'$ is expanding.
\end{enumerate}
We denote by \EMPH{$\tilde \sset_g$} the collection of all good sets in $\tilde \sset$.
Next we show that all (good) sets in $\tilde\sset_g$ lie on level at least $\hat L-O(\log r/\eps'_r)$. 
From definition of a good set and our assumption for the Unbalanced Case that every set $S\in \sset$ with $r/5\le |S|\le 4r/5$ is expanding, it is easy to see that all good sets $S$ have size at most $r/5$ (we have used the property that every non-singleton set in $\tilde \sset$ is also a set in $\sset$).

\begin{observation}
\label{obs: good set level}
Every good set in $\tilde\sset$ lies on level at least $\hat L-10\log r/\eps'_r$. 
Every terminal either forms a singleton set on level at least $\hat L-10\log r/\eps'_r$, or belongs to some good set in $\tilde\sset_g$.
\end{observation}
\begin{proof}
Denote $\hat L' \coloneqq \hat L-2\log r/\eps'_r$.
Let $S$ be a good set. Assume $S$ lies in level $i$. Let $S_{i+1},\ldots,S_{\hat L'}$ be the ancestor sets of $S$ on levels $i+1,\ldots, \hat L'$, respectively. From the definition of good sets, all sets $S_{i+1},\ldots,S_{\hat L'-1}$ are expanding, so we have
\[
1\le |S|\le |S_{i+1}|\le e^{-\eps_r}\cdot |S_{i+2}|\le \cdots\le e^{-\eps'_r\cdot(\hat L'-i-1)}\cdot |S_{L^*}|\le  e^{-\eps'_r\cdot(\hat L'-i-1)}\cdot r.
\]
Therefore, $\eps_r\cdot(\hat L'-i-1)\le \ln r$ and so $i=\hat L'-8\log r/\eps'_r=\hat L-10\log r/\eps'_r$.

Similarly, if a terminal in $\tilde S$ does not form a singleton set on level at least $\hat L-10\log r/\eps'_r$, and it does not belong to any good set in $\tilde\sset_g$, then from the inequality above, its ancestor chain has length at most $8\log r/\eps'_r$, a contradiction.
\end{proof}

Now for each good set $S$, we compute its border path set $\tilde \pset_S$ in instance $(\tilde H, \tilde U)$ in the same way as in the Balanced Case (Section~\ref{SSS:balanced}). Now define $\EMPH{$\tilde\pset$} \coloneqq \bigcup_{S\in \tilde \sset_g}\tilde\pset_S$.
We show in the next observation that the collection $\pset$ of paths is non-crossing.

\begin{observation}
The collection $\tilde\pset$ of paths is non-crossing.
\end{observation}
\begin{proof}
Assume for contradiction that the collection $\tilde\pset$ of paths is not non-crossing. Then there exist two distinct sets $S,S'\in \tilde\sset_g$, a border path $P$ connecting terminals $u_1,u_2$ in $S$ and a border path $P'$ of $S'$ connecting terminals $u'_1,u'_2$ in $S'$, such that the pairs $(u_1,u_2), (u'_1,u'_2)$ are crossing. However, from the definition of good sets, $S\cap S'=\varnothing$. Therefore, from Observation~\ref{obs: sets non-crossing}, pairs $(u_1,u_2), (u'_1,u'_2)$ are non-crossing, a contradiction.
\end{proof}

Consider now a good set $S\in \tilde\sset_{g}$.
We define $\EMPH{$S^*$} \coloneqq \check S\setminus S$, where \EMPH{$\check S$} is the parent set of $S$ in $\tilde \sset_g$.
Recall that a pair $(u,u')$ of terminals in $S$ is a border pair, if the outer-boundary of $\tilde H$ connecting $u$ to $u'$ contains no other vertices of $S$ but at least one vertex that does not lie in $S$.
Now for each border pair $(u,u')$ of terminals in $S$, let \EMPH{$P_{u,u'}$} be the $u$-$u'$ shortest path in $\tilde\pset_S$ that we have computed. We apply the algorithm from Lemma~\ref{lem: eps_cover_subset} to each vertex $u^*\in S^*$ that lies on the outer-boundary from $u$ clockwise to $u'$ with parameter $\eps_r$, and compute an $\eps_r$-cover of $u^*$ on $P_{u,u'}$. 
We then let \EMPH{$Y^S_{u,u'}$} be the union of all such $\eps_r$-covers and the endpoints of $P_{u,u'}$. We then let set \EMPH{$Y^S$} be the union of the sets $Y^S_{u,u'}$ for all border pairs $(u,u')$. 
Finally, we define \EMPH{$Y$} as the union of $\bigcup_{S\in \tilde \sset_g} Y^S$ and all branch vertices (which we denote by $Y^*$), so $Y$ is a vertex set of $V(\tilde\pset)$ that contains all branch vertices $\tilde\pset$.
Moreover, from \Cref{thm: eps_cover},
\[
\begin{split}
|Y\setminus Y^*| &  \le O\bigg(\sum_{S\in \tilde \sset_g}\frac{|S^*|}{\eps_r}\bigg)
\le O\bigg(\frac{(e^{\eps'_r}-1)\cdot \sum_{S\in \tilde \sset_g}|S|}{\eps_r}\bigg)
\le O\bigg(\frac{(e^{\eps'_r}-1)\cdot r}{\eps_r}\bigg)\\
& =
O\bigg(\frac{(1/r^{0.7})\cdot r}{\log^4 r/r^{0.1}}\bigg)=
O\bigg(\frac{r^{0.4}}{\log^4 r}\bigg).
\end{split}
\]

We now apply the algorithm $\cutpath$ to instance $(\tilde H,\tilde U)$, the path set $\tilde\pset$ and the vertex set $Y$. 
Let $\tilde \hset$ be the collection of one-hole instances we get. 
If all instances $(\hat H, \hat U)$ in $\tilde \hset$ satisfy that $|\hat U|\le (9/10)r$, then we terminate the algorithm and return $\tilde\hset$.
Assume that there is some instance $(\hat H, \hat U)$ in $\tilde \hset$ such that $|\hat U|> (9/10)r$. From similar analysis in Step 1, there can be at most one such instance. We denote such an instance by $(\hat H, \hat U)$.


We now modify the instance $(\hat H, \hat U)$ as follows. 
Denote $\EMPH{$L^*$} \coloneqq \hat L-10\log r/\eps'_r$. 
Let \EMPH{$H^*$} be the graph obtained from $\hat H$ by applying the terminal pulling operation to every terminal in $\hat U\setminus \tilde S$ via an edge of weight $\mu^{L^*-1}$. 
We then define set \EMPH{$U^*$} to be the union of $(\hat U\cap \tilde S)$ and the set of all new terminals created in the terminal pulling operation.
We use the following observation.
\begin{observation}
$\Phi(H^*,U^*)\le 2^{O(\log^2 r/\eps'_r)}$.
\end{observation}
\begin{proof}
From Observation~\ref{obs: good set level}, every pair of terminals in $U^*$ has distance at least $\mu^{L^{*}-1}$ in graph $H^*$. On the other hand, since graph $\hat H$ is a subgraph of $\tilde H$, every pair of terminals in $U^*$ has distance at most $\mu^{\hat L+1}$ in graph $H^*$. Therefore, $\Phi(H^*,U^*)\le \mu^{\hat L-L^{*}+2}=2^{O(\log^2 r/\eps'_r)}$ as $\mu=r^2$.
\end{proof}

Since $2^{O(\log^2 r/\eps'_r)}<2^{r^{0.9}\log^2 r}$ when $r$ is larger than some large enough constant, we apply the algorithm from the Small Spread Case to instance $(H^*, U^*)$ and obtain a collection $\hset_{(\hat H,\hat U)}$ of instance.
The output of the algorithm is the collection
$\Big(\tilde \hset\setminus\set{(\hat H,\hat U)}\Big)\cup \hset_{(\hat H,\hat U)}$ of instances.

\paragraph{Analysis of the Unbalanced Case.}
Recall that in this step we assume that, after Step 1, there is an instance $(H_{i^*}, U_{i^*})$ with $|U_{i^*}|>(9/10)r$, and we transformed it into another instance $(\tilde H, \tilde U)$.
We first show that it is sufficient to prove Lemma~\ref{lem: decomposing step} for instance $(\tilde H, \tilde U)$. All other conditions can be easily verified. We now show that when applying the algorithm $\gluepath$ to  $\eps$-emulators $\set{(\tilde H', \tilde U)}\cup\set{(H'_i, U_i)}_{i\ne i^*}$, we still obtain an  $(\eps+ O(\frac{\log^4 r}{r^{0.1}}))$-emulator for $(H,U)$. In fact, we only need to consider the terminal pairs $u,u'$ with $u\in S$ and $u'\notin S$. Note that such a pair $u,u'$ of terminals belongs to different level-$\hat L$ clusters in $\sset$. From the construction of $\tilde \sset$, $\dist_H(u,u')\ge \mu^{\hat L}$. Therefore, the transformation from instance $(H_{i^*}, U_{i^*})$ to instance $(\tilde H, \tilde U)$ adds at most an additive $\mu^{\hat L-1}$ to their distance, which is at most $O(\frac{1}{\mu})=O(\frac{1}{r^2})\le O(\frac{\log^4 r}{r^{0.1}})$-fraction of their distance in graph $H$.
Therefore, by gluing the  $\eps$-emulators $\set{(\tilde H', \tilde U)}\cup\set{(H'_i, U_i)}_{i\ne i^*}$, we still obtain an  $(\eps+ O(\frac{\log^4 r}{r^{0.1}}))$-emulator for $(H,U)$.

From now on, we focus on proving that the decomposition we computed for instance $(\tilde H, \tilde U)$ satisfies all properties in Lemma~\ref{lem: decomposing step}.
Recall that we have first computed a collection $\tilde \sset_g$ of good sets,  computed a path set $\tilde\pset$ and a subset $Y$ of vertices in $V(\tilde \pset)$ based on sets in $\tilde \sset_g$, and then applied the procedure $\cutpath$ to $((\tilde H, \tilde U), \tilde\pset, Y)$ and obtained a collection $\tilde \hset$ of one-hole instances. 

Assume first that all instances $(\hat H, \hat U)$ in collection $\tilde \hset$ satisfies that $|\hat U|\le (9/10)r$. Since $|Y\setminus Y^*|\le O\big(\frac{r^{0.4}}{\log^4 r}\big)$, from Claim~\ref{clm: branch pts}, we get that $\sum_{(\hat H,\hat U)\in \tilde\hset}|\hat U|\le O(r)$ and $\sum_{(\hat H,\hat U)\in \tilde\hset : |\hat U|> \lambda}|\hat U|\le r\cdot \big(1+O(1/\lambda)\big)$.
We now describe the algorithm $\algmerge$ that, takes as input, for each instance $(\hat H, \hat U)\in \hset$, an $\eps$-emulator $(\hat H', \hat U)$, computes an $\big(\eps+O(\eps_r)\big)=\big(\eps+O(\frac{\log^4 r}{r^{0.1}})\big)$-emulator for $(\tilde H, \tilde U)$.
We simply apply the algorithm $\gluepath$ to instances $\set{(\hat H',\hat U)\mid (\hat H, \hat U)\in \tilde \hset}$ and return the output instance $(\tilde H', \tilde U)$ of $\gluepath$.
The proof that instance $(\tilde H', \tilde U)$ is indeed an $\big(\eps+O(\eps_r)\big)$-emulator for $(\tilde H, \tilde U)$ and the proof that $|V(\tilde H')|\le \sum_{(\hat H, \hat U)\in \tilde \hset}|V(\hat H')|$ use identical arguments in the Balanced Case, and is omitted here.

Assume now that there exists an instance $(\hat H, \hat U)$ in collection $\tilde \hset$ with $|\hat U|> (9/10)r$. Denote $\tilde \hset'=\tilde \hset\setminus \set{(\hat H, \hat U)}$ and denote by $\overline\hset=\big(\tilde \hset\setminus\set{(\hat H,\hat U)}\big)\cup \hset_{(\hat H,\hat U)}$ the output collection of instances.
First, note that all instances $(\overline H, \overline U)$ in collection $\tilde \hset'$ satisfies that $|\overline U|\le (9/10)r$. 
Since the remaining instances in $\overline\hset$ is obtained by applying the algorithm from Case 1 to the instance $(H^*, U^*)$, that is obtained from modifying the unique large instance in $(\hat H, \hat U)$. From the algorithm in Case 1, we know that each instance in the output collection contains at most $(9/10)r$ terminals.
Second, from similar arguments, we get that 
$\sum_{(\overline H,\overline U)\in \overline\hset}|\overline U|\le O(r)$ and $\sum_{(\overline H,\overline U)\in \overline\hset : |\overline U|> \lambda}|\overline U|\le r\cdot \big(1+O(1/\lambda)\big)$.
We now describe the algorithm $\algmerge$ that, takes as input, for each instance $(\overline H, \overline U)\in \hset$, an $\eps$-emulator $(\overline H', \overline U)$, computes an $\big(\eps+O(\eps_r)\big)=\big(\eps+O(\frac{\log^4 r}{r^{0.1}})\big)$-emulator for $(\tilde H, \tilde U)$.
First, consider the instances in $\hset_{(\hat H,\hat U)}$ that are obtained from applying the algorithm in Case 1 to $(H^*, U^*)$. We simply use the algorithm $\algmerge$ described in Case 1 to compute an $\big(\eps+O(\eps_r)\big)$-emulator $(H^{**}, U^*)$ for instance $(H^*, U^*)$. 
Finally, we apply the algorithm $\gluepath$ to instances in $\set{ (\overline H',\overline U)\mid (\overline H,\overline U)\in \tilde \hset'}\cup \set{(H^{**}, U^*)}$ and denote the obtained instance by $(\tilde H', \tilde U)$.
Note that, for different sets $S,S'\in \tilde\sset_g$ such that $S\cap \hat U\ne\varnothing, S'\cap \hat U\ne\varnothing$ and $S\cap S=\varnothing$, if set $S$ lies on level $i$ and set $S'$ lies on level $i'$, then $\dist_{H}(S,S')\ge \mu^{(\max\set{i,i'}+1)}\ge \mu^{L^*}$. 
Therefore, from similar arguments at the beginning of the analysis, the terminal pulling operation only incur an multiplicative factor-$O(1/r)$ error of the distances between terminals in disjoint sets in $\tilde \sset_g$.


The rest of the proof that instance $(\tilde H', \tilde U)$ is indeed an $\big(\eps+O(\eps_r)\big)$-emulator for $(\tilde H,\tilde U)$ uses almost identical arguments in the Balanced Case, and is omitted here.

\subsection{Near-linear Time Implementation of Lemma~\ref{lem: decomposing step}}
\label{sec: 1hole_near-linear time}

Denote $n \coloneqq |V(H)|$.
In this subsection we show that the algorithm described in this section can be implemented in time $O\big( (n+r^2)\cdot\log r \cdot \log n \big)$. 

The first step of the algorithm is to split the input instance $(H,U)$ into smaller instances at cut vertices. 
The cut vertices of the plane graph $H$ are simply the vertices encountered more than once when we traverse the boundary of the outerface of $H$, and so they can be computed in $O(n)$ time. Therefore, the algorithm in \Cref{subsec: Remove All Cut Vertices} can be implemented in $O(n)$ time.

Consider now the step in \Cref{SSS:small-spread}. In this step we first compute the closest $(3/4)$-balanced pair of terminals in $U$. 
We show that this can be done in $O(n\log n+r^2\log n)$ time. In fact, we use the algorithm in \cite{kle-msppg-2005} to compute an MSSP data structure of graph $H$, which takes time $O(n\log n)$. 
We then query the distances between every pair of terminals in $U$, which takes time $O(r^2\log n)$ as the query time of the MSSP data structure is $O(\log n)$. 
We can then use the acquired information to compute the closest $(3/4)$-balanced pair of terminals in $U$ by simply dropping all the unbalanced pairs and sort.
Let this pair be $(u,u')$. Computing the $u$-$u'$ shortest-path in $H$ takes $O(n)$ time. Computing portals (vertices of $P$) takes $O(n)$ time. From \Cref{SS:split-and-glue}, the procedures $\cutpath$ and $\gluepath$ can be implemented in $O(n)$ time. Therefore, the total running time of the step in \Cref{SSS:small-spread} is $O(n\log n+r^2\log n)$.

Consider next the step in \Cref{SSS:large-spread}. In this step we first compute a hierarchical clustering of terminals in $U$, according to their distances in $H$. This can be done in $O(n\log n+r^2\log n)$ time. In fact, we can similarly use the MSSP data structure in \cite{kle-msppg-2005} and query the distances between every pair of terminals in $U$, and then consider the complete graph $K_U$ on $U$ whose edge weights are distances between pairs of its endpoints returned by the MSSP data structure. It is easy to see that, in order to construct the hierarchical clustering $\sset$, every edge of $K_U$ needs to be visited at most $O(1)$ times. Therefore, the construction of hierarchical clustering takes in total $O(n\log n+r^2\log n)$ time. Note that $\sset$ is a hierarchical clustering on a collection of $r$ elements, so $\sset$ contains at most $O(r)$ distinct sets. Since deciding whether or not a set in $\sset$ is expanding or not takes $O(1)$ time, we can tell in $O(r)$ time whether we are in the Balanced Case or the Unbalanced Case.

\begin{itemize}
\item
In the Balanced Case, the next steps are to compute border pairs, border path sets, $\eps_r$-covers and to use procedure $\cutpath$ to obtain smaller instances. From \Cref{lem: well-structured path set} and Lemma~\ref{lem: eps_cover_subset}, all these takes can be done in $O(n\log r)$ time.
\item
In the Unbalanced Case, the next steps are to first repeat apply the steps in the Balanced Case to the non-expanding set that lies on the lowest level. From the above discussion, this takes in total $O(n\log r)$ time. 
If we end up with one instance $(H_{i^*}, U_{i^*})$ with $|U_{i^*}|>(9/10)r$, we need a final step for further splitting this instance. It is easy to verify that the operation of terminal pulling can be done in $O(r)$ time. 
Constructing the new collection $\tilde \sset$ takes $O(n\log n+r^2\log n)$ time. Identifying good sets in $\tilde \sset$ takes $O(r)$ time. The remaining operations are computing border pairs, border path sets, $\eps_r$-covers and using procedure $\cutpath$ to obtain smaller instances. From the above discussion, all these takes can be done in $O(n\log r)$ time.
\end{itemize}

\noindent Altogether, the running time of the algorithm in this section is
$O\big( (n+r^2)\cdot\log r \cdot \log n \big)$.


\section{Emulator for Edge-Weighted Planar Graphs}
\label{sec: general graph}

In this section we provide the proof of Theorem~\ref{thm:main}.
%
%
In \Cref{sec: alg_O(1)hole},
we show an algorithm for computing $\eps$-emulators for $O(1)$-hole instances. 
Then in \Cref{sec: alg_general}, we complete the proof of Theorem~\ref{thm:main} using the results in \Cref{sec: alg_O(1)hole}.
We will prove in Section~\ref{SS:bootstrapping} that an $\e$-emulator of size $O_\e(k \polylog k)$ can be computed in $O_\e(n)$ time. 


\subsection{Emulator for $O(1)$-Hole Instances}
\label{sec: alg_O(1)hole}

In this subsection we present a near-linear time algorithm for constructing $\eps$-emulators for $O(1)$-hole instances. 
We first define \emph{aligned emulators} for $O(1)$-hole instances similarly as aligned emulators for one-hole instances, as follows. 
Let $(G,T)$ and $(G',T)$ be two $h$-hole instances. We denote by $\fset$ the set of holes in $G$ that contain the images of all terminals, and define $\fset'$ for $G'$ similarly, so $|\fset|=|\fset'|=h$.
We say that instances $(G,T)$ and $(G',T)$ are \emph{aligned}, if and only if there is a one-to-one correspondence between faces in $\fset$ and faces in $\fset'$, such that for every face $F\in \fset$, the set $T(F)$ of terminals that it contains is identical to the set $T(F')$ of terminals contained in its corresponding face $F'\in \fset'$, and moreover, the circular orderings in which the terminals of $T(F)$ appearing on faces $F$ and $F'$ are identical. If $(G,T)$ and $(G',T)$ aligned and $(G,T)$ is an $\e$-emulator for $(G',T)$, then we say that $(G,T)$ is an \emph{aligned $\e$-emulator} for $(G',T)$.
Throughout this section, all emulators we construct for various $O(1)$-hole instances are aligned emulators. Therefore, we will omit the word ``aligned'' and only refer to them by $\eps$-emulators or simply emulators.
The main result of this section is the following lemma.


\begin{lemma}
\label{L:emulator-constant_hole}
For any $0<\eps<1$ and any $h$-hole instance $(H,U)$ with $n \coloneqq |H|$ and $r \coloneqq |U|$, there exists an $h$-hole instance $(H',U)$ that is an $\e$-emulator for $(H,U)$ with size
$|V(H')|\le  r \cdot (c h\log r/\eps)^{c h}$ for some universal constant $c$.
Moreover, such an emulator can be computed in time $O\big((n+r^2)\cdot(h\log n/\e)^{O(h)}\big)$.
\end{lemma}


The remainder of this subsection is dedicated to the proof of Lemma~\ref{L:emulator-constant_hole}.
We first introduce basic algorithms $\cutpath_h$ and $\gluepath_h$ for splitting and gluing $h$-hole instances that are similar to the algorithms $\cutpath$ and $\gluepath$ for splitting and gluing one-hole instances in~\Cref{SS:split-and-glue}.

\begin{figure}[h!]
	\centering
	\subfigure[Graph $H$: holes $\alpha, \alpha'$ (shaded gray), terminals on $\alpha$ and $\alpha'$ (blue), path $P$ (red), vertices of $Y$ that are not endpoints of $P$ (purple). ]{\scalebox{0.5}{\includegraphics[scale=0.14]{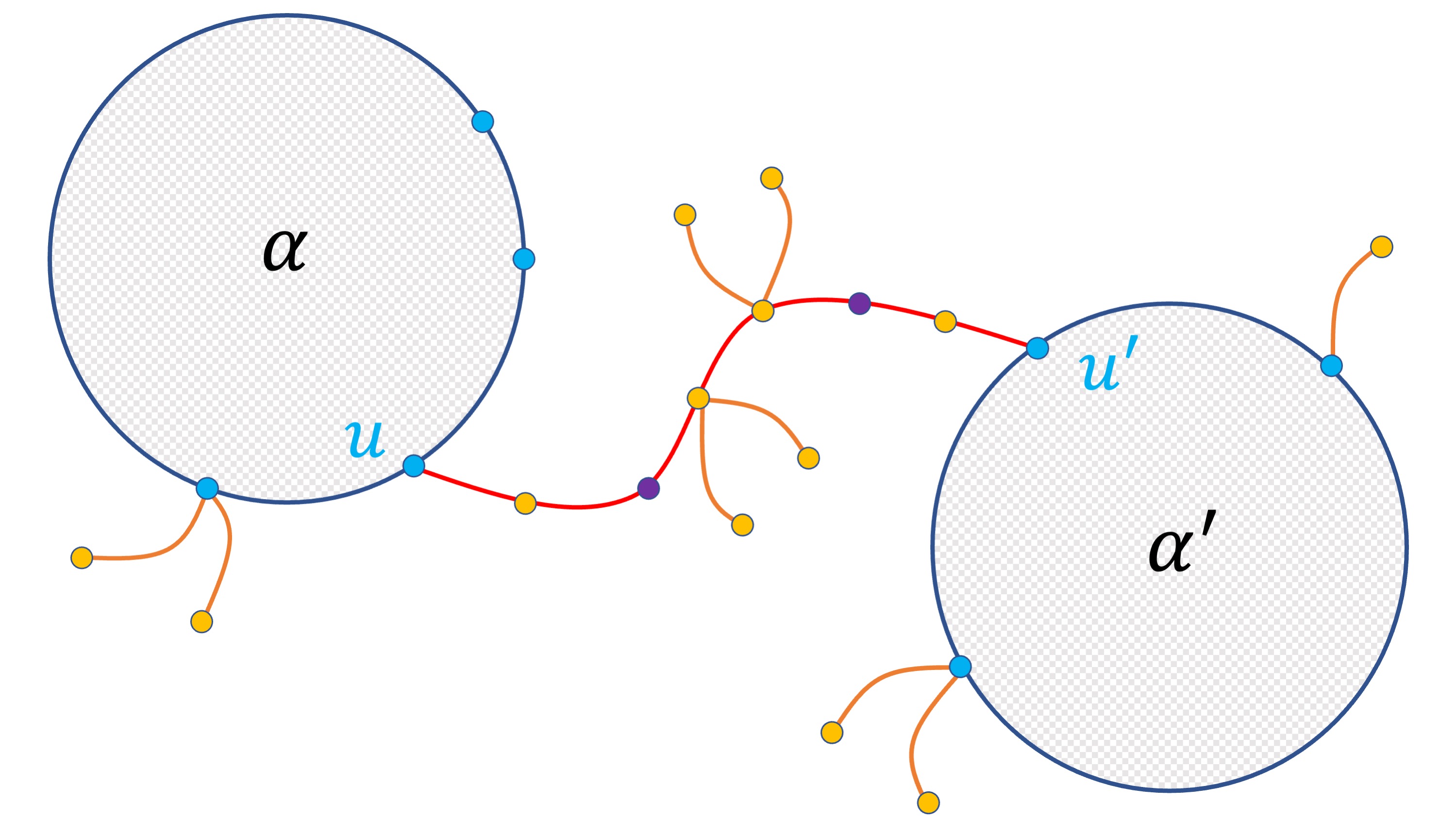}\label{fig: splitting h-hole before}}}
	\hspace{0.2cm}
	\subfigure[Graph $\tilde H$: the new hole $\beta$ (shaded gray), terminals on $\beta$ (blue and purple), and the new $u_1$-$u'_1$ path and $u_2$-$u'_2$ path (red).]{\scalebox{0.5}{\includegraphics[scale=0.14]{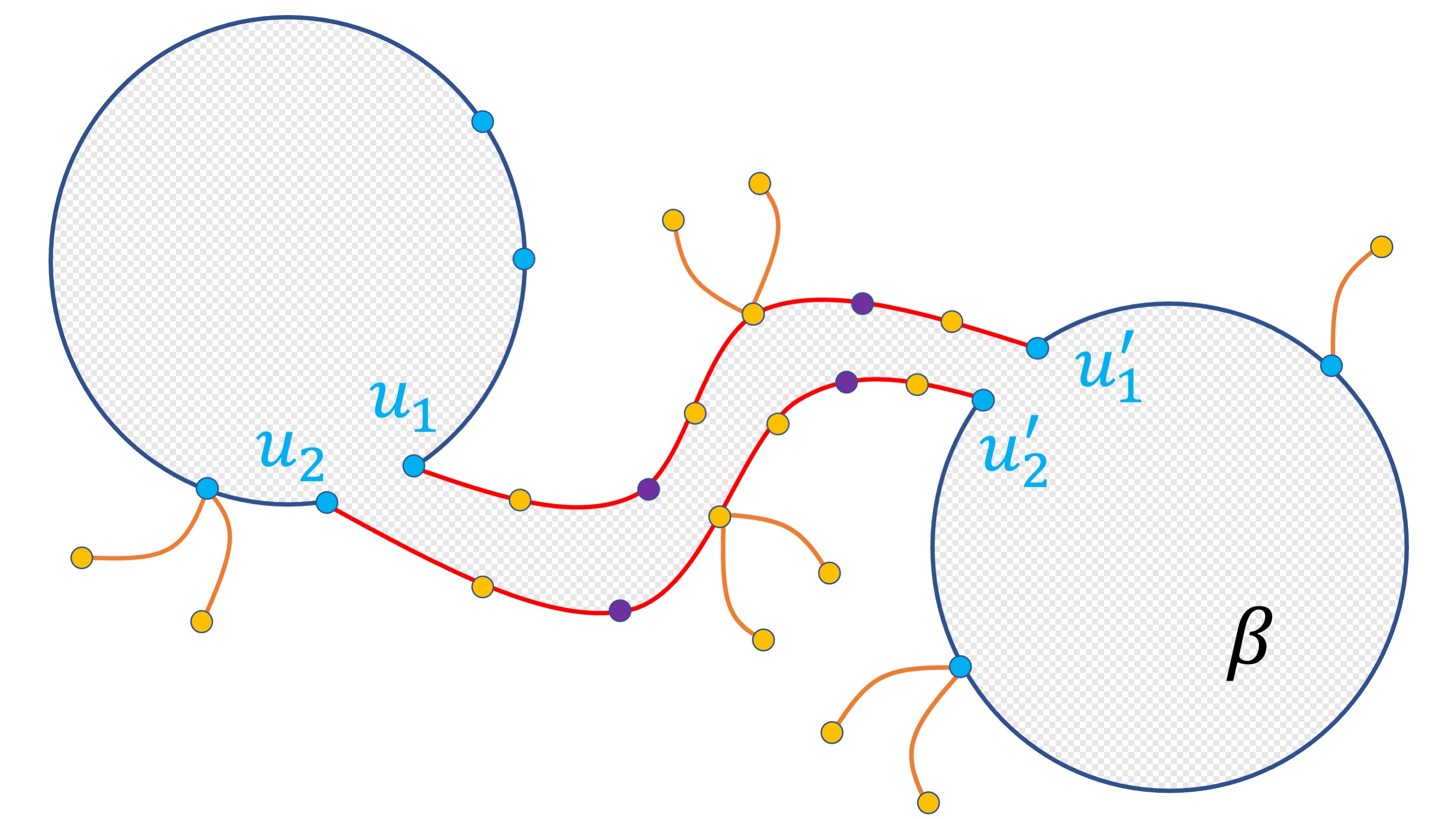}\label{fig: splitting h-hole after}}}
	\subfigure[An illustration of the output instance of $\gluepath_h$, when the input is the $(h-1)$-hole instances in \Cref{fig: splitting h-hole after}. Holes $\alpha$ and $\alpha'$ are restored.]{\includegraphics[scale=0.08]{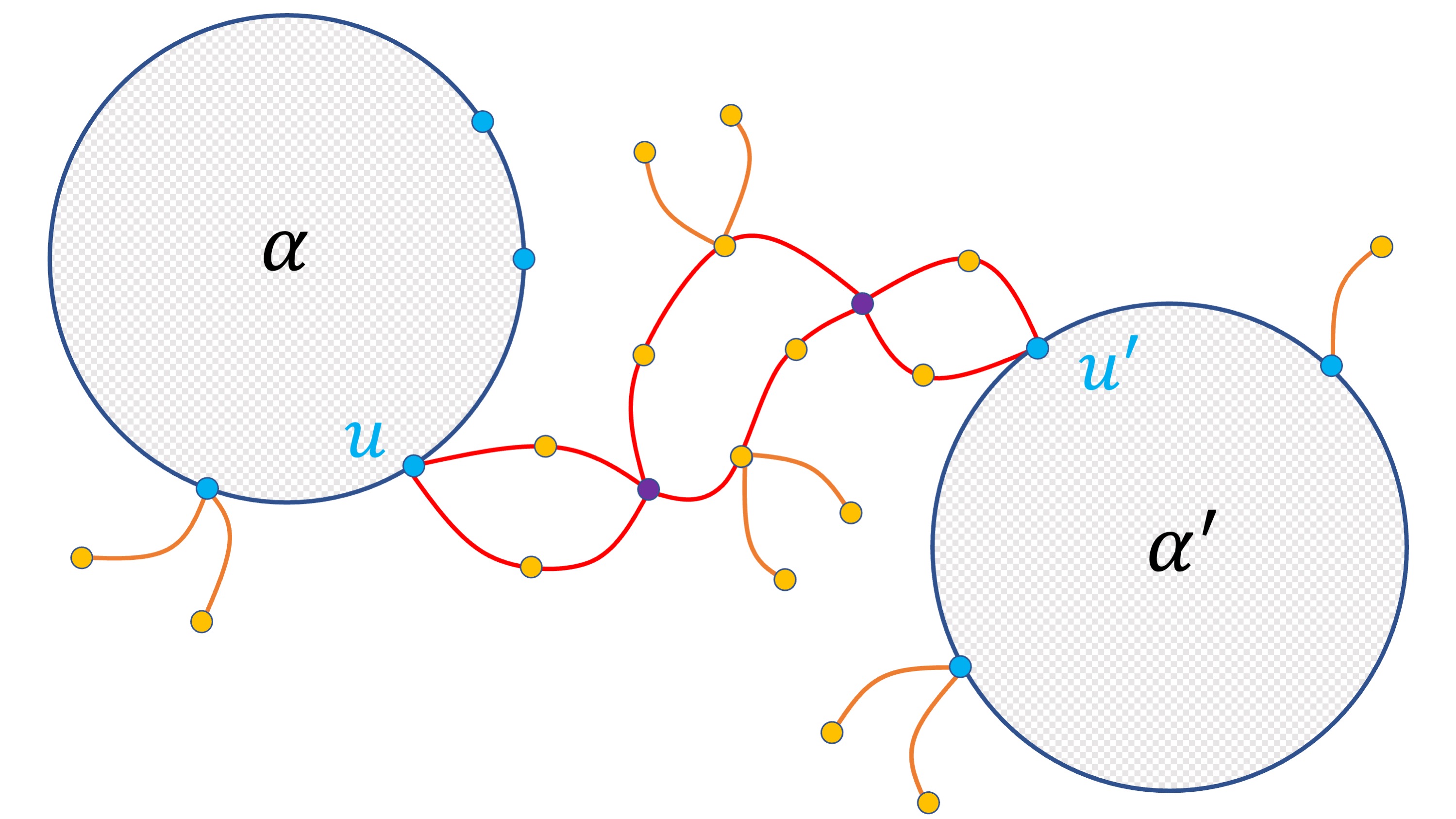}\label{fig: gluepath_2}}
	\caption{An illustration of splitting and gluing an $h$-hole instance along a path.\label{fig: splitting h-hole}}
\end{figure}

\paragraph{Splitting and Gluing.}
The input to procedure \emph{$\cutpath_h$} (for some integer $h>1$) consists of:
\begin{itemize}
\item an $h$-hole instance $(H,U)$; 
\item a path $P$ connecting a pair of terminals lying on two different holes; and
\item a set $Y\subseteq V(P)$ of vertices that contains both endpoints of $P$.
\end{itemize}
The output of $\cutpath_h$ is an $(h-1)$-hole instance.
Intuitively, \emph{$\cutpath_h$} slices the graph $H$ open along the path $P$ connecting two separate holes in the graph, as illustrated in Figure~\ref{fig: splitting h-hole after}.
We denote by $(\tilde H,\tilde U)$ the $(h-1)$-hole instance obtained by applying procedure $\cutpath_h$ to instance $(H,U)$, path $P$, and vertex set $Y$.
Intuitively, procedure \emph{$\gluepath_{h}$} takes as input an emulator for $(\tilde H,\tilde U)$,
and outputs an emulator for the original instance $(H,U)$ by identifying the two copies in $\tilde H$ of every vertex in $Y$, as illustrated in Figure~\ref{fig: gluepath_2}.
A complete description of these procedures is provided in \Cref{apd: cut and glue}.

Note that instance $(\tilde H,\tilde U)$ is also a valid input for procedure $\gluepath_h$. 
Let $(\hat H, \hat U)$ be the $h$-hole instance obtained by applying procedure $\gluepath_h$ to instance $(\tilde H,\tilde U)$. 
Clearly, $\hat U = U$.
We use the following claim, whose proof is similar to Claim~\ref{clm: glueset_emulators} and thus is deferred to Appendix~\ref{apd: Proof of split glue h holes}.

\begin{claim}
\label{clm: split glue h holes}
Let $(Z,U)$ be the instance obtained by applying procedure $\gluepath_h$ to an $\e$-emulator $(\tilde Z,\tilde U)$ of $(\tilde H,\tilde U)$.
Let $(\hat H,U)$ be the instance obtained by applying procedure $\gluepath_h$ to $(\tilde H,\tilde U)$.
Then $(Z,U)$ is an $\eps$-emulator for $(\hat H,U)$.
%
\end{claim}

We now complete the proof of Lemma~\ref{L:emulator-constant_hole} by induction on $h$. 
The base case (when $h=1$) follows from Theorem~\ref{Th:emulator-1hole}. 
Consider now the case where the input $(H,U)$ is an $h$-hole instance for $h>1$. 
We first compute a pair of terminals $(u,u')$ that lie on different holes, and a shortest path $P$ in $H$ connecting $u$ to $u'$, such that $P$ does not contain any terminal as internal vertices. 
Then for each $\hat u\in U\setminus \set{u,u'}$, we use the algorithm from \Cref{thm: eps_cover} and parameter $\EMPH{$\eps'$} \coloneqq \eps/h$ to compute an $\eps'$-cover of $\hat u$ on path $P$. Let $Y$ be the union of all such $\eps'$-covers together with the endpoints of $P$, so $Y\subseteq V(P)$. 
Note that from \Cref{thm: eps_cover} we have $|Y|\le O(|U|/\eps')\le O(rh/\eps)$, and by using the algorithm from Lemma~\ref{lem: eps_cover_subset}, $Y$ can be computed in $O(h\cdot n \log r)$ time.

Let $c$ be a large enough constant that is greater than all hidden constants in \Cref{Th:emulator-1hole}.
We then apply the procedure $\cutpath_h$ to the $h$-hole instance $(H,U)$, the path $P$ and the vertex set $Y$. Let $(\tilde H,\tilde U)$ be the $(h-1)$-hole instance $\cutpath_h$ returns. 
From procedure $\cutpath_h$,
$|\tilde U|\le |U|+2|Y|\le c\cdot rh/\eps$, since $c$ is large enough.
Recall that instance $(\hat H,U)$ is obtained by applying the procedure $\gluepath_h$ to instance $(\tilde H,\tilde U)$. We use the following claim, whose proof is similar Claim~\ref{clm: ratio loss for contracting to portals}, and is deferred to Appendix~\ref{apd: Proof of ratio loss for glue h holes}.

\begin{claim}
\label{clm: ratio loss for glue h holes}
Instance $(\hat H, U)$ is an $\eps'$-emulator for instance $(H, U)$.
\end{claim}

Consider the $(h-1)$-hole instance $(\tilde H,\tilde U)$. 
From the induction hypothesis, if we set $\EMPH{$\eps''$} \coloneqq \eps(1-\frac 1 h)$, then there is another $(h-1)$-hole instance $(\tilde H',\tilde U)$ that is an $\eps''$-emulator for $(\tilde H,\tilde U)$, such that
\[
\begin{split}
|V(\tilde H')| &
\le |\tilde U|\cdot \bigg(\frac{ch\cdot \log |\tilde U|}{\eps''}\bigg)^{c(h-1)} \\
& \le \frac{crh}{\eps}\cdot \bigg(\frac{ch\cdot \log (crh/\eps)}{\eps\cdot (1-1/h)}\bigg)^{c(h-1)} \\
&  \le r\cdot \bigg(\frac{ch}{\eps}\bigg)^{c(h-1)+1}\cdot \bigg(\frac{\log (crh/\eps)}{(1-h)}\bigg)^{c(h-1)} \\
& \le r\cdot \bigg(\frac{ch}{\eps}\bigg)^{ch}\cdot \bigg(\log r+ \log (crh/\eps)\bigg)^{c(h-1)} \\
& \le r\cdot \bigg(\frac{ch \log r}{\eps}\bigg)^{ch}.
\end{split}
\]
where we have used the fact that $(1-\frac{1}{h})^{-c(h-1)}\le e^{c}< c^{c-1}$, as $c$ is large enough.

We apply procedure $\gluepath_h$ to instance $(\tilde H',\tilde U)$, and let $(H',U)$ be the $h$-hole instance we get. From the procedure $\gluepath_h$, $|V(H')|\le |V(\tilde H')|\le r\cdot (ch\log r/\eps)^{ch}$.
On the other hand, since instance $(\tilde H',\tilde U)$ is an $\eps''$-emulator for $(\tilde H,\tilde U)$, from Claim~\ref{clm: split glue h holes}, instance $(H',U)$ is an $\eps''$-emulator for $(\hat H, U)$.
Since $(\hat H, U)$ is an  $\eps'$-emulator for instance $(H, U)$ (from Claim~\ref{clm: ratio loss for glue h holes}), using the fact that $\eps''+\eps'=\eps(1-1/h)+\eps/h=\eps$, we conclude that $(H',U)$ is an  $\eps$-emulator for instance $(H,U)$.


\medskip
Note that the above proof also gives an algorithm for constructing an $\eps$-emulator of $(H,U)$ of size at most $r \cdot (ch\log r/\eps)^{ch}$.
Specifically, if $(H,U)$ is the input $h$-hole instance, then we slice it open along some shortest path $P$ that connects a pair of terminals lying on different holes, add $\eps'$-covers of terminals in $U$ on $P$, get an $(h-1)$-hole instance $(\tilde H, \tilde U)$, and then we recursively construct an $\eps''$-emulator for $(\tilde H, \tilde U)$ and glue it along $P$ to get an $\eps$-emulator for $(H,U)$.
The following claim completes the proof of Lemma~\ref{L:emulator-constant_hole}.

\begin{claim}
The running time of the above algorithm is $O\big((n+r^2)\cdot(h\log n/\e)^{O(h)}\big)$.
\end{claim}
\begin{proof}
We prove the claim by induction on $h$.
The base case is when $h=1$. From \Cref{Th:emulator-1hole}, the running time of the above algorithm is at most $(n+r^2)\cdot (c \log n/\eps)^c$ on an $n$-vertex graph when $h=1$. 
Consider the inductive case: 
The $\cutpath_h$ and $\gluepath_h$ algorithms runs in time at most $cn$. 
Since the input to the algorithm $\cutpath_h$ is an $n$-vertex graph, $\cutpath_h$ produces a graph $(\tilde H, \tilde U)$ with at most $2n$ vertices. Therefore, from the induction hypothesis, the construction of an $\eps'$-emulator for $(\tilde H, \tilde U)$ takes at most $(2n+r^2)\cdot (c(h-1)\log n/\eps'')^{c(h-1)}$ time. Therefore, the total running time of the algorithm is at most
\[
(2n+r^2)\cdot \bigg(\frac{c(h-1)\log n}{\eps''}\bigg)^{c(h-1)}+2cn\le (n+r^2)\cdot \bigg(\frac{ch\log n}{\eps}\bigg)^{ch}.
\]
\aftermath
\end{proof}


\subsection{Algorithm for General Planar Graphs: Proof of Theorem~\ref{thm:main}}
\label{sec: alg_general}

\paragraph{Separators and recursive decomposition.}
Let $r$ be any positive integer. 
An \EMPH{$r$-division with few holes}~\cite{fre-faspp-1987,kms-srsdp-2013} of a $n$-vertex connected plane graph $G$ is a collection $\gset$ of connected subgraphs of $G$, called the \EMPH{pieces}, such that
\begin{itemize} 
\item every edge in $G$ belongs to at least one piece in $\gset$;
\item $|\gset|=O(n/r)$;
\item the number of vertices in $H$ is at most $r$ for each piece $H\in \gset$;
\item the number of \EMPH{boundary vertices} in $H$ (that is, vertices in $V(H)$ that also belong to some other piece in $\gset$) is $O(\sqrt{r})$; and
\item for each piece $H\in \gset$, there are $O(1)$ faces, called \EMPH{holes}, whose boundaries contain all boundary vertices of $H$ (when considered as a plane graph).  
\end{itemize}
We often refer to an $r$-division with few holes as an \EMPH{$r$-division}.
A standard $r$-division can be computed in linear time for any $r$~\cite{kms-srsdp-2013}.
However in our application we need to compute $r$-divisions of instances that evenly distribute the terminals among pieces. In particular, we need the following lemma, whose proof is deferred to \Cref{apd: Proof of L:r-division}.

\begin{lemma}
\label{L:r-division}
Given an instance $(G,T)$ with $n \coloneqq |V(G)|$ and $k \coloneqq |T|$ 
computing an $r$-division for graph $G$ takes in $O(n)$ time, where each piece contains $O(1 + kr/n)$ terminals.
\end{lemma}

We use the following lemma, which is crucial for the proof of Theorem~\ref{thm:main}.  

\begin{lemma}
\label{lem: size reduction step}
Given a planar instance $(H,U)$ with $n \coloneqq |V(H)|$ and $k \coloneqq |U|$, and a parameter $0<\eps<1$, 
computing an $\eps$-emulator $(H',U)$ for $(H,U)$ with $|V(H')|\le O\Paren{ \sqrt{n k}\cdot(\log n/\eps)^{c'} }$ takes $O\Paren{n\cdot (c'\log n/\eps)^{c'}}$ time for some large enough universal constant $c'$.
Furthermore, if $(H,U)$ is an $h$-hole instance, then $(H',U)$ is also an $h$-hole instance.
\end{lemma}

\begin{proof}
Let $c'$ be a constant that is greater than $c$ and all other hidden constants in Lemma~\ref{L:emulator-constant_hole}.
We first compute an $r$-division for $H$, with parameter $r \coloneqq n/k$ using the algorithm from Lemma~\ref{L:r-division}. Let $\rset$ be the collection of pieces in $H$ that we obtain. From Lemma~\ref{L:r-division},
\begin{itemize}
	\item $|\rset|=O(k)$;
	\item the number of vertices in each piece in $\rset$ is at most $O(n/k)$;
	\item the number of boundary vertices in each piece in $\rset$ is at most $O(\sqrt{n/k})$;
	\item the number of terminals in $T$ in each piece in $\rset$ is $O(1)$; and
	\item there are $O(1)$ holes in each piece in $\rset$. 
\end{itemize}
For each graph piece $R$ in $\rset$, let \EMPH{$U_R$} be the set that contains all boundary vertices of $R$ and all terminals in $U$. 
Observe that $(R,U_R)$ is an $h$-hole instance for some constant $h$. 
We apply the algorithm from Lemma~\ref{L:emulator-constant_hole} to instance $(R,U_R)$, and let $(R',U_R)$ be the $\e$-emulator we get, so $|V(R')|\le |U_R|\cdot (ch\log (n/k))/\eps)^{ch}$.
Also, such an emulator can be computed in at most $(|V(R)|+|U_R|^2)\cdot (h\log n/\eps)^{ch}$ time. Therefore, all emulators in $\set{(R', U_R)\mid R\in \rset}$ can be computed in time
\[
\sum_{R\in \rset} O\bigg(\big(|V(R)|+|U_R|^2\big)\cdot \Big(\frac{h\log n}{\eps}\Big)^{ch}\bigg)
\le O\Paren{ n\cdot \Big(\frac{h\log n}{\eps}\Big)^{ch} }
\le O\Paren{ n\cdot \Big(\frac{c'\log n}{\eps}\Big)^{c'} },
\]
as $\sum_{R\in \rset} |V(R)|\le O(k)\cdot (n/k)=O(n)$, $\sum_{R\in \rset} |U_R|^2\le O(k)\cdot (\sqrt{n/k})^2\le O(n)$, and $c'$ is large enough.
We then glue the emulators together via a process similar to $\gluepath$ and $\gluepath_h$, and eventually obtain an $\eps$-emulator $(H',U)$ for $(H,U)$, with size 
\[
|V(H')|\le \sum_{ R\in \rset}|U_R|\cdot \bigg(\frac{ch \log (n/k)}{\eps}\bigg)^{ch}
\le O\bigg(k\cdot\sqrt{\frac{n}{k}}\bigg)\cdot \bigg(\frac{ch\log n}{\eps}\bigg)^{ch}
\le O\Paren{ \sqrt{nk} \cdot \bigg(\frac{\log k}{\eps}\bigg)^{c'} }, 
\]
as both $c$ and $h$ are constants.
\end{proof}

\paragraph{Algorithm for Theorem~\ref{thm:main}.}
Let $G$ be the input $n$-vertex plane graph and let $T$ be the set of terminals of size $k$.
We first preprocess the graph $G$ into a new graph $G_0$ as follows. 
If $n<k^2$, then we set $G_0=G$. If $n\ge k^2$, we use the algorithm in {\cite[Theorem~6.9]{CGH16}} with parameter $\eps/2$ to compute an $(\eps/2)$-emulator $G_0$ for $G$ with size $O(k^2 \log^2 k/\e^2)$. 
This can be done in time $\tilde O(n/\eps^{O(1)})$ by a slight modification of the algorithm in \cite{CGH16} (in particular, we remove their preprocessing step that reduces the number of vertices to $k^4$).
Either way, we obtain an $(\eps/2)$-emulator $G_0$ for $G$, and $|V(G_0)|=O(k^2 \log^2 k/\e^2)$. 

We then set $L \coloneqq \log\log k$ and $\eps' \coloneqq \eps/2L$. 
Now sequentially for each $0\le i\le L-1$, we apply the algorithm from Lemma~\ref{lem: size reduction step} to instance $(G_i,T)$ and parameter $\eps'$ to obtain an $\eps'$-emulator $(G_{i+1},T)$ for $(G_i,T)$. Finally, we return $(G',T)=(G_L,T)$ as the output. 
Note that $\eps' L =(\eps/2)$ and thus $(G_L,T)$ is an $\eps/2$-emulator of $(G_0,L)$, and is therefore an $\eps$-emulator for $(G,T)$. 
From Lemma~\ref{lem: size reduction step}, the running time of our algorithm is $\tilde O(n /\eps^{O(1)})$. 
In order to complete the proof of \Cref{thm:main}, it suffices to show that $|V(G')|\le O(k\cdot (\log k/\eps)^{O(1)})$, which follows immediately from the next claim (by setting $i=L$).

\begin{claim}
For each $0\le i\le L$, $|V(G_i)|\le k^{1+2^{-i}}\cdot (\log k/\eps')^{2c'-c'/2^{i}}$.
\end{claim}
\begin{proof}
We prove the claim by induction on $i$. The base case is when $i=0$. From the preprocessing step, $|V(G_0)|\le O(k^2 \log^2 k/\e^2)\le k^2(\log k/\eps')^2$, so the claim holds, as $c'$ is large enough.
Consider the inductive case. 
From Lemma~\ref{lem: size reduction step},
\[
\begin{split}
|V(G_{i})| & \le \sqrt{|V(G_{i-1})|\cdot k}\cdot \bigg(\frac{\log k}{\eps'}\bigg)^{c'}\\
& \le \sqrt{\big(k^{1+2^{-(i-1)}}\cdot (\log k/\eps')^{2c'-c'/2^{(i-1)}}\big)\cdot k}\cdot \bigg(\frac{\log k}{\eps'}\bigg)^{c'}\\
& \le k^{(1+2^{-(i-1)}+1)/2}\cdot \bigg(\frac{\log k}{\eps'}\bigg)^{(2c'-c/2^{(i-1)})/2+c'}\\
& = k^{1+2^{-i}}\cdot (\log k/\eps')^{2c'-c'/2^{i}}.
\end{split}
\]
Therefore the claim holds for all $i$. 
\end{proof}

\subsection{Bootstrapping}
\label{SS:bootstrapping}


Perhaps surprisingly, we can further reduce the running time for constructing an $\e$-emulator to be linear to the size of the graph whenever $k$ is ``sufficiently'' sublinear and the range of the edge weights (that is, the ratio between the smallest and largest weights) are polynomially bounded, using the idea of \emph{bootstrapping} combining with a precomputed look-up table.

\begin{theorem}
\label{Th:bootstrap}
Given any parameter $0<\eps<1$ and any instance $(H,U)$ with $n \coloneqq |H|$ and $k \coloneqq |U|$ satisfying $k \le n / \log^{D} n$ for some big enough constant $D$, and the range of the edge weights are bounded by polynomial in $n$,
computing an emulator $(Z,U)$ for $(H,U)$ of size $|V(Z)| \le O(k \polylog k/\e^{O(1)})$ takes $O_\e(n)$ time.
Furthermore, if $(H,U)$ is an $h$-hole instance, then $(Z,U)$ is an $h$-hole instance.
\end{theorem}

\begin{proof}
We apply $r$-division iteratively with exponentially-growing values of $r$; intuitively each time we shrink the graph by a very small amount, just enough to absorb the logarithmic terms required to compute the emulators.
%
\begin{itemize}
\item
First compute $r$-division of $H$ for $r \coloneqq (\log\log\log n)^{6C}$ that evenly distribute the terminals in $U$ using Lemma~\ref{L:r-division}, where $C$ is bigger than the number of logs we need in the running time of Theorem~\ref{thm:main}.
Replace each piece in the $r$-division by an $\e$-emulator with respect to the boundary vertices and terminals using Theorem~\ref{thm:main}; every piece contains $O(r^{1/2} + k(\log\log\log n)^{6C}/n) \le O(r^{1/2})$ boundary vertices and terminals.
The total time on the emulator construction is 
\[
O\Paren{ r \cdot \Paren{\frac{\log r}{\e}}^{O(1)} } \cdot O\Paren{\frac{n}{r}} \le O\Paren{ \frac{n \cdot (\log\log\log\log n)^{O(1)}}{\poly\e} };
\]
and the new graph $H'$ has size 
\[
O\Paren{ r^{1/2} \Paren{\frac{\log r^{1/2}}{\e}}^{C} } \cdot O\Paren{\frac{n}{r}} \le O \Paren{ \frac{n}{\e^C (\log\log\log n)^{2C}} }.
\]

\item
Now the graph is about $(\log\log\log n)^{2C}$-factor smaller than original, we can compute another $r'$-division for $r' \coloneqq (\log\log n)^{6C}$, and replace each piece in the $r'$-division by an $\e$-emulator with respect to the boundary vertices and terminals; every piece contains $O(r'^{1/2} + k(\log\log n)^{6C}/n) \le O(r'^{1/2})$ boundary vertices and terminals.
This way, instead of spending $O_\e(n (\log\log\log n)^{O(1)})$ time if we perform $r'$-division directly on the original graph, now it takes time
\[
O_\e \Paren{ \frac{n}{(\log\log\log n)^{2C}} \cdot (\log\log\log n)^{C} } \le O_\e(n).
\]
The new graph $H''$ has size about $O_\e(n/(\log\log n)^{2C})$. 

\item
Now the graph is about $(\log\log n)^{2C}$-factor smaller than original, we can compute another $r''$-division for $r'' \coloneqq (\log n)^{6C}$, and replace each piece in the $r''$-division by an $\e$-emulator with respect to the boundary vertices and terminals; every piece contains $O(r''^{1/2} + k(\log n)^{6C}/n) \le O(r''^{1/2})$ boundary vertices and terminals,
and this takes time
\[
O_\e \Paren{ \frac{n}{(\log\log n)^{2C}} \cdot (\log\log n)^{C} } \le O_\e(n).
\]
The new graph $H'''$ has size about $O_\e(n/(\log n)^{2C})$. 

\item 
Finally, compute an $\e$-emulator for $H'''$ with respect to the terminals.
This takes time
\[
O_\e\Paren{ \frac{n}{(\log n)^{2C}} \cdot (\log n)^C } \le O_\e(n)
\]
The final emulator has size $O(k \polylog k / \e^{O(1)})$.
\end{itemize}

\noindent The accumulated distortion in distance is $4\e$. 
Overall the bottleneck is to compute the first set of emulators for pieces in the $r$-division, which takes $O(n \cdot (\log\log\log\log n)^{O(1)})$ time.
We can avoid spending super-linear time to compute the first set of emulators; instead, we precompte a look-up table for every graph up to size $r = (\log\log\log n)^{6C}$, every possible subset of terminals, and every edge-weight functions rounded to the closest power of $1+\e$.

\paragraph{Look-up table.}
Now we can describe the construction of the look-up table.
\begin{itemize}
\item There are $2^{O(r)}$ plane graphs $K$ up to size $r$.
\item There are $2^{r}$ possible choices for the terminal subset $U_K$.
\item The spread of any instance $(K,U_K)$ is at most $n^{O(1)}$ because the range of the edge weights is polynomial in $n$; so if we round the weight of each edge to the closest power of $1+\e$, there are $\log_{1+\e} n^{O(1)} \le O(\log n / \e)$ possible weight values per edge, and thus ${O(\log n / \e)}^{2^{O(r)}}$ many different (rounded) edge-weight functions (because $\e$ is a constant).
\item Computing an $\e$-emulator for each instance $(K,U_K)$ takes $r^{O(1)}$ time.
\end{itemize}
Overall, it takes 
\[
2^{O(r)} \cdot 2^r \cdot O(\log n / \e)^{2^{O(r)}} \cdot r^{O(1)} \le 2^{2^{O_\e((\log\log\log n)^{6C})}} \le o_\e(n)
\]
time to precompute a look-up table, so that for any instance $(K,U_K)$ from the pieces of the first $r$-division, one can round the edge weights of $K$ and find the $\e$-emulator for $(K,U_K)$ directly from the look-up table.  Rounding the edge-weights to the closest powers of $(1+\e)$ will introduce at most $O(\e)$ distortion. 
As a result, an $\e$-emulator of size $O(k \polylog k/\e^{O(1)})$ for $(H,U)$ can be computed in $O_\e(n)$ time.
\end{proof}

\section{Applications}
\label{S:applications}

In this section we present efficient $\e$-approximate algorithms to several optimization problems on planar graphs that beat their exact counterparts, inclusing multiple-source shortest paths, minimum $(s,t)-$cut, graph diameter, and offline dynamic distance oracle.
To put emphasis on the new ideas presented, we assume the readers are familiar with the various tools for optimization on planar graphs and only provide citations to the earlier literature.

\subsection{Approximate Multiple-Source Shortest Paths}
\label{SS:mssp}


The approximate multiple-source shortest paths data structure (\emph{$\e$-MSSP}) can achieve the following task:
Preprocess a plane graph $P$ and a set of terminals $U$ on the outerface of $P$ (that is, a one-hole instance $(P,U)$),
and answer distance queries between terminal pairs within $(1+\eps)$-approximation.


To prove Theorem~\ref{Th:mssp},
apply Theorem~\ref{Th:bootstrap} on $(P,U)$ to construct another one-hole instance $(P',U)$ that is an $\e$-emulator of $(P,U)$, which has size 
\[
O\Paren{ \frac{ ({n}/{\log^C n}) \cdot \poly\log n}{\e^{O(1)}} }
= O\Paren{ \frac{n}{\e^{O(1)} \poly\log n} }
\]
and takes $O_\e(n)$ time.  
Now construct the MSSP data structure on $P'$ using Klein's algorithm~\cite{kle-msppg-2005}, which takes $O\Paren{ \frac{n}{\e^{O(1)} \poly\log n} \cdot \log n} = O(n/ \e^{O(1)})$ time; MSSP answers queries in time $O(\log n)$, which is an $\e$-approximation to the actual distance between the pairs due to the fact that $(P',U)$ is an $\e$-emulator.
This proves Theorem~\ref{Th:mssp}.

\subsection{Approximate Minimum Cut}


Here we briefly summarize the minimum $(s,t)$-cut algorithm on planar graphs with non-negative weights by Italiano, Nussbaum, Sankowski, and Wulff-Nilsen~\cite{insw-iamcm-2011}.  
Many details and edge-cases are omitted for the clarity of presentation.
Let $G$ be the input plane graph, and two vertices $s$ and~$t$.
\begin{enumerate}
\item Compute the dual graph $G^*$ of $G$; it is sufficient to compute a shortest cycle in $G^*$ that separates the \emph{faces} $s^*$ and $t^*$.
Find a shortest $s^*$-$t^*$ path $\pi$ in $G^*$.  This step takes $O(n)$ time~\cite{hkrs-fsapg-1997}.

\item Construct $r$-division in $G^*$ respecting $\pi$ where $r \coloneqq \log^6 n$.  Cut $\pi$ open; now each vertex on $\pi$ has a copy.  
This step takes $O(n)$ time~\cite{kms-srsdp-2013}. 

\item Compute MSSP~\cite{kle-msppg-2005} for each piece in the $r$-division with respect to the boundary vertices.  Prepare the Monge heap data structures~\cite{fr-pgnwe-2006}, and represent each piece as a \emph{dense distance graph}.  
This step takes $O(n \log r) = O(n \log\log n)$ time for the MSSP~\cite{kle-msppg-2005}, and $O(n \log\log n)$ time to set up the Monge heap data structures and dense distance graphs~\cite{fr-pgnwe-2006}. 

\item Denote the length of $\pi$ as $p$.
Compute $p/\log p$ shortest paths between the two copies of each evenly spaced points on $\pi$, using Reif's divide-and-conquer strategy~\cite{rei-mscpu-1981}; each shortest path is computed by FR-Dijkstra~\cite{fr-pgnwe-2006} on the dense distance graphs. 
Now the graph is cut into $p/\log p$ \emph{slabs}.  This step takes $\tilde{O}(n/\sqrt{r} \cdot \log (p/\log p)) \le O(n)$ time.

\item Apply Reif's strategy directly on each slab which now has only $O(\log p)$ vertices from $\pi$, so it takes $O(n \log p) = O(n \log\log n)$ time.
\end{enumerate}
Overall the algorithm takes $O(n \log\log n)$ time, with Step~3 being the bottleneck.

We can safely truncate the edge weights to have polynomial range in linear time when solving the minimum $(s,t)$-cut problem.
Now by simply choosing $r \coloneqq \log^C n$ with a bigger $C$ and replacing Step~3 with an $\e$-emulator per piece using Theorem~\ref{Th:bootstrap}, the new graph has size $O(\frac{n}{r} \cdot \sqrt{r} \poly\log r/\e^{O(1)}) = O(n / \e^{O(1)} \poly\log n)$.
We can now compute $p$ shortest paths (instead of $p/\log p$) in Step~4 without recursion in Step~5 using Reif's divide-and-conquer strategy directly on the emulators without preparing the MSSP and Monge heap data structures in Step~3 and FR-Dijkstra in Step 4.
Therefore the total running time is now $O_\e(n)$, proving Theorem~\ref{Th:minimum-cut}. 



\subsection{Approximate Diameter}


Here we summarize the $(1+\e)$-approximate algorithm to compute the diameter of planar graphs with non-negative edge weights by Weimann-Yuster~\cite{wy-adpgl-2016} and Chan-Skrepetos~\cite{cs-faddo-2019}.
Again we omit some details about marking/unmarking vertices in the actual algorithm to emphasize on core concepts.
Let $G$ be the input planar graph.
Given three graphs $H$, $H'$ and $H''$, denote \EMPH{$\diam_{H}(H',H'')$} the longest shortest-path distance with respect to $H$ between a vertex in $H'$ and a vertex in $H''$.
\begin{enumerate}
\item Compute a \emph{shortest-path} cycle separator $C$ in $G$ and splits $G$ into $A$ and $B$, where $A \cup B = G$ and $A \cap B = C$, using the algorithm by Thorup~\cite{tho-corad-2004}.  
This step takes $O(n)$ time.

\item Construct an auxillary graph $G^+$ by selecting $O(1/\e)$ evenly-spaced \emph{portals} on $C$; run single-source shortest path algorithm on each portal $p$ to get maximum distance out of all paths from $p$, denoted as $\ell$; add edges from every vertex in $A$ and $B$ to the portals, with the edge-weights being their distances rounded to multiples of $\e\ell$.  This step takes $O(n\cdot (1/\e))$ time using the linear-time single-source shortest path algorithm by Henzinger-Klein-Rao-Subramanian~\cite{hkrs-fsapg-1997}.

\item Approximate $\diam_{G^+}(A,B)$. 
This step takes $O(n/\e) + 2^{O(1/\e)}$ time using brute-force~\cite{wy-adpgl-2016}, or $O(n\cdot(1/\e)^5)$ time using the farthest Voronoi diagram~\cite{cs-faddo-2019}.

\item Build another auxillary graph $A^+$ from $G$ by first adding \emph{denser portals} on $C$, computing shortest paths between denser portals on $C$ with respect to $B$, then planarizing the union of all the shortest paths between dense portal pairs so that $A^+$ remains planar.  
Following Chan-Skrepetos~\cite{cs-faddo-2019},
the number of denser portals can be set to $|G|^{1/8}/\e$; compute all-pairs shortest paths between dense portals in $B$ takes $O(|B|\log n + \log n\cdot \sqrt{|B|}/\e^4)$ time using MSSP~\cite{kle-msppg-2005}; $A^+$ has size $|A| + O(|A|^{1/2}/\e^4)$.
Build the graph $B^+$ similarly by switching the roles of $A$ and $B$. 

\item Approximate $\diam_{A^+}(A,A)$ and $\diam_{B^+}(B,B)$ recursively; the recursion depth is $O(\log n)$.

\item Return the maximum of $\diam_{G^+}(A,B)$, $\diam_{A^+}(A,A)$, and $\diam_{B^+}(B,B)$.
\end{enumerate}
%
Overall the algorithm takes $O(n \log^2 n + n\log n \cdot (1/\e)^5)$ time.

Again we can safely truncate the edge weights to have polynomial range when solving the diameter problem.
Now we can substitute the construction of $A^+$ and $B^+$ using planarized shortest paths in Step 4 with two $\e$-emulators using Theorem~\ref{Th:bootstrap}, which only takes $O_\e(|A|+|B|)$ time to construct and has size $O_\e((|A|^{1/8}+|B|^{1/8})\poly\log n)$.
Thus we improve the total running time to $O_\e(n \log n)$, proving Theorem~\ref{Th:diameter}.




\subsection{Offline Dynamic Approximate Distance Oracle}

Here we describe the crucial step in the algorithm by Chen \etal~\cite{cgh+-fdcde-2020a} to construct an offline dynamic $(1+\e)$-approximate distance oracle with $O(\poly\log n)$ query and update time, assuming that a $(1+\e)$-\emph{distance-approximating minor} of size $\tilde{O}(k)$ for a planar graph of size $n$ and $k$ terminals can be computed in $O(n \poly(\log n, \e^{-1}))$ time.
Given a sequence of graphs $G_0 \subseteq G_1 \subseteq \dots \subseteq G_\ell$, denote $H_p \coloneqq G_p \setminus G_{p-1}$ for any $p \in \set{1,\dots,\ell}$.
The proof of Theorem~4.15 in Chen \etal~\cite{cgh+-fdcde-2020a} iteratively constructs graphs $G'_1, \dots, G'_\ell$ in the following way:
\[
G'_{p} \coloneqq \textsc{Emulator}(G'_{p-1} \cup H_{p}, T_{p})
\]
for some terminal set $T_p$ (irrelevant to the discussion here),
where $\textsc{Emulator}(G, T)$ returns an $\e$-emulator of $G$ with respect to terminal set $T$.
When $\textsc{Emulator}(G, T)$ guarantees to return a minor of the input graph $G$, one can argue that $G'_{p}$ must be a minor of $G'_{p-1} \cup H_{p}$, which by induction is a minor of $\bigcup_{1 \le k \le p-1} H_k \cup H_{p} = G_p$ which must be planar {\cite[Lemma~4.16]{cgh+-fdcde-2020a}}.

\smallskip
To prove Theorem~\ref{Th:dynamic-oracle}, we follow the algorithm by Chen \etal~\cite{cgh+-fdcde-2020a} almost verbatim; the only missing piece is to prove that $G'_{p}$ remains planar in our setting.
Observe that our emulator construction solely relies on the $\cutpath$ and $\gluepath$ procedures introduced in Section~\ref{SS:split-and-glue}.
(The base case from Theorem~\ref{thm: quartergrid enumator} can be replaced by the $O(k^4)$-size distance-approximating minor~\cite{KNZ14}.)
While the emulator $G'$ produced by split-and-glue is technically not a minor of the input graph $G$, there is another planar supergraph $\hat G$ modified from $G$ such that $G'$ is a minor of $\hat G$.
%
Now we can proceed to prove that $G'_{p}$ is planar using our construction for $\textsc{Emulator}(G, T)$.

\begin{claim}
For any $p \in \set{1,\dots,\ell}$, $G'_{p}$ is planar when $\textsc{Emulator}(G, T)$ is implemented using Theorem~\ref{thm:main}.
\end{claim}

\begin{proof}
We will prove the following stronger statement by induction on $p$:  
there is a planar graph $\hat G_p$ constructed from $G_p$ by vertex spitting 
(the reverse operation to edge contraction), edge subdivision (by breaking an edge into two using a degree-2 node), and edge duplications (by creating multiedges from an existing edge), and contains $G'_p$ as a minor. 
We say a plane graph $H$ is a \EMPH{topological minor} of some graph $\hat H$ if $\hat H$ is constructed from $H$ by vertex spitting, edge subdivision and edge duplications.  (Notice that this is difference from the standard terminology; in fact it is a topological minor \emph{in the dual}.)
Notice the crucial property that if plane graph $H$ is a topological minor of $\hat H$, then $\hat H$ must also be a plane graph.

First we introduce an operation that we will later use in the construction of $\hat G_p$.
Recall that we can \emph{slice} a graph $H$ open along some path $P$ by duplicating every vertex and edge of $P$ to create another path $P'$ identical to $P$.
The set of edges incident to each vertex on $P$ are split into two sides naturally based on their cyclic order around the vertex.
Now we also add an edge between each vertex on $P$ and its copy in $P'$.
We call this operation a \EMPH{pizza slice}.
A pizza slice of a graph $H$ must contain $H$ as a topological minor.
Every graph constructed from slice-and-gluing $H$ along a set of paths is a minor of some pizza slice of $H$. 

By induction hypothesis, there is a planar graph $\hat G_{p-1}$ containing $G'_{p-1}$ as a minor and $G_{p-1}$ as a topological minor.  
Now because the endpoints of all edges in $H_p$ can still be found in $G'_{p-1}$ and $\hat G_{p-1}$, $G'_{p-1} \cup H_{p}$ is a minor of $\hat G_{p-1} \cup H_{p}$.
We know by induction hypothesis that $\hat G_{p-1}$ contains $G_p$ as a topological minor, so edges in $H_p$ can be safely added to $\hat G_{p-1}$ without destroying planarity; therefore $\hat G_{p-1} \cup H_{p}$ is still planar, and so does $G'_{p-1} \cup H_{p}$.
Therefore $G'_{p} \coloneqq \textsc{Emulator}(G'_{p-1} \cup H_{p}, T_{p})$ is also planar from the emulator construction.

Now we describe the construction of $\hat G_p$ from $G_p$ and $G'_p$.
As $G'_p$ is constructed using split-and-glue from $Z_p \coloneqq G'_{p-1} \cup H_{p}$ by Theorem~\ref{thm:main}, 
there is a pizza slice $\hat Z_p$ of $Z_p$ that contains $G'_p$ as a minor.
Using the lifting property that a topological minor commutes with a minor, 
there is another plane graph $\hat G_p$ that contains $\hat G_{p-1} \cup H_{p}$ as a topological minor; one can indeed construct $\hat G_p$ from $\hat G_{p-1} \cup H_{p}$ using pizza slices on a set of paths mimicking the one used during the slice-and-glue operations to obtain $G'_p$ from $Z_p$.
Now $\hat G_p$ contains $G'_p$ as a minor because $\hat G_p$ contains $\hat Z_p$ as a minor and $\hat Z_p$ contains $G'_p$ as a minor by construction.
$\hat G_p$ also contains $G_p$ as a topological minor because $\hat G_p$ contains $\hat G_{p-1} \cup H_p$ as a topological minor, which by induction contains $G_{p-1} \cup H_p$ as a topological minor.
Therefore the existence of $\hat G_p$ is established. 

The base case is clear: Define $\hat G_1$ to be the pizza slice of $G'_0 \cup H_1 = G_0 \cup H_1 = G_1$ that contains $G'_1$ as a minor from the emulator construction.  Thus the claim is proved.
\end{proof}

\newpage
\section*{Acknowledgements}
We thank the anonymous reviewers for their helpful comments, as well as pointing out the result by Chen \etal~\cite{cgh+-fdcde-2020a} on the offline dynamic approximate distance oracles.
\small
\bibliographystyle{alphaurl}
\bibliography{planar-sketch,robi}

\newpage


\newpage

\appendix
\normalsize

\section{Missing Proofs in \Cref{sec:prelim} and \Cref{sec: planar emulator}}

\subsection{Proof of Lemma~\ref{lem: well-structured path set}}
\label{apd: Proof of well-structured path set}

Let $w: E(G)\to \mathbb{R}^+$ be the edge weight function of graph $G$. We slightly perturb $w$ to obtain another function $w': E(G)\to \mathbb{R}^+$, such that for every pair $P,P'$ of distinct paths in $G$: $w'(P)\ne w'(P')$; and if $w'(P)>w'(P')$, then $w(P)\ge w(P')$. Therefore, for each pair $v,v'$ of vertices in $G$, there is a unique $v$-$v'$ shortest path in $G$ under the weight function $w'$, and this path is also a $v$-$v'$ shortest path in $G$ under the weight function $w$ \cite{mvv-memi-1987,cab-mdpg-2012}.

The algorithm uses the technique of divide-and-conquer. We now describe the recursive step.

We first construct an auxiliary planar graph $H$ as follows. Its vertex set is $V(H)=T$, and its edge set $E(H)$ contains, for each pair $(t_1,t_2)\in \mset$, an edge connecting $t_1$ to $t_2$. 
Graph $H$ inherits a planar embedding from $G$ and is therefore an outerplanar graph.
%
Denote by $\fset$ the set of bounded faces in $H$ lying inside a disc $D$. 
We construct a graph $R$ as follows. Its vertex set is $V(R)=\set{u_F\mid F\in \fset}$, and its edge set $E(R)$ contains, for every pair $F,F'\in \fset$, an edge $(u_F,u_{F'})$ if and only if faces $F$ and $F'$ share a segment of non-zero length on their boundaries. It is easy to verify that $R$ is a tree, and $|V(R)|=|\mset|+1$. 
(In other words, $R$ is the \emph{weak-dual} of an outerplanar graph.)
We can now efficiently compute a vertex $u_F$ of $R$, such that every connected component of graph $R\setminus \set{u_F}$ contains no more than $|V(R)|/2$ vertices. Denote this vertex by $u_{F^*}$.
Consider now the face $F^*$ of $H$. Since in graph $R$, every connected component of graph $R\setminus \set{u_F}$ contains no more than $|V(R)|/2$ vertices, it is easy to see that we can find a pair $t_i,t_j$ of terminals on the intersection of the boundary of $D$ and the boundary of $F^*$, such that, if we draw a straight line segment connecting $t_i,t_j$, and denote by $D_1,D_2$ the discs obtained by cutting $D$ along this segment, then each edges of $H$ is drawn either inside $D_1$ or inside $D_2$, and each of $D_1,D_2$ contains the image of at most $3/4$-fractions of edges in $H$.

Consider now the one-hole instance $(G,T)$. We compute a $t_i$-$t_j$ shortest path $P$ in $G$, and cut the graph $G$ into two subgraphs $G_1,G_2$ along path $P$ (so $G_1\cap G_2=P$). Define $\mset_1$ to be the subset of $\mset$ that contains all pairs whose corresponding edge in graph $H$ is drawn inside $D_1$ in $H$, and we define subset $\mset_2$ similarly, so sets $\mset_1,\mset_2$ partition $\mset$, and $|\mset_1|,|\mset_2|\le (3/4)\cdot |\mset|$.
We now recurse on graph $G_1$ for computing the shortest paths connecting pairs of $\mset_1$ and graph $G_2$ for computing the shortest paths connecting pairs of $\mset_2$. This completes the description of the algorithm.

It is easy to verify that the running time of the algorithm is $O(\log |\mset|\cdot |E(G)|)$, since in every recursive layer, every edge of the original graph $G$ appears in at most two of the graphs that lie on this layer. To complete the proof of Theorem~\ref{lem: well-structured path set}, it suffices to show that, in a recursive step described above, for every pair $(t_1,t_2)\in \mset_1$, the unique shortest path in $G$ under $w'$ lies entirely in graph $G_1$ (the case for $\mset_2$ and $G_2$ is symmetric), and the set of resulting shortest paths that we computed is well-structured.

Assume for contradiction that the $t_1$-$t_2$ shortest-path $P'$ in $G$ does not lie entirely in $G_1$. We view $P'$ as being directed from $t_1$ to $t_2$. Let $v$ ($v'$, resp.) be the first (last, resp.) vertex of $P'$ that lies on $P$ and denote by $\hat P$ ($\hat P'$, resp.) the subpath of $P$ ($P'$, resp.) between $v$ and $v'$. Therefore, some inner vertex of $\hat P'$ does not belong to $G_1$ and therefore does not belong to $P$, and so $\hat P\ne \hat P'$. However, since both $P$ and $P'$ are shortest paths under $w'$, $w'(\hat P)=w'(\hat P')$, a contradiction to the fact that every pair of distinct paths have different weight in $w'$.
Via similar arguments we can also show that set of resulting shortest paths that we computed is well-structured.

\subsection{Proof of Theorem~\ref{thm: one-hole lower bound}}
\label{apd: Proof of one-hole lower bound}

In this subsection we provide the proof of Theorem~\ref{thm: one-hole lower bound}. Our example is inspired by the hard example constructed in \cite{KNZ14}. Assume that $1/\eps$ is an integer and $k$ is a multiple of $1/\eps$. This will only cause an additional constant factor in the size bound and will not influence the bound in Theorem~\ref{thm: one-hole lower bound}.

We first construct a circular ordering $\sigma$ and a metric $d$ on the terminals. From \cite{co-pemm-2020}, if $d$ satisfies the Monge property (under the circular ordering $\sigma$), then there exists a one-hole instance $(G,T)$ with terminals in $T$ appearing on the boundary in the order $\sigma$.

\begin{figure}[h]
	\centering
	\subfigure[An illustration of ordering $\sigma$.]{\scalebox{0.5}{\includegraphics[scale=0.18]{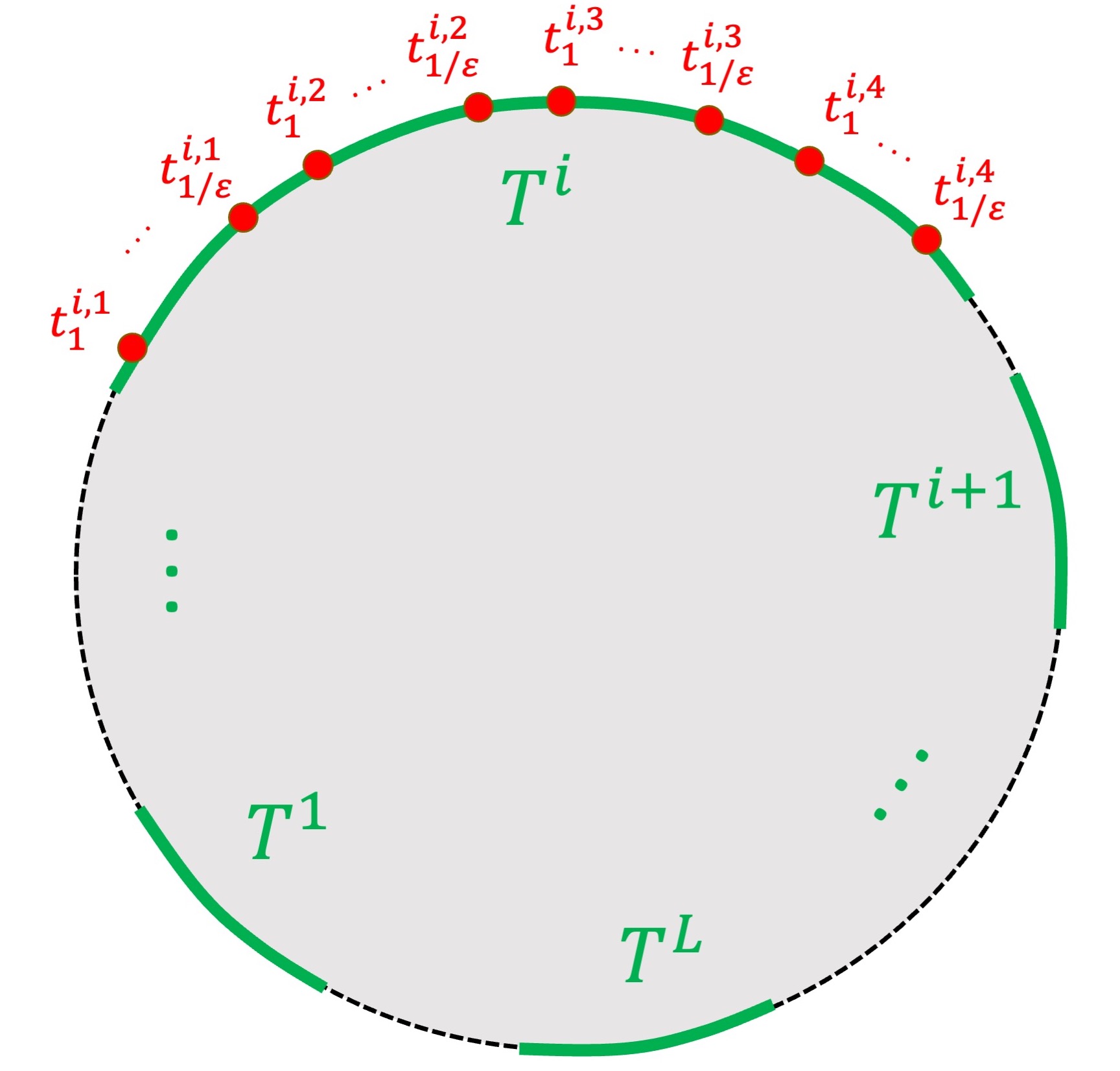}}\label{fig: onehole_lower1}}
	\hspace{0.3cm}
	\subfigure[An illustration of metric $d$ within a group $T^i$ of terminals.]{\scalebox{0.5}{\includegraphics[scale=0.18]{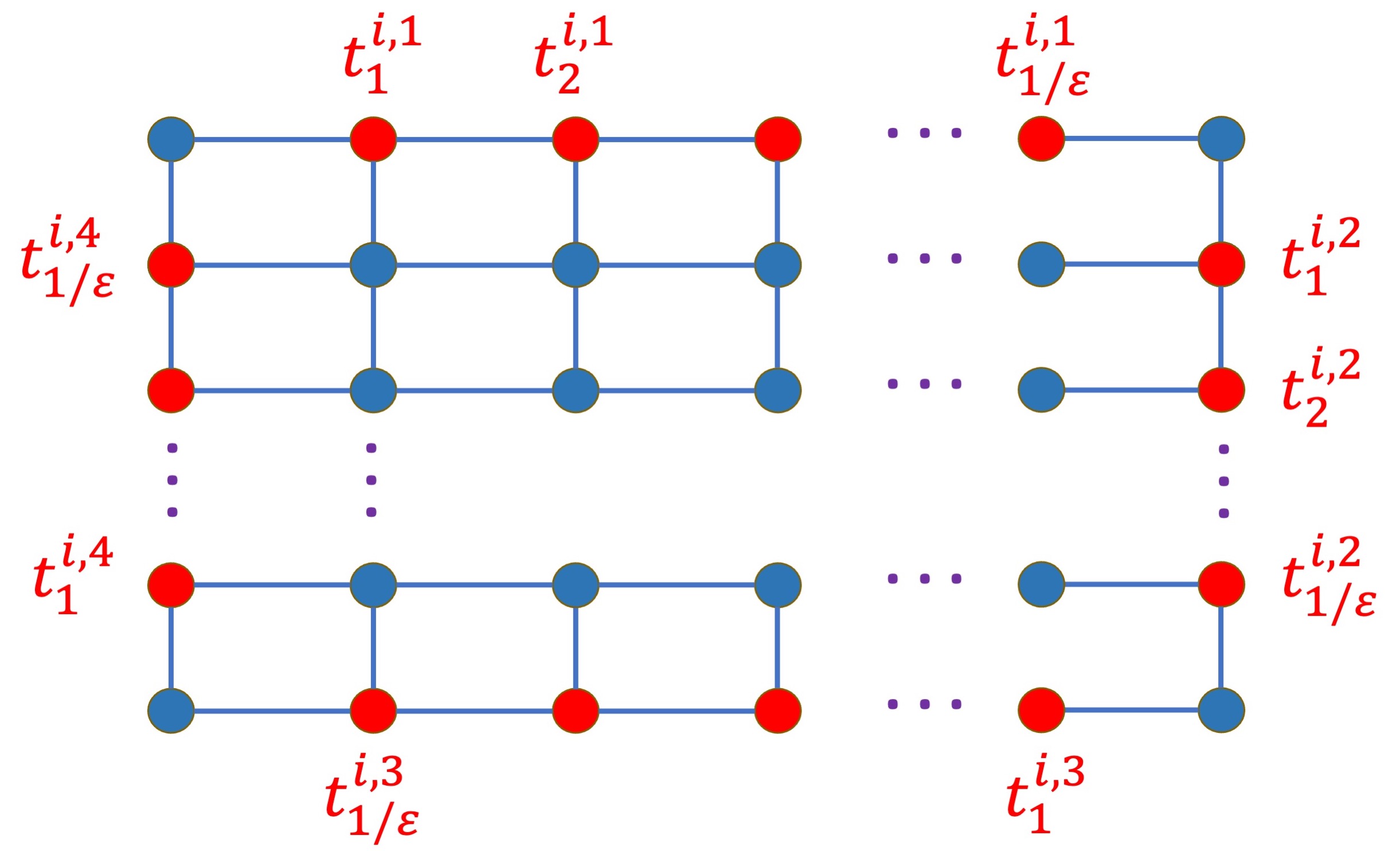}}\label{fig: onehole_lower2}}
	\caption{Illustrations of circular ordering $\sigma$ and metric $d$ within a group $T^i$ of terminals.}
\end{figure}

The set $T$ is partitioned into $L=\eps k/4$ groups $T=\bigcup_{1\le i\le L}T^i$, where each group contains $4/\eps$ terminals.
Each group $T^i$ is then partitioned into four subgroups $T^i=T^{i,1}\cup T^{i,2}\cup T^{i,3}\cup T^{i,4}$, each containing $1/\eps$ terminals.
We denote $T^{i,j}=\set{t^{i,j}_1, \ldots,t^{i,j}_{1/\eps}}$, for each $1\le j\le 4$.
The circular ordering $\sigma$ on terminals of $T$ is defined as follows. The groups $T^1,\ldots, T^{L}$ appear clockwise in this order; within each group $T^i$, the subgroups $T^{i,1}, T^{i,2}, T^{i,3}, T^{i,4}$ appear clockwise in this order; and within each subgroup $T^{i,j}$, the vertices $t^{i,j}_1, \ldots,t^{i,j}_{1/\eps}$ appear clockwise in this order. See Figure~\ref{fig: onehole_lower1} for an illustration.
The metric $d$ on $T$ is defined as follows. For every pair $t,t'$ of terminals that belong to different groups, $d(t,t')=1/\eps^2$.
Consider now a group $T_i$. The metric between terminals in $T_i$ is defined as follows. Consider the $(\frac{1}{\e}+2)\times (\frac{1}{\e}+2)$ grid with unit edge weight. We place each terminal in $T$ at a boundary vertex of $H$, in the way shown in Figure~\ref{fig: onehole_lower2}. Now for each pair $t^{i,j}_r, t^{i,j'}_{r'}$ of terminals in $T^i$, we define $d(t^{i,j}_r,t^{i,j'}_{r'})=\dist_H(t^{i,j}_r,t^{i,j'}_{r'})$. It is easy to verify that $d$ is a metric and satisfies the Monge property.

Consider now a one-hole instance $(G',T)$ such that the circular ordering in which terminals in $T$ appear on the outer boundary of $G'$ is $\sigma$ and for each pair $t,t'\in T$, $e^{-\e/3}\cdot \dist_{G'}(t,t')\le d(t,t') \le e^{-\e/3}$. For each $1\le i\le L$, we define $G'_i$ to be the subgraph of $G'$ induced by the set of all vertices in $G'$ that have distance at most $10/\eps$ from terminal $t^{i,1}_1$. Since in $d$, the distance between every pair of terminals in $\set{t^{1,1}_1,\ldots,t^{L,1}_1}$ is $1/\eps^2$, it is easy to see that the graphs $\set{G'_1,\ldots,G'_L}$ are mutually vertex-disjoint. On the other hand, it is easy to verify that, for every $1\le i\le L$ and every pair $t,t'$ of terminals in $T^i$, the shortest path in $G'$ connecting $t$ to $t'$ is entirely contained in $G'_i$. Therefore, for each $1\le i\le L$, $(G'_i, T^i)$ is an aligned $\eps/3$-emulator for $(G,T^i)$. From similar arguments in \cite{KNZ14}, we get that $|V(G'_i)|\ge \Omega(|T^i|^2)=\Omega(1/\eps^2)$. Therefore, $|V(G')|\ge \sum_{1\le i\le L}|V(G'_i)|\ge L\cdot \Omega(1/\eps^2)= \Omega(k/\eps)$.
This shows that any aligned $(\e/3)$-emulator for $(G,T)$ has size at least $\Omega(k/\eps)$. Theorem~\ref{thm: one-hole lower bound} now follows by scaling.

\subsection{Calculations for size and error bounds in \Cref{sec: planar emulator}}
\label{apd: calculations}

For convenience, we denote $\lambda=\lambda^*$. We prove the following observations.
\begin{observation}
	\label{Obs:size}
	Let $r_1,\ldots,r_t$ be a sequence of integers, such that $r_1\le k$, $r_t\ge \lambda$, and for each $1\le i\le t-1$, $r_i\ge (10/9)\cdot r_{i+1}$. Then $\sum_{1\le i\le t}(\log_{(10/9)} r_i)^{-2}\le 1/(\log_{(10/9)} \lambda -1)$.
\end{observation}

\begin{proof}
	Since for each $1\le i\le t-1$, $r_{i+1}\le (9/10)\cdot r_i$, $\log_{(10/9)} r_{i+1}\le \log_{(10/9)} r_i-1$. Therefore, 
	\[
	\sum_{1\le i\le t}\frac{1}{(\log_{(10/9)} r_i)^{2}}\le 
	\sum_{j\ge \log_{(10/9)}\lambda}\frac{1}{j^2}\le 
	\sum_{j\ge \log_{(10/9)}\lambda}\bigg(\frac{1}{j-1}-\frac{1}{j}\bigg)\le 
	\frac{1}{\log_{(10/9)} \lambda -1}.  
	\]
	\aftermath
\end{proof}

\begin{observation}
	\label{Obs:error}
	Let $r_1,\ldots,r_t$ be a sequence of integers, such that $r_1\le k$, $r_t\ge \lambda$, and for each $1\le i\le t-1$, $r_i\ge (10/9)\cdot r_{i+1}$. Then $\sum_{1\le i\le t}(\log r_i)^{4}/r_i^{0.1}\le 101(\log \lambda)^{4}/\lambda^{0.1}$.
\end{observation}

\begin{proof}
	Consider any index $1\le i\le t-1$. Denote $x=\log r_i/\log r_{i+1}$, so $r_i=(r_{i+1})^x$. Assume first that $x<1+10^{-100}$, then since $r_i\ge (10/9)\cdot r_{i+1}$, we get that 
	\[\bigg(\frac{(\log r_{i})^4}{r_i^{0.1}}\bigg)/\bigg(\frac{(\log r_{i+1})^4}{r_{i+1}^{0.1}}\bigg)
	=\frac{r_{i+1}^{0.1}}{r_{i}^{0.1}}\cdot \bigg(\frac{\log r_{i}}{\log r_{i+1}}\bigg)^4
	\le \frac{99}{100}\cdot x^4\le \frac{100}{101}.\]
	Assume now that $x\ge 1+10^{-100}$, then since $r_{i+1}\ge \lambda$ and from the definition of $\lambda$,
	\[\bigg(\frac{(\log r_{i})^4}{r_i^{0.1}}\bigg)/\bigg(\frac{(\log r_{i+1})^4}{r_{i+1}^{0.1}}\bigg)
	=\frac{r_{i+1}^{0.1}}{r_i^{0.1}}\cdot \bigg(\frac{\log r_{i}}{\log r_{i+1}}\bigg)^4
	= \frac{x^4}{(r_{i+1})^{\frac{x-1}{10}}}
	\le \frac{x^4}{\lambda^{\frac{x-1}{10}}}\le \frac{100}{101}.\]
	and so
	\[
	\sum_{1\le i\le t}\frac{(\log r_i)^{4}}{r_{i}^{0.1}}< 
	\frac{(\log \lambda)^{4}}{\lambda^{0.1}}\cdot\bigg(1+\frac{100}{101}+\big(\frac{100}{101}\big)^2+\cdots \bigg)\le 
	\frac{101(\log \lambda)^{4}}{\lambda^{0.1}}. 
	\]
	\aftermath
\end{proof}

\subsection{Proof of Claim~\ref{clm: branch pts}}
\label{apd: Proof of branch pts}


\paragraph{Item~\ref{p1} of Claim~\ref{clm: branch pts}.}
We define the graph $\tilde H$ as the union of (i) all paths in $\pset$; and (ii) the cycle that connects all vertices of $U$ in the order that they appear on the outer-boundary of the drawing associated with $H$, so $\tilde H$ is a planar graph, and the drawing of $H$ naturally induces a planar drawing of $\tilde H$. Let $\tilde H'$ be the graph obtained from $\tilde H$ by suppressing all degree-$2$ vertices, so the planar drawing of $\tilde H$ naturally induces a planar drawing of $\tilde H'$. 
Since $\tilde H'$ has no degree-$2$ vertices, the number of faces, edges and vertices are all within a constant factor. Therefore, to show that the number of branch vertices is $O(|U|)$, it suffices to show that the number of vertices in $\tilde H'$ is $O(|U|)$, and therefore it suffices to show that the number of faces in the planar drawing of $\tilde H'$ is $O(|U|)$.

We first construct an outerplanar graph $X$ on $U$ as follows. The edge set of $X$ is the union of (i) all edges of the cycle that connects all vertices of $U$ in the order that they appear on the outerface; and (ii) for each path in $\pset$, an edge connecting its endpoints in $U$.
Clearly, $X$ has $|U|$ vertices and $O(|U|)$ edges. The circular ordering on vertices of $U$ naturally defines a drawing of $X$. Clearly, the number of faces in this drawing is $O(|U|)$, and moreover, the total size of all faces is $O(|U|)$ (where the size of a face is the number of vertices that lie on the boundary of the face).

Let $F$ be a face in the drawing of $X$ defined above. We denote by $|F|$ the number of vertices that lie on the boundary of $F$. We now show that this face gives birth to at most $O(|F|)$ faces in $\tilde H'$. 
Let $Y$ be a graph defined as follows. The vertex set $V(Y)$ contains, for each boundary edge $e$ of $F$, a node $y_e$ representing $e$. The edge set $E(Y)$ contains, for every pair $y_e,y_{e'}$ of vertices, an edge connecting them iff the corresponding paths (in $\pset$) of edge $e$ and $e'$ either share an edge or share an internal vertex that does not belong to any other path in $\pset$. Since $\pset$ is well-structured and non-crossing, the graph $Y$ is an outerplanar graph, and so $|E(Y)|=O(|V(Y)|)=O(|F|)$. Since the number of faces in $\tilde H'$ that $F$ gives birth to is at most the number of edges in $Y$ plus one, we get that the number of faces in $\tilde H'$ that $F$ gives birth is at most $O(|F|)$.

Therefore, the total number of faces in $\tilde H'$ in at most a constant times the total size of all faces in $X$, which is $O(|U|)$. 
This completes the proof of \ref{p1}.

\bigskip
\paragraph{Item~\ref{p2} of Claim~\ref{clm: branch pts}.}
For convenience, we rename $Y\setminus Y^*$ as $Y$. In other words, set $Y$ only contains vertices that belong to exactly two paths of $\pset$, so each vertex of $Y$ is contained in at most two instances in $\hset$, contributing at most $2$ to the sum $\sum_{(H_R,U_R)\in \hset:\text{ } |U_R|\ge \lambda}|U_R|$.

We denote by $\rset$ the set of regions in $H$ obtained by the procedure $\cutpath$.
Recall that, for each region $R\in \rset$, set $U_R$ contains all branch vertices and vertices of $U\cup Y$ that lie on the boundary of $R$. 
Therefore, if we denote by $U'_R$ the set that contains all branch vertices and vertices of $U$ lying on the boundary of $R$, then it suffices to show that 
\begin{equation}
\label{eq: main}
\sum_{R\in \rset:\text{ } |U'_R|\ge \lambda/2}|U'_R|\le |U|\cdot\bigg(1+O\bigg(\frac{1}{\lambda}\bigg)\bigg). 
\end{equation}
This is because, for each $R\in \rset$, if $|U'_R|< \lambda/2$ while $|U_R|\ge \lambda$, then $|Y\cap U_R|\ge \lambda/2 \ge |U'_R|$ and so $|U_R|\le 2\cdot |Y\cap U_R|$, and since every vertex of $Y$ appears on the boundaries of at most two regions in $\rset$, we get that
$$\sum_{R\in \rset:\text{ } |U'_R|< \lambda/2,\text{ } |U_R|\ge \lambda}|U_R|\le \sum_{R\in \rset:\text{ } |U'_R|< \lambda/2,\text{ } |U_R|\ge \lambda}2\cdot|Y\cap U_R|\le O(|Y|).$$
Combined with Inequality~\ref{eq: main} and the above discussion, this completes the proof of Claim~\ref{clm: branch pts}.

The remainder of this section is dedicated to the proof of Inequality~\ref{eq: main}.
Using similar arguments in the proof of Claim~\ref{clm: ratio loss for contracting to portals}, we can show that it suffices to prove Inequality~\ref{eq: main} when no vertex of $U$ is a cut vertex of $H$.
In other words, when we traverse the outerface of graph $H$, every terminal in $U$ will be visited once, and so we get a circular ordering on terminals in $U$.

Denote $\lambda' \coloneqq \lambda/2$. We say that a region $R\in \rset$ is \emph{big} if $|U'_R|\ge \lambda'$, otherwise we say it is \emph{small}.
We need the following observation: if all regions in $\rset$ are big, then Claim~\ref{clm: branch pts} holds.

\begin{observation}
\label{obs: all big face}
Let $\hat \lambda>10$ be any integer. If for all $R\in \rset$, $|U'_R|\ge \hat \lambda$, then 
$$\sum_{R\in \rset}|U'_R|\le |U|\cdot\big(1+O(1/\hat \lambda)\big).$$
\end{observation}
\begin{proof}
Denote $U=\set{u_1,\ldots,u_r}$, where the terminals are indexed according to the circular ordering in which they appear on the outerface of $H$. 
We define a graph $W$ as follows. We start from the graph obtained by taking the union of all paths in $\pset$. We then suppress all degree-$2$ non-terminals. Finally, we add the cycle $(u_1,\ldots,u_r,u_1)$. 
Clearly, $W$ is a planar graph, and the planar drawing of $H$ naturally defines a drawing of $W$: start with the planar drawing of all paths in $\pset$ induced by the planar drawing of $H$, contracting degree-$2$ non-terminals, and finally draw every edge $(u_i,u_{i+1})$ along the boundary of the disc in which the one-hole instance $(H,U)$ lies in. Note that each region $R\in \rset$ corresponds to a face in the planar drawing of $W$, that we denote by $F_R$. 
Moreover, the vertices lying on the boundary of $F_R$ are exactly the vertices of $U'_R$.

Consider now the dual graph $W^*$ of $W$ with respect to the planar drawing defined above. Clearly, every node in $W^*$ corresponds to a region in $\rset\cup \set{R_{\infty}}$, where $R_{\infty}$ is the region outside the disc in which the one-hole instance $(H,U)$ lies in. We denote $V(W^*)=\set{v_R\mid R\in \rset}\cup \set{v_{\infty}}$.

On the one hand, for each $R\in \rset$, $|U'_R|$ is equal to the number of edges on the boundary of face $F_R$, which is then equal to the degree of vertex $v_R$ in $W^*$. Therefore, 
$$\sum_{R\in \rset}|U'_R|=\sum_{v\in V(W^*), v\ne v_{\infty}} \deg_{W^*}(v).$$
Recall that every region $R\in \rset$ satisfies that $|U'_R|\ge \hat \lambda$, so $\deg_{W^*}(v)\ge \hat \lambda$ for all $v\in V(W^*), v\ne v_{\infty}$.

On the other hand, since the paths in $\pset$ are well-structured and non-crossing, and we have suppressed all degree-$2$ vertices, it is easy to observe that the subgraph of $W^*$ induced by all vertices of $\set{v_R\mid R\in \rset}$ is a simple graph. In other words, all edges that have a parallel copy in $W^*$ must be incident to $v_{\infty}$.

Since the number of edges in $W^*$ incident to $v_{\infty}$ is $|U|$, if we subdivide every edge incident to $v_{\infty}$ by a new vertex, then the resulting graph, which we denote by $\hat W^*$, is a planar simple graph, and so $|E(\hat W^*)|\le 3\cdot |V(\hat W^*)|$. Therefore,
\[
|U|+2\cdot |U|+\sum_{v\in V(W^*), v\ne v_{\infty}}\deg_{W^*}(v)\le 2\cdot |E(\hat W^*)|\le 6\cdot |V(\hat W^*)|\le 6\cdot (|U|+|V(W^*)|),
\]
so $3|U|+(|V(W^*)|-1)\cdot \hat \lambda\le 6(|U|+|V(W^*)|)$, and so $|V(W^*)|\le (3|U|+\hat \lambda)/(\hat \lambda-6)\le O(|U|/\hat \lambda)$.

Altogether, we get that
\[
\sum_{R\in \rset}|U'_R|=\sum_{v\in V(W^*), v\ne v_{\infty}} \deg_{W^*}(v)=|U|+\sum_{v\in V(W^*), v\ne v_{\infty}} \deg_{W^*\setminus v_{\infty}}(v)
\le |U|+O(|U|/\hat \lambda).
\] 
\aftermath
\end{proof}

\noindent We now proceed to prove Inequality~\ref{eq: main} using \Cref{obs: all big face}.
Let $W$ be the plane graph defined in the proof of \Cref{obs: all big face}, and we say that graph $W$ is \emph{generated} by the set $\pset$ of paths. We prove the following observation.
\begin{observation}
\label{obs: path face}
Let $P$ be a path in $\pset$, let $F$ be a face, and let $C$ be the boundary cycle of $F$. Then either $P\cap C=\emptyset$, or the intersection between $P$ and $C$ is a subpath of both $P$ and $C$.
\end{observation}
\begin{proof}
Assume that $P\cap C\ne\emptyset$; and furthermore, $P\cap C$ contains at least two vertices (since otherwise a single vertex is a subpath of both $P$ and $C$, and we are done). Assume for contradiction that $P\cap C$ is not a subpath of $P$. It is easy to verify that there are two vertices $u,u'$, such that $u,u'\in V(P)\cap V(C)$, but every vertex in $P$ between $u$ and $u'$ does not belong to $C$. Denote by $P'$ the subpath of $P$ connecting $u$ to $u'$. Note that $u,u'$ separates $C$ into two path, that we denote by $C_1, C_2$. Assume without loss of generality that the region surrounded by $P'\cup C_1$ does not contain the outerface. Let $e$ be an edge of $C_1$. Since graph $W$ is generated by paths in $\pset$, edge $e$ must belong to some path $P''\in \pset$. However, since both endpoints of $P''$ lie outside of the region surrounded by $P'\cup C_1$, and since $C_1$ is a segment of a face, path $P''$ must contain two vertices of $P'$, and the subpath of $P''$ between these two vertices contains the edge $e$, which does not belong to $P'$. Therefore, paths $P'$ and $P''$ are not well-structured, a contradiction.
\end{proof}

Let $P$ be a path and let $F$ be a face, such that $P$ and the boundary cycle $C_F$ of $F$ intersect, and the intersection $P\cap C_F$ is a subpath of both $P$ and $C_F$. Let $u,u'$ be the endpoints of this subpath. We define $P_{\oplus F}$ as the path obtained from $P$ by replacing the subpath between $u$ and $u'$ with the other subpath of $C_F$ connecting $u$ to $u'$ that does not belong to $P$.

\newcommand{\bs}{\textsf{bs}}

Recall that we only need to prove Inequality \ref{eq: main} for $W$, where the left hand side $\sum_{R\in \rset:\text{ } |U'_R|\ge \lambda'}|U'_R|$, which we denote by $\bs(W)$, is the sum of the sizes of all big faces.
We will first iteratively modify $W$ until we are unable to do so, such that the value $\bs(W)$ never decreases. Then we will show that the value $\bs(\tilde W)$ of the resulting graph $\tilde W$ is bounded by $|U|\cdot(1+O(1/\lambda'))$ using \Cref{obs: all big face}.

We now describe the algorithm that iteratively modifies the graph $W$. Throughout, we maintain a plane graph $\hat W$, that is initialized to be $W$, and a set $\hat\pset$ of paths, that is initialized to be $\pset$. We will always ensure that $\hat\pset$ is a well-structured set of paths, and graph $\hat W$ is generated by $\hat \pset$. 
When the algorithm proceeds, the plane graph $\hat W$ evolves, and so does the set of faces in its planar drawing. We say that a face is big (small, resp.) iff its boundary contains at least (less than, resp.) $\lambda'$ vertices.

We say that a tuple $(e,F_1,F_2)$ \emph{critical}, iff (i) $e$ is an edge in $\hat W$, $F_1$ is a small face, and $F_2$ is a big face, such that $e$ is incident to $F_1$ and $F_2$; and (ii) no vertex of $F_1$ is incident to any other big face than $F_2$.
We say that a pair $(P,P')$ of paths in $\hat \pset$ is a \emph{blocking pair} for a critical tuple $(e,F_1,F_2)$, iff (i) $e\in E(P)$, $e\notin E(P')$; and (ii) the pair $P_{\oplus F_1}, P'$ of paths are not well-structured. 
We distinguish between the following cases.

\begin{figure}[h]
	\centering
	\subfigure[Before: Faces $F_1$ and $F_2$ share an edge $e$. Two paths containing $e$ in $\hat \pset$ are shown in green and red.]{\scalebox{0.11}{\includegraphics{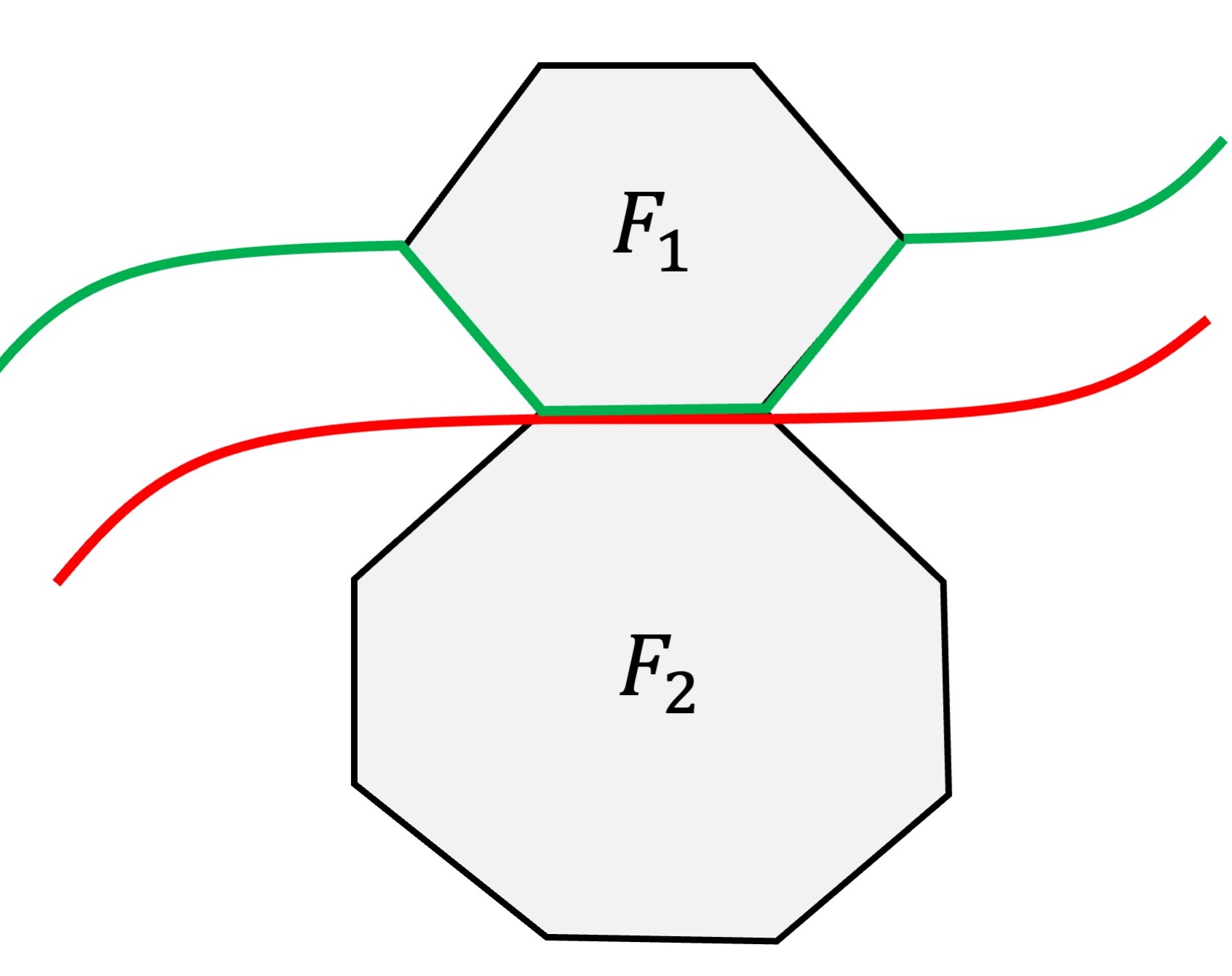}}}
	\hspace{1.0cm}
	\subfigure[After: Faces $F_1$ and $F_2$ are merged into $F$. The modified segment of two paths are shown in dashed lines.]{\scalebox{0.11}{\includegraphics{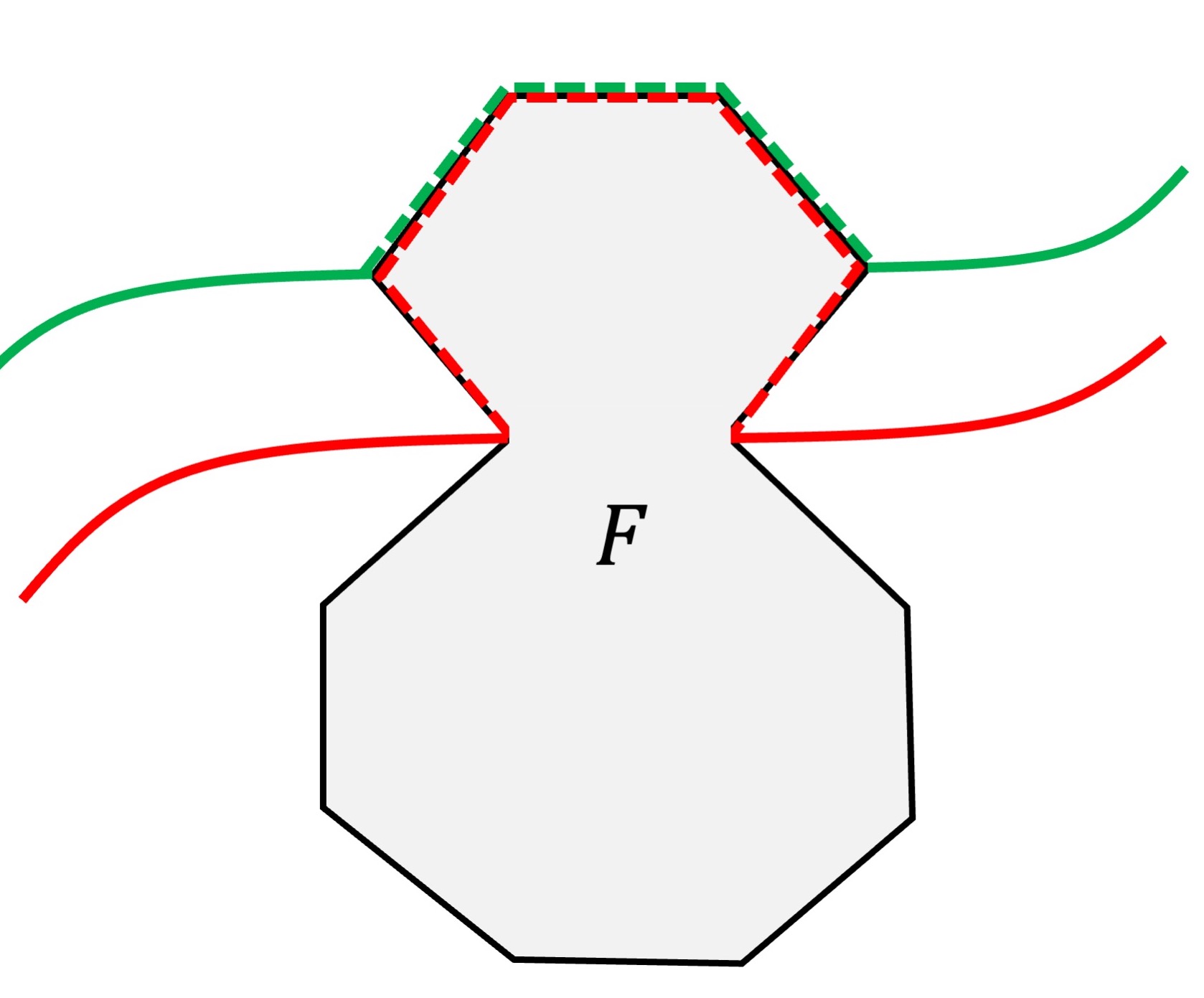}}}
	\caption{An illustration of graph and path modification in Case 1.\label{fig: case1}}
\end{figure}

\medskip
\noindent \textbf{Case 1:} There is a critical tuple $(e,F_1,F_2)$ with no blocking pairs, and the degree of at least one endpoint of $e$ is at least $4$.
In this case, we simply replace each path $P\in \hat\pset$ that contains the edge $e$ with path $P_{\oplus F_1}$, and then update $\hat W$ to be the graph generated by the resulting set $\hat\pset$ of paths. See \Cref{fig: case1}.

It is clear that the invariant that $\hat W$ is generated by $\hat{\pset}$ still holds in this case. Also, since there is no blocking pair for the critical tuple $(e,F_1,F_2)$, the resulting path set $\hat \pset$ is still well-structured.
Moreover, since no path in the resulting set $\hat \pset$ contains the edge $e$, the resulting graph $\hat W$ no longer contains the edge $e$, either.
Since the resulting graph $\hat W$ may not contain any new edge, the number of faces in $\hat W$ decreases by at least $1$ (as faces $F_1$ and $F_2$ are merged into a single face). We now show that the value $\bs(\hat W)$ does not decrease.

First, since the modification of paths in $\hat\pset$ only involves edges and vertices in $C_{F_1}$, the boundary cycle of face $F_1$, the graph $\hat W\setminus C_{F_1}$ remain unchanged, so every big face other than $F_2$ remain unchanged as well, and so is their contribution to $\bs(\hat W)$. Second, consider the resulting face $F$ into which $F_1$ and $F_2$ are merges. Note that $F$ contains all original vertices of $F_2$ as branch vertices. This is because all vertices of $F_2\setminus F_1$ remain unchanged, and since at least one of the endpoints of $e$ has degree at least $4$ in $\hat W$ before this iteration, this endpoint remain as branch vertices in the resulting graph $\hat W$, and the face $F$ contains at least one more branch vertex. Therefore, face $F$ contains at least the same number of branch vertices as the previous big face $F_2$. It follows that the value $\bs(\hat W)$ does not decrease.

\medskip
\noindent \textbf{Case 2:} There is a critical tuple $(e,F_1,F_2)$ with no blocking pairs, where $F_1$ contains more than $3$ vertices, and the degrees of both endpoints of $e$ are $3$.
In this case, we update the path set $\hat \pset$ and graph $\hat W$ in the same way as the previous case. Via similar arguments, we can show that the number of faces decreases by at least $1$, and the value $\bs(\hat W)$ does not decrease.

\medskip
\noindent \textbf{Case 3:} There is a critical tuple $(e,F_1,F_2)$ and a blocking pair $(P,P')$ for it.
Since paths $P$ and $P'$ are well-structured, but paths $P_{\oplus F_1}$ and $P'$ are not, from \Cref{obs: path face}, there must be two disjoint subpaths $P'_1, P'_2$ of $P'$, such that $P'_1=P\cap P'$ and $P'_2=C_{F_1}\cap P'$.
We first give both paths $P$ and $P'$ a direction, such that $P'_1$ appears before $P'_2$ on $P'$, and $P'_1$ appears before edge $e$ on $P$. Let $u$ be the last vertex of $P'_1$, let $v'$ be the first vertex of $P'_2$, and let $v$ be the first vertex of $C_{F_1}\cap P$ that appears on $P$.

\begin{figure}[h]
	\centering
	\subfigure[Before: Vertex $u$ is shown in brown. Paths $P,P'$ are shown in red, green respectively. Face $F'$ is shown in orange.]{\scalebox{0.11}{\includegraphics{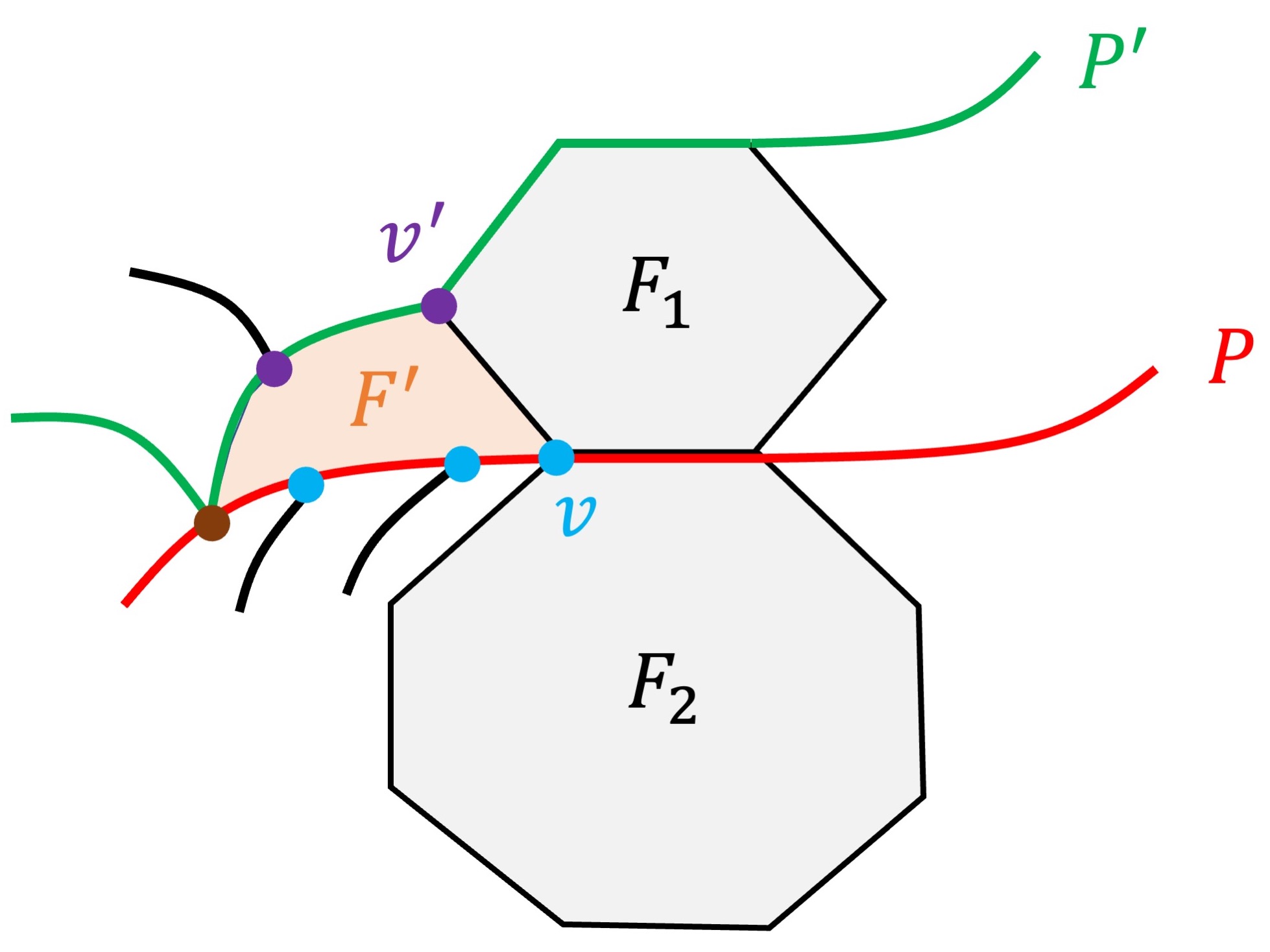}}\label{fig: case3_before}}
	\hspace{1.0cm}
	\subfigure[After: Face $F'$ is suppressed, vertices $v,v'$ are contracted into $v''$, and the two subpaths are identified.]{\scalebox{0.11}{\includegraphics{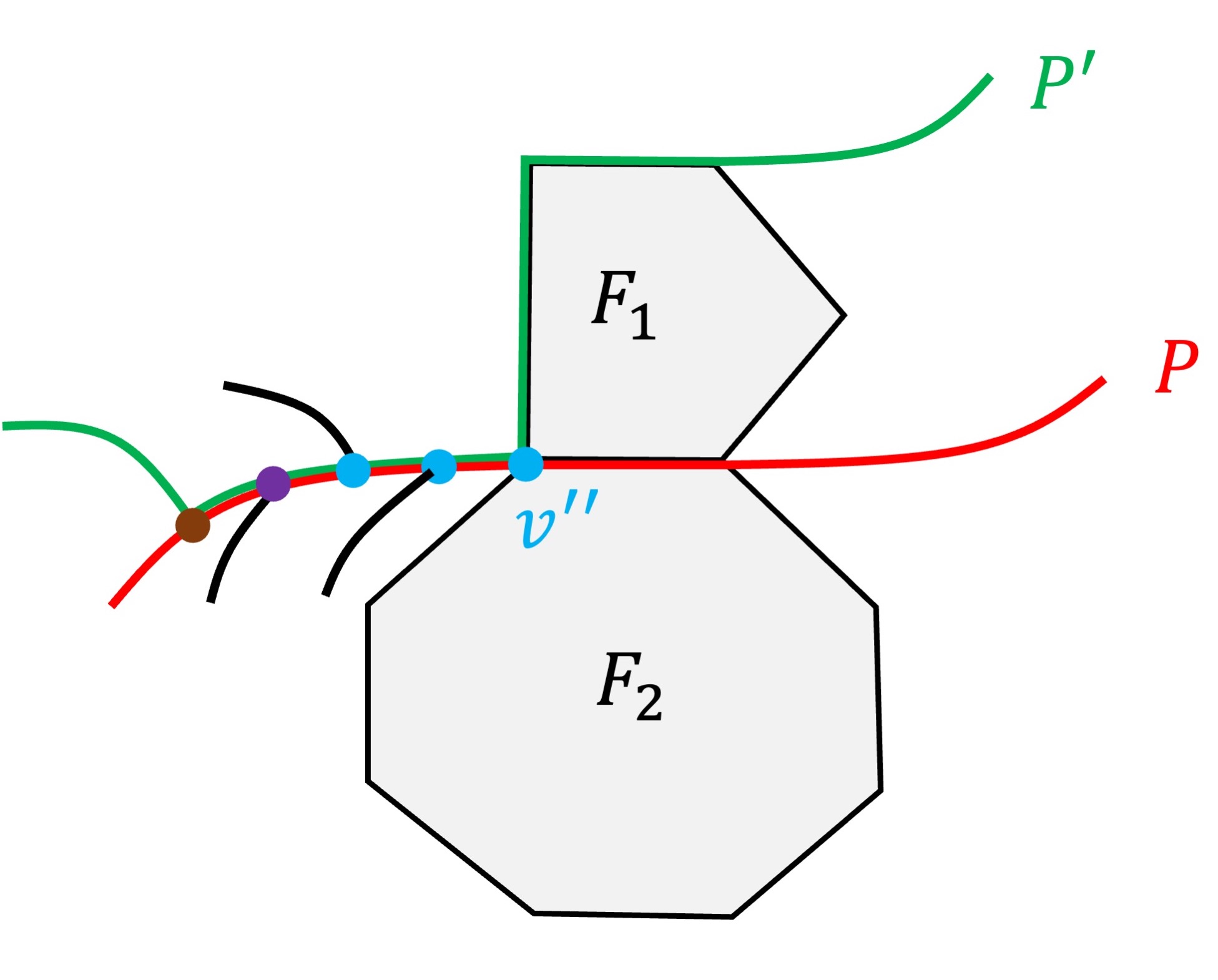}}\label{fig: case3_after}}
	\caption{An illustration of graph and path modification in Case 3.\label{fig: case3}}
\end{figure}

We first show that $v$ and $v'$ must be adjacent on $C_{F_1}$. Assume not, let $X$ be the segment of $C_{F_1}$ between $u$ and $v'$ that does not contain $e$, and let $x$ be an inner vertex of $X$. Since $\deg(x)\ge 3$, we let $e_x$ be an edge incident to $x$, such that $e_x\notin C_{F_1}$. Consider the region $R$ surrounded by (i) the subpath of $P$ between $u$ and $v$; (ii) the subpath of $P'$ between $u$ and $v'$; and (iii) path $X$. It is clear that $e_x$ must lie entirely in $R$. On the other hand, let $P_x$ be a path in $\hat\pset$ that contains the edge $e_x$, so both endpoints of $P_x$ lie outside $R$. Since paths in $\hat \pset$ are non-crossing and well-structured, path $P_x$ must exit region $R$ at $v$ and $v'$, but since $e_x\notin E(C_{F_1})$, the intersection between $P_x$ and $C_{F_1}$ is neither a subpath of $C_{F_1}$ nor a subpath of $P_x$, a contradiction to \Cref{obs: path face}.
Via similar arguments, we can show that no edge may lie inside the interior of region $R$. In other words, region $R$ is in fact a face, which we denote by $F'$ (see \Cref{fig: case3_before}).
Moreover, since vertices $v,v'$ are not incident to any other big faces, $F'$ is a small face.

We now ``suppress'' the face $F'$ as follows. We first contract the edge $(v,v')$ of $C_{F_1}$, while identifying vertices $v$ and $v'$ into a single vertex $v''$. We then ``identify'' the subpath of $P$ between $u$ and $v$ (which we denote by $\tilde P$) with the subpath of $P'$ between $u$ and $v'$ (which we denote by $\tilde P'$). Specifically, if originally $\tilde P=(u,y_1,\ldots,y_s,v)$ and $\tilde P'=(u,y'_1,\ldots,y'_t,v')$, then we replace these two paths with a new path $\tilde P''=(u,y_1,\ldots,y_s,y'_1,\ldots,y'_t,v'')$, and we do not modify the incident edges of any $y_i$ or $y'_j$ (see \Cref{fig: case3_after}). We update $\hat W$ to be the resulting graph after this step.

This face suppression naturally defines a way of modifying the paths in $\hat \pset$, as follows.
Denote by $C_{F'}$ the boundary cycle of face $F'$.
For every path $P\in \hat \pset$:
\begin{itemize}
\item if $P\cap C_{F'}=\emptyset$, then we do not modify it;
\item if $P\cap V(C_{F'})\subseteq \set{v,v'}$, then we let it contain the new vertex $v''$ at the same location;
\item if $P\cap C_{F'}$ is a subpath of $\tilde P$ or a subpath of $\tilde P'$, then we replace that subpath of $P$ with the corresponding subpath of $\tilde P''$.
\end{itemize}

It is easy to verify that the resulting set $\hat\pset$ is non-crossing and well-structured, and it still generates the resulting graph $\hat W$. 
Also, the number of faces in $\hat W$ decreases by $1$ in this case.
We now show that the value $\bs(\hat W)$ does not decrease.
Note that the degree of every vertex except for $v,v'$ does not change, and the degree of the new vertex $v''$ obtained from contracting $(v,v')$ has degree at least $3$ in the resulting graph, so all big faces remain unchanged, and so are their contribution to $\bs(\hat W)$.

\bigskip

We denote by $\tilde W$ the graph $\hat W$ when  none of the Cases 1-3 described above happens. We are then guaranteed that, for each small face $F$ in $\tilde W$, either 
\begin{itemize}
\item it does not share a vertex with any big faces; or
\item it contains exactly $3$ vertices, it shares a vertex with exactly one big face, and both endpoints of the edge that it shares with that big face has degree exactly $3$; or
\item it shares a vertex with at least two big faces (in this case we call it a \emph{bridge} face).
\end{itemize}

We call vertices that are shared by a bridge face and a big face \emph{bridge vertices}, and we call vertices that belong to at least two big faces \emph{interface vertices}. Clearly, bridge vertices and interface vertices must be branch vertices. Consider now any big face $F$, and let $V'_F$ be the set of its bridge vertices and interface vertices. We prove the following observation.

\begin{observation}
\label{obs: O(1) between interface}
Let $F$ be a big face and let $u,u'$ be a pair of vertices in $V'_F$ that appear consecutively on  $C_F$. That is, there is a subpath $Q$ of $C_F$ connecting $u$ to $u'$ that does not contain any other vertex of $V'_F$. Then the number of branch vertices that is an internal vertex of $Q$ is at most $2$.
\end{observation}
\begin{proof}
Consider any edge $e$ in path $Q$ that is not incident to $u$ or $u'$. Let $F'$ be the other face that $e$ is incident to, so $F'$ is a small face. Since both endpoints of $e$ are not in $V'_F$, face $F'$ do not share vertex with any other big faces. From the above discussion, face $F'$ has to contain exactly three vertices, and the degrees of both endpoints of $e$ are exactly $3$. Let $z_e$ be the other vertex of face $F'$. Note that, via similar arguments we can show that all internal vertices of $Q$ have degree exactly $3$. Therefore, the vertex $z_{e'}$ defined for every other edge $e'$ of $Q$ that is not incident to $u$ or $u'$ has to coincide with $z_e$. But if the number of branch vertices that is an internal vertex of $Q$ is greater than $2$, then there exists a vertex $u''\in V(Q)$ that is not adjacent to either $u$ or $u''$. Now the existence of edge $(z_e,u'')$ can be shown to cause a contradiction to the well-structuredness of $\hat \pset$, using similar arguments in the proof of \Cref{obs: path face}.
\end{proof}

Similarly, we can prove the following observation.

\begin{observation}
Let $F,F'$ be a pair of big faces, and let $\hat F, \hat F'$ be a pair of bridge faces, such that both $\hat F, \hat F'$ share vertices with both $F, F'$. Then if we denote by $R$ the region outside $F,F',\hat F, \hat F'$ surrounded by the boundaries of $F,F',\hat F, \hat F'$ that does not contain the outerface, then the boundary of $R$ contains at most $8$ bridge vertices.
\end{observation}

Consider now the dual graph $\tilde W^*$ of the resulting graph $\tilde W$. From similar arguments in the proof of \Cref{obs: all big face}, we know that in order to show 
$\sum_{R\in \rset}|U'_R|\le |U|\cdot\big(1+O(1/ \lambda')\big)$, it suffices to show that
$\sum_{v\in V(\tilde W^*), v\ne v_{\infty}} \deg_{\tilde W^*\setminus v_{\infty}}(v)
\le O(|U|/ \lambda')$.
We denote by $\check W$ the subgraph of $\tilde W^*$ induced by all nodes corresponding to big faces and bridge faces.
From the above two observations, we know that, it suffices to show that
$\sum_{v\in V(\check W)} \deg_{\check W}(v)
\le O(|U|/\lambda')$.

Let $\hat F$ be a bridge face. We denote by $F_1,\ldots, F_t$ the big faces that share a vertex with $\hat F$, where the faces are indexed according to the circular ordering in which they intersect with $\hat F$. Then, it is easy to see that, if we replace, for each bridge face $\hat F$, all edges incident to node $v_{\hat F}$ (the node in $\check W$ that corresponds to face $\hat F$) with edges $(v_{F_1},v_{F_2}),\ldots, (v_{F_t},v_{F_1})$, then the resulting graph $\check{W}$ is still a planar graph, with each every having at most one parallel copy. Using similar arguments in the proof of \Cref{obs: all big face}, we can show that $\sum_{v\in V(\check W)} \deg_{\check W}(v)
\le O(|U|/\lambda')$. This completes the proof of Claim~\ref{clm: branch pts}.

\subsection{Proof of Claim~\ref{clm: glueset_emulators}}
\label{apd: Proof of glueset_emulators}

Let $u,u'$ be terminals in $U$. We will show that $e^{-\eps}\cdot \dist_Z(u,u')\le \dist_{\hat H}(u,u')\le e^{\eps}\cdot\dist_Z(u,u')$. 

On the one hand, let $Q$ be the $u$-$u'$ shortest path in $\hat H$. We view path $Q$ as being directed from $u$ to $u'$. Let $\set{u_1,\ldots,u_k}$ be the set of all inner vertices of $Q$ that belongs to $V^*\cup Y$ (recall that $V^*$ is the set of branch vertices), where the vertices are indexed according to the order in which they appear on $Q$. Therefore, if we set $u_0=u$ and $u_{k+1}=u'$, then for each $0\le i\le k$, either one of $u_i,u_{i+1}$ is a branch vertex and so $\dist_Z(u_i,u_{i+1})=\dist_{\hat H}(u_i,u_{i+1})$, or $u_i,u_{i+1}$ are both vertices of $Y$ and belong to the same instance in $\hset$ and so $\dist_{\hat H}(u_i,u_{i+1})\ge e^{-\eps}\cdot\dist_{Z}(u_i,u_{i+1})$. Thus, if we set, for each $0\le i\le k$, $H_{R_i}$ to be the graph in $\hset$ that vertices $u_i,u_{i+1}$ belong to, then
\[
\begin{split}
\dist_{\hat H}(u,u') & = \sum_{0\le i\le k}\dist_{\hat H}(u_i,u_{i+1})
\ge \sum_{0\le i\le k}\dist_{H_{R_i}}(u_i,u_{i+1})\\
& \ge \sum_{0\le i\le k}e^{-\eps}\cdot \dist_{Z_{R_i}}(u_i,u_{i+1})
\ge \sum_{0\le i\le k}e^{-\eps}\cdot \dist_{Z}(u_i,u_{i+1})\ge e^{-\eps}\cdot \dist_{Z}(u,u').
\end{split}
\]

On the other hand, let $Q'$ be the $u$-$u'$ shortest path in $Z$. We view path $Q'$ as being directed from $u$ to $u'$. Let $\set{u'_1,\ldots,u'_k}$ be the set of all inner vertices of $Q'$ that belongs to $V^*\cup Y$ (recall that $V^*$ is the set of branch vertices), where the vertices are indexed according to the order in which they appear on $Q'$. Therefore, if we set $u'_0=u$ and $u'_{k+1}=u'$, then for each $0\le i\le k$, either one of $u'_i,u'_{i+1}$ is a branch vertex and so $\dist_Z(u'_i,u'_{i+1})=\dist_{\hat H}(u'_i,u'_{i+1})$, or $u'_i,u'_{i+1}$ are both vertices of $Y$ and belong to the same instance in $\hset$ and so $\dist_Z(u'_i,u'_{i+1})\ge e^{-\eps}\cdot\dist_{\hat H}(u'_i,u'_{i+1})$. Thus, if we set, for each $0\le i\le k$, $H_{R_i}$ to be the graph in $\hset$ that vertices $u'_i,u'_{i+1}$ belong to, then
\[
\begin{split}
\dist_Z(u,u') & = \sum_{0\le i\le k}\dist_Z(u'_i,u'_{i+1})
\ge \sum_{0\le i\le k}\dist_{Z_{R_i}}(u'_i,u'_{i+1})\\
& \ge \sum_{0\le i\le k}e^{-\eps}\cdot \dist_{H_{R_i}}(u'_i,u'_{i+1})
\ge \sum_{0\le i\le k}e^{-\eps}\cdot \dist_{\hat H}(u'_i,u'_{i+1})\ge e^{-\eps}\cdot \dist_{\hat H}(u,u').
\end{split}
\]

\subsection{Proof of Claim~\ref{clm: ratio loss for gluepathset}}
\label{apd: Proof of ratio loss for gluepathset}


We denote by $\ell$ the level that set $S$ belongs to.
We use the following simple observations.

\begin{observation}
	\label{obs: S' easy}
	For every pair $(u,u')$ with $u\in S$ and $u'\in S'$, $\dist_{H}(u,u')\ge \mu^{\ell+1}$. For every pair $(u,u')$ of terminals in $S'$ that do not belong to the same graph in $\hset$, $\dist_{H}(u,u')\ge \mu^{\ell+1}$.
\end{observation}
\begin{proof}
	From the construction of the collection $\sset$ and the definition of sets $S,S',S^*$, if $u\in S$ and $u'\in S'$, then $u,u'$ do not belong to the same $(\ell+1)$-level set, and so $\dist_H(u,u')>\mu^{\ell+1}$.
	Consider now a pair $u,u'$ of terminals in $S'$ that do not belong to the same graph in $\hset$. From the construction of the graphs in $\hset$, there must exist a pair $\hat u,\hat u'$ of terminals in $S$, such that the pairs $(\hat u,\hat u')$ and $(u, u')$ are crossing. Therefore, from Monge property,
	\[
	\dist_H(u,u')\ge \dist(u,\hat u)+\dist(u',\hat u')- \dist(\hat u,\hat u') \ge \mu^{\ell+1}+\mu^{\ell+1}-2r\mu^{\ell}>\mu^{\ell+1},
	\]
	where we have used the fact (from Observation~\ref{obs: diameter}) that $\dist(\hat u,\hat u')\le 2r\mu^{\ell}$.
\end{proof}

Let $u,u'$ be terminals in $U$. 
We will show that $\dist_{\hat H}(u,u')\le \dist_H(u,u')\le e^{\eps_r}\cdot\dist_{\hat H}(u,u')$.
If vertices $u,u'$ belong to the same instance in $\hset$, then since the instances in $\hset$ is obtained by cutting along shortest paths in $H$, it is easy to see that $\dist_H(u,u')=\dist_{\hat H}(u,u')$. Therefore, we assume from now on that that terminals $u,u'$ do not belong to the same instance in $\hset$.
We denote by $Y$ the set of all vertices that belongs to more than one instances in $\hset$.

Recall that, in the procedure $\cutpath$, we have sliced $H$ open along a set of shortest paths in $H$.
Let $\rset$ be the collection of regions (of $H$) that we get.
Recall that each instance in $\hset$ corresponds to a region in $\rset$. 
We say that an instance $(H_R,U_R)\in \hset$ is a \EMPH{regular} instance if the corresponding region $R$ is surrounded by (i) a contiguous segment of the outer-boundary of $H$ and (ii) the image of a single path in $\pset$.
Since the paths are well-structured, when we consider a $u$-$u'$ shortest path $Q$ in $H$, we can assume that, for each regular instance $(H_R,U_R)\in \hset$ with $u,u'\notin V(H_R)$, the intersection between $Q$ and $H_R$ is a subpath of the path in $\pset$ that surrounds the region $R$ and both endpoints of this subpaths are branch vertices.

Consider now the $u$-$u'$ shortest path $Q$ in $H$. Assume that $u\in H_{R}$ and $u'\in H_{R'}$. 
We view path $Q$ as being directed from $u$ to $u'$. Let $v$ be the last vertex of $Q$ that belongs to $H_R$, and let $v'$ be the first vertex of $Q$ after $v$ that belongs to $H_{R'}$. We distinguish between the following cases.

\medskip
\noindent\textbf{Case 1. $v\ne v'$.} 
From the construction of graph $\hat H$ and the above discussion, it is easy to verify that the entire path $Q$ is also contained in graph $\hat H$, so $\dist_{\hat H}(u,u')\le \dist_{H}(u,u')$. On the other hand, it is easy to verify that any shortest path in $\hat H$ connecting $u$ to $u'$ is also entirely contained in $H$, so 
$\dist_{\hat H}(u,u')\ge \dist_{H}(u,u')$. Therefore, $\dist_{\hat H}(u,u')= \dist_{H}(u,u')$.


\medskip
\noindent
\textbf{Case 2. $v=v'$.} 
This means that path $Q$ only touches two regions, $R$ and $R'$. If one of $u,u'$ belongs to set $S'$, then from Observation \ref{obs: S' easy} and the fact (from Observation~\ref{obs: diameter}) that the boundary path of $R$ and $R'$ have total length at most $2r\mu^{\ell}$, it is easy to verify that \[\dist_{H}(u,u')\le \dist_{\hat H}(u,u')\le (1+O(1/r))\cdot\dist_{H}(u,u')\le e^{\eps_r}\cdot\dist_{H}(u,u').\]
If both $u,u'$ belong to $S$, then from the construction of $\hat H$, $\dist_{H}(u,u')\le \dist_{\hat H}(u,u')$.
It remains to consider the case where at least one of $u,u'$ belongs to set $S^*$.
Assume without loss of generality that $u\in S^*$. Since the set $Y\cap U_R$ contains an $\eps_r$-cover of $u$ on the boundary path of $R$, there exists a vertex $\hat v\in Y\cap U_R$, such that $\dist_H(u,\hat v)+\dist_H(\hat v,v)\le e^{\eps_r}\cdot \dist_H(u,v)$.
In this case we denote by $v_1$ the copy of $v$ in $H_R$ and by $v_2$ the copy of $v$ in $H_{R'}$, then
\[
\begin{split}
\dist_{H}(u,u')\le \dist_{\hat H}(u,u')\le & \text{ }
\dist_{\hat H}(u,\hat v)+\dist_{\hat H}(\hat v,v_2)+\dist_{\hat H}(v_2,u')\\
\le & \text{ }\dist_{H}(u,\hat v)+\dist_{H}(\hat v,v_2)+\dist_{H}(v',u')\\
\le & \text{ } e^{\eps_r}\cdot \dist_{H}(u,v)+\dist_{H}(v,u') \le e^{\eps_r}\cdot \dist_{H}(u,u').
\end{split}
\]

\section{Missing Proofs in \Cref{sec: general graph}}

\subsection{Complete Description of Procedures $\cutpath_h$ and $\gluepath_h$}
\label{apd: cut and glue}

\paragraph{Splitting.}
The input to procedure \emph{$\cutpath_h$} consists of
\begin{itemize}
\item an $h$-hole instance $(H,U)$; 
\item a path $P$ connecting a pair of its terminals that lie on different holes; and
\item a set $Y$ of vertices in $P$ that contains both endpoints of $P$.
\end{itemize}
The output of procedure $\cutpath_h$ is an $(h-1)$-hole instance $(\tilde H, \tilde U)$ that is constructed as follows.
Let $u,u'$ be the endpoints of $P$.
We denote by $\gamma$ the curve representing the image of path $P$ in $H$, and view it as being directed from $u$ to $u'$. 
For each $v\in V(P)$, we define $\delta_1(v)$ ($\delta_2(v)$, resp.) as the set of all incident edges of $v$ in graph $H$, whose image lie on the left (right, resp.) side of $\gamma$, as we traverse along $\gamma$ from $u$ to $u'$. 
We now modify the graph $H$ as follows. 
Replace each vertex $v\in V(P)$ by two new vertices $v_1$ and $v_2$, where $v_1$ is incident to all edges in $\delta_1(v)$ and $v_2$ is incident to all edges in $\delta_2(v)$. 
Then we add, for each edge $(v,v')$ of path $P$, an edge $(v_1,v'_1)$ and an edge $(v_2,v'_2)$. The resulting graph is denoted by $\tilde H$.

We naturally construct a planar drawing of graph $\tilde H$, as follows. 
We start from the drawing $\phi$ associated with instance $(H,U)$. We first erase from it the images of all vertices and edges of $P$.
Denote by $\alpha$ ($\alpha'$, resp.) the hole in $\phi$ whose boundary contains the image of $u$ ($u'$, resp.). Let $S$ be a thin strip around the curve $\gamma$. 
We draw the new vertices $u_1,u_2$ at the intersections of $S$ and the boundary of hole $\alpha$, where $u_1$ lies on the left of $\gamma$ and $u_2$ lies on the right of $\gamma$. Similarly, we draw the new vertices $u'_1,u'_2$ at the intersections of $S$ and the boundary of hole $\alpha'$, where $u'_1$ lies on the left of $\gamma$ and $u'_2$ lies on the right of $\gamma$. 
Now for every other vertex $v\in V(P)$, we draw the new vertex $v_1$ ($v_2$, resp.) on the boundary of $S$ just to the left (right, resp.) of the old image of $v$ in $\phi$. The images of other vertices remain the same as in $\phi$. 
For each vertex $v\in V(P)$ and each edge $e\in \delta_1(v)$ ($\delta_2(v)$, resp.), we slightly modify the image of $e$ to make it direct to $v_1$ ($v_2$, resp.). 
Lastly, for each edge $(v,v')\in P$, we draw the image of new edge $(v_1,v'_1)$ ($(v_2,v'_2)$, resp.) as the segment of the boundary of strip $S$ between the points representing the images of $v_1,v'_1$ ($(v_2,v'_2)$, resp.). This completes the construction of a planar drawing of $\tilde H$, that we denote by $\tilde \phi$. 
See \Cref{fig: splitting h-hole before} and \Cref{fig: splitting h-hole after} for an illustration.

We now define $\tilde U$ to be the set obtained from $U$ by replacing 
for each vertex $y\in Y$, two new vertices $y_1$ and $y_2$ (since such a vertex $y$ belongs to path $P$), so $|\tilde U|=|U|+2|Y|$.
The instance $(\tilde H,\tilde U)$ is the output of procedure $\cutpath_h$. We now show that it is indeed an $(h-1)$-hole instance.

We define area $\beta=\alpha\cup S\cup \alpha'$.
It is easy to observe that no vertices or edges are drawn inside the interior of area $\beta$, and if we denote by $U(\alpha)$ the set of terminals in $H$ that lie on the boundary of $\alpha$, and define set $U(\alpha')$ similarly, then in $\tilde H$, the boundary of $\beta$ contains the images of terminals in $(U(\alpha)\setminus \set{u})\cup (U(\alpha')\setminus \set{u'})\cup \set{y_1,y_2\mid y\in Y}$.
Therefore
$(\tilde H,\tilde U)$ is a valid $(h-1)$-hole instance.



\paragraph{Gluing.}
We next describe the procedure \emph{$\gluepath_{h}$}, which is intuitively a reverse process of procedure called $\cutpath_{h}$.
Assume that we have applied the procedure $\cutpath_h$ to some $h$-hole instance $(H,U)$, some path $P$ connecting a pair $u,u'$ of terminals in $U$ that lie on holes $\alpha,\alpha'$ respectively, and a subset $Y$ of vertices in $P$. Let $(\tilde H,\tilde U)$ be the $(h-1)$-hole instance that the procedure $\cutpath_h$ outputs, where holes $\alpha,
\alpha'$ are merged into hole $\beta$.
We then denote, for each $y\in Y$, by $y^1$ and $y^2$ the two terminals in $\tilde U$ obtained by splitting $y$. 
The procedure $\gluepath_{h}$ takes as input an  emulator $(\tilde H',\tilde U)$ for instance $(\tilde H,\tilde U)$, and works as follows.

We let graph $H'$ be obtained from graph $\tilde H'$ by identifying,
for each $y\in Y$, vertex $y^1$ with vertex $y^2$ (and name the obtained vertex $y$). Denote $\tilde Y=\set{y^1,y^2\mid y\in Y}$. 
We then set $U'=(\tilde U\setminus\tilde Y)\cup \set{u,u'}$. Clearly, $U'=U$. The output of algorithm $\gluepath_h$ is instance $(H',U)$.

We associate with instance $(H',U)$ a planar drawing with terminals of $U$ drawn on the boundary of $h$ holes as follows.
We denote by $\gamma$ the boundary segment of hole $\beta$ from $u^2$ to $u^1$ that does not contain any other vertex of $\tilde Y$, and denote by $\gamma'$ the boundary segment of hole $\beta$ from $(u')^1$ to $(u')^2$ that does not contain any other terminal of $\tilde Y$. 
We now compute,
for each $y\in Y$, a curve $\gamma_y$ connecting $y^1$ to $y^2$, such that the curves $\set{\gamma_y\mid y\in Y}$ all lie in hole $\beta$ and are mutually disjoint. 
We now move, for each $y\in Y$, the images of $y^1$ and $y^2$ along the curve $\gamma_y$ towards each other until they are identified. Now $\gamma$ becomes a closed curve that surrounds a region which does not contain the image of any vertices or edges in its interior. We designate this region by hole $\alpha$. We define hole $\alpha'$ for the closed curve $\gamma'$ similarly. It is easy to verify that all terminals of $U'$ that previously lied on the boundary of hole $\beta$ now lie on the boundary of either hole $\alpha$ or hole $\alpha'$. 
See \Cref{fig: gluepath_2} for an illustration. Therefore, $(H',U)$ is a valid $h$-hole instance, and it is easy to verify that instance $(H',U)$ is aligned with instance $(H,U)$.

\subsection{Proof of Claim~\ref{clm: split glue h holes}}
\label{apd: Proof of split glue h holes}

For convenience, we rename the selected terminals $u,u'$ by $\hat u, \hat u'$, respectively. Throughout the proof, we will use
$u,u'$ to denote some pair of terminals in $U$, and we will show that $e^{-\eps'}\cdot\dist_{Z}(u,u')\le \dist_{\hat H}(u,u')\le e^{\eps'}\cdot\dist_{Z}(u,u')$.

On the one hand, let $Q$ be the shortest path in $\hat H$ connecting $u$ to $u'$. We view the path $Q$ as being directed from $u$ to $u'$.
Recall that in graph $\hat H$, for each vertex $y\in Y$, we have denoted by $\delta_1(y)$ the incident edges of $y$ that lie on one side of path $P$, and denote by $\delta_2(y)$ the incident edges of $y$ that lie on the other side of path $P$.
We denote $E_1=\bigcup_{y\in Y}\delta_1(y)$ and $E_2=\bigcup_{y\in Y}\delta_2(y)$.  
If either $E(Q)\cap E_1=\varnothing$ or $E(Q)\cap E_2=\varnothing$ holds, then it is immediate to verify that path $Q$ is entirely contained in graph $\tilde H$. Since $(\tilde Z,\tilde U)$ is an $\eps$-emulator for instance $(\tilde H,\tilde U)$, we get that
\[\dist_{\hat H}(u,u')=\dist_{\tilde H}(u,u')\ge e^{-\eps}\cdot\dist_{\tilde Z}(u,u')\ge e^{-\eps}\cdot\dist_{Z}(u,u').\]
Assume now that $E(Q')\cap E_1\ne \varnothing$ or $E(Q')\cap E_2\ne \varnothing$. 
Recall that graph $\hat H$ contains two copies $P_1,P_2$ of path $P$ that corresponds to the sides of $E_1, E_2$, respectively.
We can assume without loss of generality that path $Q$ is the concatenation of (i) a path $Q_1$ connecting $u$ to some vertex $x_1\in V(P_1)$, that is internally disjoint from $P_1$; (ii) a subpath $P'_1$ of $P_1$ connecting $x_1$ to some vertex $y\in Y$; (iii) a subpath $P'_2$ of $P_2$ connecting $y$ to some vertex $x'_2\in V(P_2)$; and (iv) a path $Q'_2$ connecting $x'_2$ to some vertex $u'$, that is internally disjoint from $P_2$. Recall that $(\tilde Z,\tilde U)$ is an $\eps$-emulator for instance $(\tilde H,\tilde U)$, and instance $(Z,U)$ is obtained by applying the procedure $\gluepath_h$ to instance $(\tilde Z,\tilde U)$. We denote by $y_1,y_2$ the copies of $y$ in graph $\tilde H$, where $y_1\in V(P_1)$ and $y_2\in V(P_2)$. Then
\begin{align*}
\dist_{\hat H}(u,u') &= \dist_{\hat H}(u,x_1)+ \dist_{P_1}(x_1,y)+\dist_{P_2}(y,x'_2)+\dist_{\hat H}(x'_2,u)\\
&\ge \dist_{\tilde H}(u,x_1)+ \dist_{\tilde H}(x_1,y_1)+\dist_{\tilde H}(y_2,x'_2)+\dist_{\tilde Z}(x'_2,u)\\
&\ge e^{-\eps}\cdot (\dist_{\tilde Z}(u,x_1)+ \dist_{\tilde Z}(x_1,y_1)+\dist_{\tilde Z}(y_2,x'_2)+\dist_{\tilde Z}(x'_2,u))\\
&\ge e^{-\eps}\cdot (\dist_{Z}(u,x_1)+ \dist_{Z}(x_1,y)+\dist_{Z}(y,x'_2)+\dist_{Z}(x'_2,u))\\
&\ge e^{-\eps}\cdot \dist_{Z}(u,u').
\end{align*}

On the other hand, let $Q'$ be the shortest path in $Z$ connecting $u$ to $u'$. We view the path $Q'$ as being directed from $u$ to $u'$. Via similar analysis, we can easily show that, if $Q'$ does not contain vertices of $Y$, then
\[\dist_{Z}(u,u')=\dist_{\tilde Z}(u,u')\ge e^{-\eps}\cdot\dist_{\tilde H}(u,u')\ge e^{-\eps}\cdot\dist_{\hat H}(u,u').\]
We assume from on now that $Q'$ contains some vertices of $Y$. 
In graph $\tilde Z$, we denote by $\tilde E_1$ the set of edges incident to some vertex of $Y_1=\set{y_1\mid y\in Y}$, and define set $\tilde E_2$ for set $Y_2=\set{y_2\mid y\in Y}$ similarly.
Let $y^1\ldots,y^r$ be the vertices of $Y\cap V(Q)$, where the vertices are indexed according to their appearance on $Q$. For each $0\le j\le r$, we denote by $Q_j$ the subpath of $Q$ between vertices $y^j$ and $y^{j+1}$ (where we set $y^0=u$ and $y^{r+1}=u'$). 
For each $0\le j\le r$, we set $a(j)$ to be $1$ ($2$, resp.) if the first edge of $Q_j$ belongs to $\tilde E_1$ ($\tilde E_2$, resp.), and set set $b(j)$ to be $1$ ($2$, resp.) if the last edge of $Q_j$ belongs to $\tilde E_1$ ($\tilde E_2$, resp.).
Since $(\tilde Z,\tilde U)$ is an $\eps$-emulator for instance $(\tilde H,\tilde U)$, and instance $(Z,U)$ is obtained by applying the procedure $\gluepath_h$ to instance $(\tilde Z,\tilde U)$, we get that
\begin{align*}
\dist_{Z}(u,u') & = \sum_{0\le j\le r}\dist_{Z}(y^j,y^{j+1}) 
    = \sum_{0\le j\le r}\dist_{\tilde Z}(y^j_{a(j)},y^{j+1}_{b(j)}) \\
    &\ge \sum_{0\le j\le r} e^{-\eps}\cdot \dist_{\tilde H}(y^j_{a(j)},y^{j+1}_{b(j)})
    \ge  \sum_{0\le j\le r} e^{-\eps}\cdot\dist_{\hat H}(y^j,y^{j+1})
    \ge e^{-\eps}\cdot\dist_{\hat H}(u,u').
\end{align*}
Altogether, we get that
$e^{-\eps'}\cdot\dist_{Z}(u,u')\le \dist_{\hat H}(u,u')\le e^{\eps'}\cdot\dist_{Z}(u,u')$.

\subsection{Proof of Claim~\ref{clm: ratio loss for glue h holes}}
\label{apd: Proof of ratio loss for glue h holes}

For convenience, we rename the selected terminals $u,u'$ by $\hat u, \hat u'$, respectively. Throughout the proof, we will use
$u,u'$ to denote some pair of terminals in $U$, and we will show that $e^{-\eps'}\cdot\dist_{\hat H}(u,u')\le \dist_H(u,u')\le \dist_{\hat H}(u,u')$.

On the one hand, let $Q$ be a shortest path in $H$ connecting $\hat u$ to $\hat u'$. We view $Q$ as being directed from $u$ to $u'$.
If $V(Q)\cap V(P)= \varnothing$, then it is immediate to verify that path $Q$ is entirely contained in graph $\hat H$, so $\dist_{\hat H}(u,u')\le \dist_{H}(u,u')$.
Assume now that $V(Q)\cap V(P)\ne \varnothing$.
Since $Q$ and $P$ are shortest paths in $H$, $Q\cap P$ is a subpath of both $Q$ and $P$.
Let $v,v'$ be the endpoints of this path where $v$ is closer to $u$ and $v'$ is closer to $u'$ on $Q$ (note that it is possible that $v=v'$).
Since set $Y$ contains an $\eps'$-cover of $u$ on $P$, there exists some vertex $y\in Y$, such that $\dist_H(u,y)+\dist_H(y,v)\le e^\eps\cdot \dist_H(u,v)$; and similarly since set $Y$ contains an $\eps'$-cover of $u'$ on $Q$, there exists some vertex $y'\in Y$, such that $\dist_H(u',y')+\dist_H(y',v')\le e^\eps\cdot \dist_H(u',v')$. From the construction of graph $\hat H$, we get that
\[
\begin{split}
\dist_{H}(u,u') = & \text{ } \dist_{H}(u,v)+\dist_{H}(v,v')+\dist_{H}(u',v')\\
\ge & \text{ } e^{-\eps'}\cdot(\dist_H(u,y)+\dist_H(y,v))+\dist_{H}(v,v')+e^{-\eps}\cdot(\dist_H(u',y')+\dist_H(y',v'))\\
\ge & \text{ } 
e^{-\eps'}\cdot(\dist_{\hat H}(u,y)+\dist_{\hat H}(y,v)+\dist_{\hat H}(v,v')+\dist_{\hat H}(u',y')+\dist_{\hat H}(y',v'))\\
\ge & \text{ }  e^{-\eps}\cdot \dist_{\hat H}(u,u').
\end{split}
\]

On the other hand, let $Q'$ be a shortest path in $\hat H$ connecting $\hat u$ to $\hat u'$. We view $Q'$ as being directed from $u$ to $u'$.
Recall that in graph $H$, for each vertex $y\in Y$, we have denoted by $\delta_1(y)$ the incident edges of $y$ that lie on one side of path $P$, and denote by $\delta_2(y)$ the incident edges of $y$ that lie on the other side of path $P$.
We denote $E_1=\bigcup_{y\in Y}\delta_1(y)$ and $E_2=\bigcup_{y\in Y}\delta_2(y)$.  
If either $E(Q')\cap E_1=\varnothing$ or $E(Q')\cap E_2=\varnothing$ holds, then it is immediate to verify that path $Q'$ is entirely contained in graph $H$, so $\dist_{H}(u,u')\le \dist_{\hat H}(u,u')$.
Assume now that $E(Q')\cap E_1\ne \varnothing$ or $E(Q')\cap E_2\ne \varnothing$. 
Recall that graph $\hat H$ contains two copies $P_1,P_2$ of path $P$ that corresponds to the sides of $E_1, E_2$, respectively.
We can assume without loss of generality that path $Q'$ is the concatenation of (i) a path $Q'_1$ connecting $u$ to some vertex $x_1\in V(P_1)$, that is internally disjoint from $P_1$; (ii) a subpath $P'_1$ of $P_1$ connecting $x_1$ to some vertex $y\in Y$; (iii) a subpath $P'_2$ of $P_2$ connecting $y$ to some vertex $x'_2\in V(P_2)$; and (iv) a path $Q'_2$ connecting $x'_2$ to some vertex $u'$, that is internally disjoint from $P_2$.
Let $x$ be the original copy of $x_1$ in graph $H$, and let $x'$ be the original copy of $x_1$ in graph $H$. From the construction of graph $\hat H$, we get that
\[
\begin{split}
\dist_{\hat H}(u,u') = & \text{ } \dist_{\hat H}(u,x_1)+\dist_{P_1}(x_1,y)+\dist_{P_2}(y,x'_2)+\dist_{\hat H}(u',x'_2)\\
\ge & \text{ } \dist_{H}(u,x)+\dist_{H}(x,x')+\dist_{H}(u',x') \ge \dist_{H}(u,u').
\end{split}
\]

\subsection{Proof of \Cref{L:r-division}}
\label{apd: Proof of L:r-division}

	Similar to Frederickson~\cite{fre-faspp-1987} and Klein-Mozes-Sommer~\cite{kms-srsdp-2013}, we recursively find balanced cycle separators to subdivide the input graph.
	To control the number vertices, boundary vertices, holes, and terminals within each piece simultaneously, we ask the cycle separator to balance these quantities in rounds.  Specifically, at recursive level $\ell$:
	\begin{itemize} \itemsep=0pt
		\item If $\ell \bmod  4 = 0$, balance the vertices.
		\item If $\ell \bmod  4 = 1$, balance the boundary vertices.
		\item If $\ell \bmod  4 = 2$, balance the holes by inserting one \emph{supernode} per hole.
		\item If $\ell \bmod  4 = 3$, balance the terminals.
	\end{itemize}
	We terminate the recursion four rounds after a piece has size at most $r$.  
	The depth of the recursion tree is $\log(n/r)$, and a similar analysis as in Klein-Mozes-Sommer~\cite{kms-srsdp-2013} shows that the number of terminals within each piece is $O(kr/n)$.

\end{document}